\documentclass{apa}
\usepackage{apacite}
\usepackage{amsmath}
\usepackage{amssymb}
  \usepackage{graphicx}

\newcommand{\cIN}[1]{%
\ensuremath{%
		\vec{c}^{IN}_{#1}
}%
}

\newcommand{\fIN}[1]{%
\ensuremath{%
		\vec{f}^{IN}_{#1}
}%
}

\newcommand{\Lk}{\ensuremath{\mathbf{L}^{\textrm{\scriptsize{-1}}}_{\textrm{k}}}}

\newcommand{\taustar}{\ensuremath{\overset{*}{\tau}}}

\newcommand{\ftilde}{\ensuremath{\tilde{f}}}

\begin{document}
  \author{Marc W.~Howard}
  
  \title{Formal models of memory based on temporally-varying representations}
\leftheader{Formal models of Memory}
\shorttitle{Formal Models of Memory}
\rightheader{Formal models of Memory}
\affiliation{Boston University}
\note{In press,  In F. G. Ashby, H. Colonius, \& E. Dzhafarov (Eds.), The new
handbook of mathematical psychology, Volume 3. Cambridge University Press.}
 	\abstract{
The idea that memory behavior relies on a gradually-changing internal state
has a long history in mathematical psychology.  This chapter traces this line
of thought from statistical learning theory in the 1950s, through distributed
memory models in the latter part of the 20th century and early part of the
21st century through to modern models based on a scale-invariant temporal
history. We discuss the neural phenomena consistent with this form of
representation and sketch the kinds of cognitive models that can be
constructed using it and connections with formal models of various memory
tasks. 
}

\maketitle

Human babies, while adorable, are remarkably incompetent.  They know
essentially no facts about the world and are unable to perform any but the
simplest motor actions, and perform very poorly on behavioral assays of memory.  
Memory researchers evaluate  memory in adults with a variety of behavioral paradigms,
such as cued recall, in which the participant is given a series of pairs,
e.g., \textsc{absence-hollow}, \textsc{pupil-river}, \textsc{campaign-helmet}.
The participants' task is to produce the correct associate when given a cue
word.  For instance, after being probed with \textsc{pupil} the correct
response is \textsc{river}.  After being presented with a list of words for a
cued recall test, a human baby is more likely to emit curdled milk than a
correct response.   Over the course of a lifetime, normally-developing humans
learn many facts about their world, acquire complicated motor skills and can
bring to mind vivid recollections of many events from their lives.  Because
all of these abilities must be learned, they can be understood as forms of
memory.   

Viewed in this light, the task of a memory theorist seems daunting.  How
can one possibly construct a theory that can make sense of the ability to
recall that Paris is the capitol of France, the ability to ride a bike without
falling over, \emph{and} the ability to vividly remember a birthday party well enough
to bring a smile to one's face after decades?  The strategy taken by cognitive
neuroscientists in the latter part of the 20th century (and continuing to the
present day) is to carve up the set of abilities and skills that differentiate
a baby from an adult into different ``kinds'' of memory, each associated with
distinct parts of the brain.  For instance, many memory researchers would
say that retrieving facts about the world depends on semantic memory,
being able to ride a bicycle is a consequence of implicit memory and
vivid recollection of specific events from ones life relies on episodic
memory.  This strategy of dividing learning and memory phenomena into
different ``kinds of memory'' has been extremely productive. 
However, throughout the history of psychology, there has been an urge towards
developing unified theories of learning and memory.

\subsection{Associations in the mind and brain}
Radical behaviorists (most famously B.~F.~Skinner) attempted to understand the rich
repertoire of memory phenomena as special cases of stimulus-response
associations.  Pavlov's dogs learned to associate the sound of a bell with the
delivery of food,  so that the sound of the bell by itself leads to an overt
response (salivation).  Experimentalists learned that animals (in particular
rats and pigeons) can be trained to perform complex sequences of behaviors in
response to appropriate training experiences.  According to behaviorists' conception
of learning, even complex behaviors could be described as complex chains of
associations.  

Mathematical psychologists have developed formal models  of association to
provide quantitative models of behavior in a variety of experimental
paradigms.  Early work focused on animal conditioning experiments.  In this
case the behavioral measure is typically a scalar value that describes the
probability or magnitude of a conditioned response; for instance, the amount
of saliva produced by Pavlov's dog (or, more typically, the proportion of time
the animal spends freezing in a fear conditioning experiment).  But later work
also applied similar ideas to memory experiments with humans using lists of
words as stimuli.  In the cued recall task described above, it is
straightforward to write down a model that constructs simple associations
between neural representations of the words (e.g., associate \textsc{absence}
to \textsc{hollow}) such that probing the memory with the stimulus
\textsc{absence} causes a pattern like \textsc{hollow} to be produced as an
output.  These models can produce many distinct responses in responses to many
different cues.

Associations  can be understood neurally as a consequence of changes in the
connection strength between neurons.  The mammalian brain contains a great
number of specialized cells called neurons.  
Neurons are known to communicate
information between one another by means of their electrical activity.   The
connections between individual neurons are referred to as synapses.
The
strength of synapses can be modified by experience.  These facts are
sufficient to write down a very crude neural model of Pavlovian conditioning.
If one identifies the set of neurons that changes its firing in response to
the sound of the bell, and the set of neurons responsible for salivation,
one could in principle understand the association learned by Pavlov's dog as
an increase in the strength of the synapses connecting the ``bell''  
neurons to the ``drool'' neurons.  These assumptions can be formalized in
tractable mathematical models that are (at least) neurally reasonable. Extending
this idea to models of more elaborate tasks, such as human cued recall
requires mapping each of the stimuli that will be part of the
experiment -- i.e., each of the words in the list -- to a pattern of activation
over neurons.  This is typically done by mapping each word to a vector in a
space of neurons.  In this case, the synapses between the neurons can be
understood as a matrix.  With appropriate assumptions, many results can be
derived and a particular set of assumptions can be compared to behavior.

\subsection{Cognitive models of memory}
The basic theoretical stance of behaviorism is that we should construct
psychological theory without reference to the internal state of the organism.
This approach is difficult to reconcile with many human laboratory memory tasks.
For instance, a radical behaviorist model of the free recall task is
untenable.   In free recall, participants are presented with a sequential
experiece (e.g., a list of words) and later asked  to verbally report their
memory for the experience.  What is the ``cue'' in free recall?  Participants
can report many different experiences and can report on different aspects of
their experience.  It is difficult to make sense of these phenomena without
simply assuming that the participant has some internal experience of their
memory that they then describe.  

Cognitive models make a hypothesis about the internal state of the organism
and use that hypothesis to predict behavior.  Radical behaviorists explicitly
eschewed any reference to the internal experience of the behaving organism
under the belief that such theorizing was underconstrained and can not lead to
a satisfactory scientific theory.  However, advances in modern neuroscience
have made this concern largely obsolete.  In principle, cognitive models can
simultaneously describe the observable behavior of an organism and neural
observables from the brain during performance of that behavior.  In this way,
cognitive models can be constrained by comparison to activity of neurons in
the brain.  

A broad class of cognitive models proceed by building simple associations
between stimuli mediated by a hypothesized internal state.  For instance,
short-term memory models hypothesize the existence of a short-term store that
holds information about recently presented stimuli.  According to one
influential approach, associations between stimuli can only be formed among
stimuli that are simultaneously active in the short-term store
\cite{AtkiShif68,RaaiShif80}.  Another
widely-used approach assumes the existence of a ``temporal context'' that
mediates associations between items \cite{SedeEtal08,PolyEtal09}.  Temporal
context models assume that the brain maintains a representation at each moment
of the recent past.  This temporal context  changes gradually.  When a person
remembers a specific instance from their past (like vividly remembering a
particular event such as a birthday), this cognitive event is accompanied by a
recovery of temporal context.  These models make specific neural predictions.
Short-term memory models and temporal context models predict that it ought to
be possible to examine the activity of neurons in the brain (using electrodes
or non-invasive methods such as EEG or fMRI) and decode the content of recent
experiences.  Cognitive models of this class are introduced in
Section~\ref{sec:ststcms}.

\subsection{Beyond associations: Representing temporal relationships in the
mind and brain}

Although associations have been an extremely productive idea in the
mathematical psychology of memory, there is no question that simple
associations as understood by behaviorists are
insufficient to describe the richness of human memory.  Associations that can
be described by a scalar value are extremely limited.  If the association
between stimulus \textsc{x} and stimulus \textsc{y} is some specific number, say 2.38, and the
association between \textsc{x} and \textsc{z} is 0.35, we can say that the
\textsc{x} $\rightarrow$ \textsc{y} association is stronger than the 
\textsc{x} $\rightarrow$ \textsc{z} association.  Operationally, if we probe memory
with \textsc{x}, memory returns ``more'' \textsc{y} than \textsc{z}.
However,  human memory can learn and express many different \emph{kinds} of
relationships.  For instance, \textsc{x} might be two meters to the East of
\textsc{y},  or \textsc{x} might be a member of the category \textsc{z}, or
\textsc{y} and \textsc{z} might be married to one another.  In order to
express these kinds of relationships, a richer formalism is required.

\begin{figure}
	\includegraphics[width=0.8\textwidth]{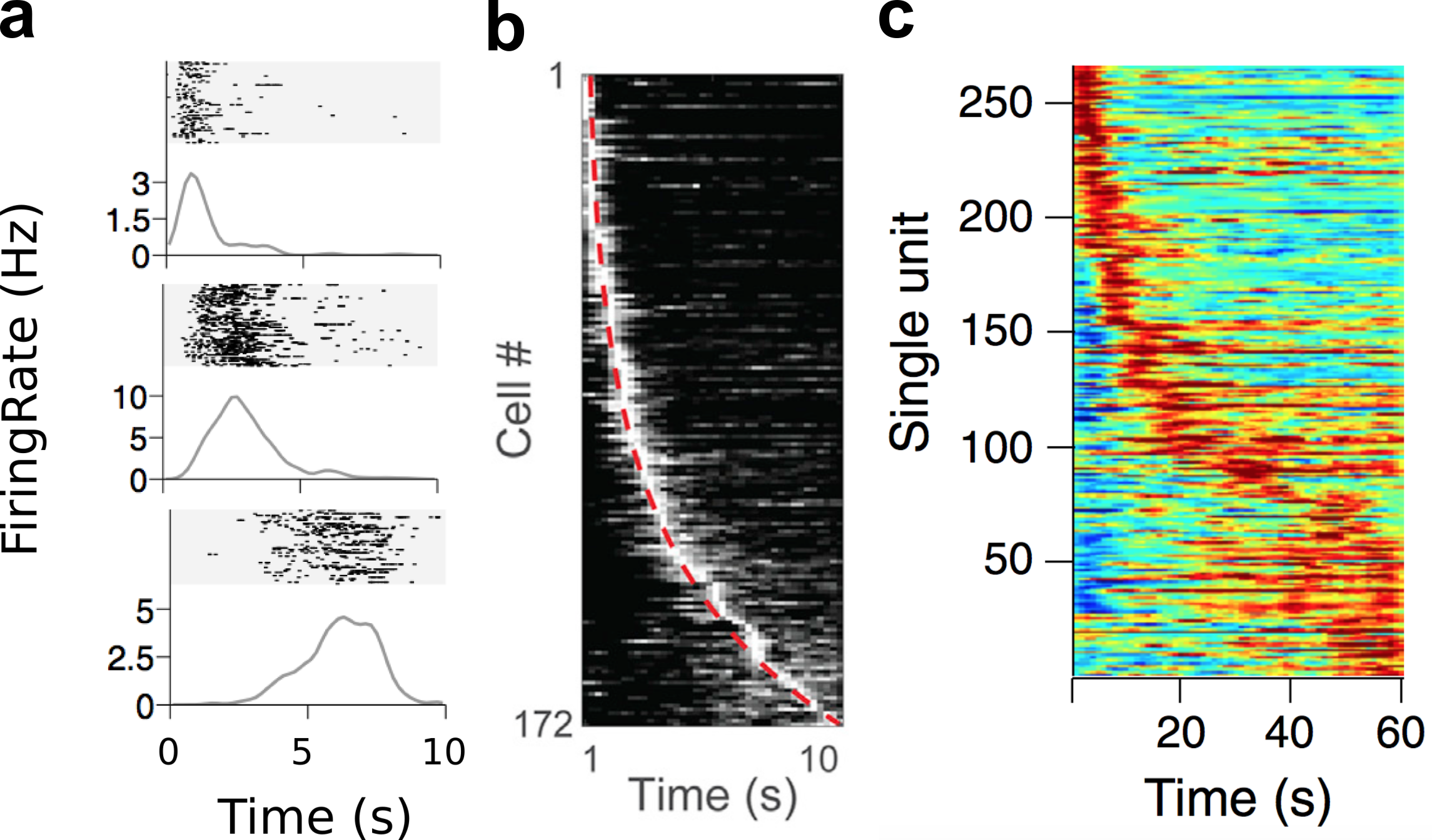}
	\caption{So-called ``time cells'' are neurons that fire in sequence
	following a triggering stimulus.  \textbf{a.}   Three time cells
	recorded from the hippocampus following the beginning of the delay
	period in a memory experiment.  The top cell fires consistently over
	trials early in the delay.  The middle and bottom cell also fire
	consistently, but at progressively later delays. After
	MacDonald, et al., (2011).
	\textbf{b.} A set of time cells in the hippocampus recorded during the
	delay interval sorted on their time of peak firing.  Note that the
	population tiles the delay.  This set of time cells could be used to
	determine the time within the delay.  Note
	further that more cells fire earlier in the delay than later.  This
	implies that there is greater resolution to the representation of time
	within the delay
	early in the delay period rather than later in the delay period. After
	Mau, et al., (2018).  \textbf{c.}  Time cells from the medial
	prefrontal cortex (mPFC).  Note the scale of the $x$-axis extends out
	60~s.  After Bolkan, et al., (2017).
	\label{fig:timecells}}
\end{figure}

The mammalian brain contains neurons that can express metric relationships
between stimuli.   For instance, consider neurons 
referred to as ``time cells'' in the rodent hippocampus during performance of
a behavioral task \cite{Eich17a}.   After presentation of a stimulus, e.g.,
ringing a bell,
these time cells fire in a sequence such that each neuron fires for a
circumscribed period of time (e.g., Figure~\ref{fig:timecells}).  Because the
sequence is reliable across different presentations of the same stimulus, it
is possible to look at which time cell is firing and decode how far in the
past the triggering stimulus was experienced.  As we will see, the information
about the time in the past at which the bell was presented written across this
population of neurons can be used to learn temporal \emph{relationships}
between the presentation of the bell and other stimuli.  
This class of models has been used to develop cognitive models of relatively
complex behavioral tasks and at the same time the properties of time cells can
be evaluated against experiments recording from populations of neurons in
mammals.   To the extent that this hypothesis is consistent with both
behavioral and neurophysiological data, it makes sense to take the equations
seriously.  As we will see, the formalism is quite rich, providing an
opportunity to do meaningful theoretical work on physical models of memory.

\subsection{\label{sec:history}  A brief history of  mathematical
models of memory}

This chapter covers a tiny proportion of the work in mathematical models of
human memory.  To provide at least pointers to the topics that are missing,
and to properly contextualize the topics that are covered, this subsection
provides a very concise history of mathematical models of memory.

Descriptive quantitative models of behavior date back to the very beginning of
modern memory research.
\citeA{Ebbinghaus} conducted early empirical studies of human memory, testing
himself on serial recall of nonsense syllables.   \citeA{Ebbinghaus} included
quantitative descriptions of many of the phenomena he studied.  For instance,
Ebbinghaus introduced the power law of forgetting to describe his findings
relating the persistence of memory to the passage of time.  In the early part
of the 20th century, radical behaviorism led many researchers to focus on simple
stimulus-response associations.   Quantitative models of these data
attempted to describe observable phenomena with as few assumptions as
possible.  \citeA{Hull39} provides an excellent example of the spirit of this
work, fitting equations to observed empirical relationships.

The 1950s saw the first process models of memory.  Process models, in contrast
to descriptive models, make hypotheses about internal mechanisms that cause
observable behavior.    Stimulus sampling theory
\cite{Este50,BushMost51}, provides an early example
of such a process model.  Stimulus sampling theory introduced a number of
ideas that are still extremely influential today (see section~\ref{sec:SST}). 

The 1960s and 1970s saw memory research divide into a set of subfields as the
cognitive revolution dramatically changed the kinds of theories that were
acceptable in psychology.  There were two major developments in mathematical
models of memory during this era that had long-lasting effects over the next
several decades.  First, building on a long tradition of mathematical models
of conditioning,  the Rescorla-Wagner model  
\cite<see Chapter 5>{RescWagn72} successfully accounted for essentially
everything that was known about classical conditioning up to that time.
The Rescorla-Wagner model is built on a really simple idea -- that change in an association between a cue and a response
depends on how well the  outcome is predicted.
Second, the 1960s saw the development of the first models of short-term memory
building on early ideas from \citeA{Mill56}.  The two-store memory model of
\citeA{AtkiShif68} provided a conceptually simple description of an immense amount of data (see Section~\ref{sec:ststcms}).  This was also perhaps the first influential mathematical model of memory to make use of computer simulations to test its predictions.  These two very different models spawned entire fields of research in psychology and neuroscience that continue to this day.

The Rescorla-Wagner model led directly to reinforcement learning \cite{SuttBart81}.  Reinforcement learning has been extremely influential in neuroscience, where the connection between these models and the dopamine
system in the brain \cite{SchuEtal97} has spawned an immense amount of work
that continues to the present (e.g., see Ashby, Crossley, Inglis, this
volume).  Reinforcement learning has also been extremely influential in
artificial intelligence research, including very high profile papers building
models to achieve human-level performance in video games and the game of go
\cite{MnihEtal15,SilvEtal16}.

The \citeA{AtkiShif68} model also led to a great deal of work in psychology
and neuroscience.  The model coincided with the discovery of patients with
brain damage that showed problems with short-term memory but not long-term
memory, and \emph{vice versa}.  \citeA{BaddHitc77} further subdivided
short-term store and mapped these components onto distinct brain circuits.
This kind of model -- with many components that map onto different parts of the
brain -- was well-suited for posing the kinds of questions that could be
answered with early cognitive neuroimaging techniques such as PET and
univariate fMRI. Mathematical models of short-term memory continue to be
influential in contemporary cognitive neuroscience (see \citeNP{TrutEtal21} for a recent
review).

In the 1980s and 1990s,  a great deal of attention was focused on a class of
mathematical models of memory that were collectively known as distributed
memory models.  These models focused on human memory experiments, primarily
experiments that would be understood today as episodic memory tasks. 
Models that fall into this class include TODAM \cite{Murd82}, CHARM \cite{Metc85}, SAM
\cite{GillShif84},  MINERVA-2 \cite{Hint87}, the matrix model
\cite{HumpEtal89b}, and REM \cite{ShifStey97}.     
Although these models differed in many details, there were some common
assumptions.   First, they represented studied items as a distributed set of
features, building on early work by  \citeA{Ande72,Ande73}.
Section~\ref{sec:SST} also adopts this convention.   Second,  the distributed
memory models were all associative.  It was implicit that short-term memory
controlled which items and associations were stored in memory.  An important
conceptual contribution of these models was the introduction of quantitative
models for context \cite<see especially>{Murd97} that we build on in
Section~\ref{sec:ststcms}.  The temporal context models discussed in
section~\ref{sec:ststcms} grew out of this tradition.  

The early distributed memory models did not make a connection to neuroscience.
In contrast, connectionist models of memory (see \citeNP{HassMcCl99} for a review of early
work) paid close attention to neuroscience.  
For the most part, these models did not
focus on detailed behavioral data from human memory experiments  \cite<but
see>{HassWybl97,NormORei03}.  Rather, these models focused more on problems,
such as amnesia and sleep, that had a clear connection to neural processes.
For instance, in one very influential paper \citeA{McClEtal95} postulated that
behavioral patterns observed in amnesia patients -- for instance the ability to
remember events from early in ones life but not the ability to remember more
recent events -- were attributable to separate memory stores that learned
associations with different statistics.  Connectionist memory models were
developed in parallel with advances in artificial neural networks that are
fundamental to contemporary AI. 

One very important development in the early part of the 20th century was that models of conditioning made contact with models of timing behavior.  Scalar expectancy theory \cite{Gibb77} provided an excellent model of behavioral experiments where animals had to use their sense of time to receive reward
\cite<see also>{KillFett88}.  \citeA{GallGibb00} constructed a mathematical model out of scalar expectancy theory that  described a range of findings
from conditioning experiments.  The hypothesis was that  behavioral associations fundamentally result from learning about the temporal relationships between the stimulus and response.   \citeA{BalsGall09}
provide an elegant overview of this idea.   Notably, because timing behavior
has the same properties over a range of time scales, models of conditioning
built on this assumption can naturally accommodate scale-invariance in memory,
which is discussed further in section~\ref{sec:scaleinv}.

Section~\ref{sec:sith} draws on work over the last decade or so that
synthesizes aspects of many of these approaches.   The scale-invariant temporal
history was originally proposed to address limitations in temporal context
models \cite{ShanHowa10}.  As such  it is continuous with the distributed memory
models and can be used to build models of similar tasks.  At the same time,
because neuroscientific considerations place such strong constraints on these
models, it is similar in spirit to the connectionist models of memory.
Finally, because memory traces are formed using a population that contains information about the time at which events took place, this approach is closely related to (and in actual fact was very much inspired by) work pursuing a close relationship between timing and conditioning.   

\section{\label{sec:SST} ``Simple'' associations in the mind and brain}
In this section we will introduce a formalism to describe mathematical models based on simple associations. We will suppose that learning consists of forming and accessing associations between a set of ``items.''   These items can correspond to words in a cued recall experiment, in which we attempt to describe the association between two words (e.g., \textsc{absence--hollow} above).  Or we could use the same formalism to describe the association between a tone that serves as a conditioned stimulus and an unconditioned response, such as salivation in the case of Pavlov's dog.  

Distributed memory models (DMM) assume that each item is described by a vector over some high dimensional
space.  We will write vectors as lower-case bold letters,  $\mathbf{v} =
\left\{v_1, v_2, \ldots v_n\right\}$, where $n$ is some ``large'' integer.
We can envision the vector as a  list of numbers that describes the activity
over a large population of neurons.   If a particular item $\mathbf{v}$
represents a word we might understand $\mathbf{v}$ as the ``pattern of
activity'' over a population of neurons that are caused by presentation of
that word.  A different word would produce a different pattern of activity.
If a particular item corresponds to a response, such as salivation, we might
understand $\mathbf{v}$ as the pattern of activity in a particular population
that is necessary for salivation rather than some other response (such as
freezing).

\newcommand{\mytranspose}{\ensuremath{^\textnormal{T}}}

\subsection{Hebbian learning}
As an illustration of the distributed-memory model approach, let us consider a simple model of cued recall.
We map all of the words that could possibly be presented in an experiment onto
a set of vectors within the same space.  We assume further that the
overwhelming majority of entries
$v_i$ are zero and the remainder are some small positive number and that the number of
entries $n$ is large.   Suppose that we randomly choose vectors corresponding to two
different words $\mathbf{v}_i$ and $\mathbf{v}_{j\neq i}$.  We can take the
inner product between any two vectors as a measure of their ``similarity.''
With these assumptions, the inner product of a vector with itself,
$\mathbf{v}_i\mytranspose \mathbf{v}_i$ or 
$\mathbf{v}_j\mytranspose \mathbf{v}_j$ will tend to be much greater than the 
inner product between different words $\mathbf{v}_i\mytranspose \mathbf{v}_j$, because
the entries of these are not perfectly correlated.  We might even suppose that
related words (e.g., \textsc{couch} and \textsc{sofa}) correspond to vectors
that are more similar to one another than unrelated words (e.g.,
\textsc{couch} and \textsc{rutabaga}).  To keep the arithmetic simple, let us
suppose that we have chosen the entries in the vectors to ensure that the
expected value of $\mathbf{v}_i\mytranspose \mathbf{v}_j$ is 1 if $i=j$ and close to
zero otherwise.

\newcommand{\myassoc}{\ensuremath{\mathbf{M}}}
\newcommand{\thestim}{\ensuremath{\mathbf{f}}}
\newcommand{\theresp}{\ensuremath{\mathbf{g}}}
\newcommand{\learnparam}{\ensuremath{\alpha}}
\newcommand{\forgetparam}{\ensuremath{\rho}}

Let us flesh this model out sufficiently to model a simple cued recall
experiment.  Let us describe a list of word pairs by denoting the cue of the
pair presented at time $t$ with a vector $\thestim_t$ and the response member
of the pair with a vector $\theresp_t$.   So, if we had a list of two pairs,
\textsc{absence--hollow} and \textsc{pupil--river}, we would refer to the
vector corresponding to \textsc{absence} as $\thestim_1$, the vector
corresponding to \textsc{hollow} as $\theresp_1$, the vector correspoding to
\textsc{pupil} as $\thestim_1$ and \textsc{river} as $\theresp_2$.

Now, we can model associations between the words as an outer product matrix
between the vectors corresponding to the cue and response of each pair.  Let
us assume that the matrix $\myassoc$ is initialized as an $n\times n$ matrix
of zeros before the list.  Then as each item is presented, $\myassoc$ is
updated as: 
\begin{equation}
		\Delta \myassoc_t = \theresp_t \thestim_t\mytranspose  
		\label{eq:Hebb}
\end{equation}
so that after learning the entire list,
\begin{equation}
		\myassoc = \sum_t \theresp_t \thestim_t\mytranspose  
		\label{eq:Hebblist}
\end{equation}
where the sum is over all of the pairs presented in the experiment.

\begin{figure}
\centering
	\includegraphics[width=0.3\textwidth]{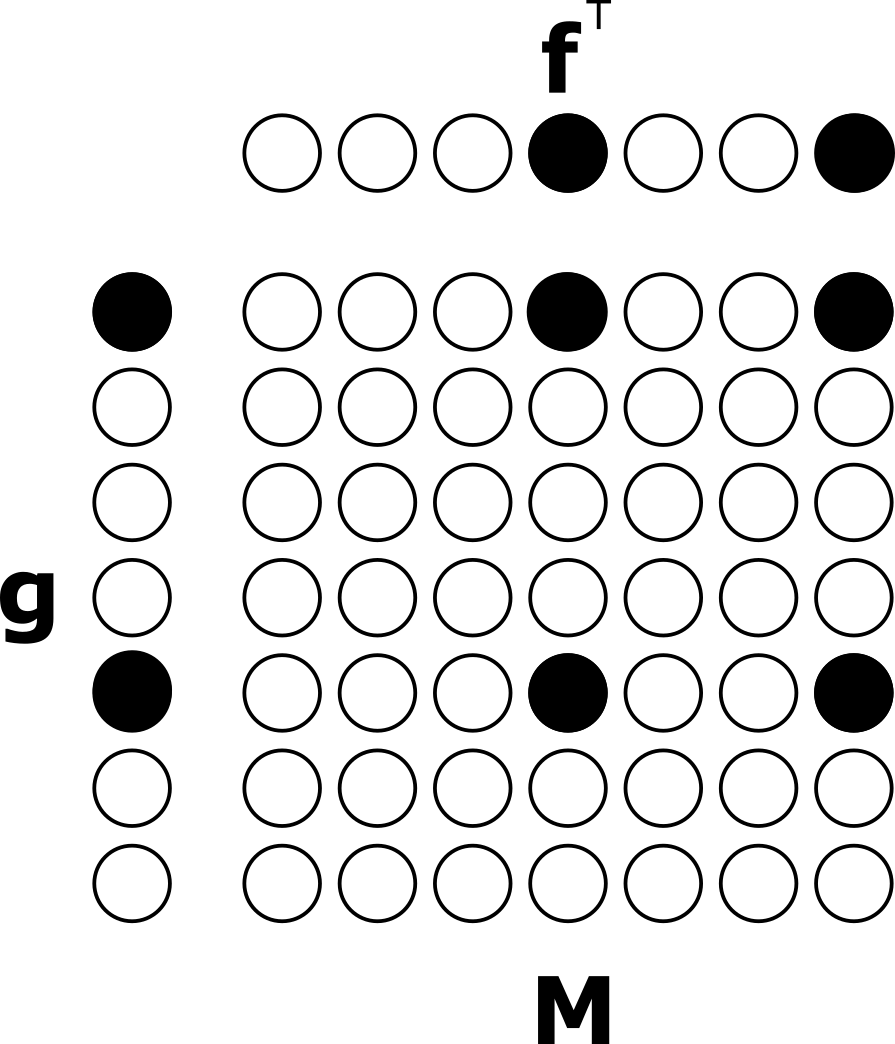}
	\caption{Graphic illustration of the equation  $\myassoc = \theresp
	\thestim\mytranspose $.  Here $\thestim$ is a vector that is zero except for
	entries $4$ and $7$; $\theresp$ is a vector that is zero except for
	entries $1$ and $5$.  The outer product matrix $\myassoc$ is zero
	except for entries where both $\thestim$ and $\theresp$ had nonzero
	values.  Probing as $\myassoc \thestim$ gives back $\theresp$
	multiplied by the squared length of $\thestim$. 
	\label{fig:outerproduct}}
\end{figure}

To understand the role of the outer product, let us imagine we have a one-pair
list so that $\myassoc = \theresp \thestim\mytranspose $ (Fig.~\ref{fig:outerproduct}).
Any particular entry $M_{ij} = g_i f_j$ gives the product of the activity in
``neuron $i$'' in pattern $\theresp$ and ``neuron $j$'' in pattern $\thestim$.
The product is non-zero if both $g_i$ and $f_j$ are non-zero.  The anatomical
structure that connects the axon of one neuron to the dendrite of another is
referred to as a synapse.  These connections can be strengthened or weakened
based on the activity of the pre- and post-synaptic neurons through a variety
of molecular processes.  Hebbian learning (originally proposed by Donald Hebb
in 1948) is a learning rule in which synapses are strengthened if both the
pre- and post-synaptic neurons are active at the same time (see Ashby et al.,
this volume, for more details).   Informally, Hebbian learning is often
summarized by the slogan ``neurons that fire together, wire together.''
Hebbian learning has been demonstrated experimentally in a number of brain
regions and a number of species.

To understand why this is referred to as an association, let us probe
$\myassoc$ with a probe word, which we denote as $\thestim_p$.  Then we find
that $\myassoc \thestim_p = \left(\theresp \thestim\mytranspose \right) \thestim_p = 
\theresp \left(\thestim\mytranspose  \thestim_p \right)$.  That is, probing $\myassoc$
with a probe vector $\thestim_p$ returns $\theresp$ weighted by the similarity
between the probe vector and the studied cue vector.  If the probe vector
$\thestim_p$ is the same as the studied cue $\thestim$, the output is
$\theresp$ multiplied by a large number.  If $\thestim_p$  is not the same as
$\thestim$, the output is $\theresp$ mutiplied by a small number.  Returning
to the situation where there are many pairs in the list, we find, exploiting
the linearity of matrix addition and commutativity of multiplication by a
scalar,
\begin{eqnarray}
	\myassoc \thestim_p &=& \left[\sum_t \theresp_t \thestim_t\mytranspose \right]
	\thestim_p \\
	&=& \sum_t  \left(\thestim_t\mytranspose  \thestim_p\right) \theresp_t
		\label{eq:Hebblistprobe}
\end{eqnarray}
That is, after probing memory with a specific word $\thestim_p$, the output is
the vector sum of the response words $\theresp_t$ weighted by the similarity of the
probe word to the cue that was paired with that response.   
Because the similarity of the probe words to themselves is much greater than
between different words, this sum gives a large number for the appropriate
response and much smaller numers for the other possible responses.  If one
probes $\myassoc$ with $\thestim_{\textnormal{ABSENCE}}$, the output is
``mostly'' $\theresp_{\textnormal{HOLLOW}}$;  if one probes with
$\thestim_{\textnormal{PUPIL}}$, the output is mostly
$\theresp_{\textnormal{RIVER}}$.   By adding assumptions that map the output
of the associative memory onto a probability of successfully recalling the
appropriate response, one can construct relatively elaborate models of
behavior.

If each component of $\thestim$ and $\theresp$ can be thought of as a neuron,
then each entry in $\myassoc$ can be understood as a synapse.  The entire
matrix $\myassoc$ can thus be understood as the set of synapses connecting the two
populations.  The outer product learning rule in Eq.~\ref{eq:Hebb} can thus be
understood as a simple hypothesis for how populations of neurons can store
information \emph{via} Hebbian learning.  Although this is undoubtedly a
grotesque oversimplification of what happens in the brain, this framework is
sufficiently simple that one can write out tractable models of behavioral
experiments.  

To actually compare this model to behavioral data, it's necessary to specify
some means to map the strength of the association onto behavioral observables,
for instance probability of recall.  Having said that, this simple Hebbian
mechanism responds appropriately to many experimental manipulations in a
sensible way.   For instance, suppose that some pairs in the list are
repeated.  Adding additional terms with the same vectors to
Eq.~\ref{eq:Hebblist} results in a stronger association between those items
(this follows from linearity).\footnote{One can easily construct a similar
argument for the effect of increasing the study time for some of the pairs in
the list.}  Similarly, one can compare recall of a
particular pair in lists of various lengths.  Examining
Eq.~\ref{eq:Hebblistprobe}, we see that the effect of including additional
pairs is to add noise to the output of memory.  That is, after probing with
$\thestim_i$, the output of memory is $\theresp_j$ times a big number plus all
of the other items in the list weighted by small numbers.  As one adds pairs
to the list, this second component grows more important, acting like
background noise for retrieval of the target response.  
Similarly, one could imagine that attention
fluctuates from moment-to-moment and model that by multiplying
Eq.~\ref{eq:Hebb} by a factor that estimates the current amount of attention.
Distributed memory models pursued questions along these lines and carefully
compared the results to behavioral experiments.

\subsection{Forgetting} 
The Hebbian outer product model sketched above has several problems, many of which are
addressed by subsequent work described in the remainder of this chapter.
Here we discuss ways to enable the model to \emph{forget}.   We discuss two
approaches to forgetting.  Perhaps the most obvious way to implement
forgetting is to allow the weights to decrease in amplitude.  A less obvious
way to implement forgetting is to assume that the cue itself is not constant
over time.   That is, although an experimenter may take care to present the
word \textsc{absence} several times in the same font, in the same location of
the screen, for precisely the same duration of time, this does not ensure that
this stimulus activates the same set of neurons in the brain on each
presentation.  There are many other possible approaches to forgetting and
different mechanisms may contribute differentially to forgetting in different
experimental paradigms. This
chapter focuses on these two mechanisms for forgetting because they lend
themselves to concise mathematical descriptions and are conceptually distinct
from one another. 

\subsubsection{Forgetting \emph{via} changes in the weight matrix}
One simple way
to augment Eq.~\ref{eq:Hebb} to enable forgetting is to allow the weights to
decay exponentially as a function of time: 
\begin{equation} 
	\myassoc_{t+1} = \forgetparam \myassoc_t + \theresp_t
	\thestim_t\mytranspose   
	\label{eq:Hebbforget}
\end{equation}
where $0 < \forgetparam < 1$.   Each additional time step results in an
additional power of $\forgetparam$, so that the output caused by a memory
probe decreases the longer it has been available in memory.
After studying $L$ items, we find 
\begin{equation}
	\myassoc_L \thestim_p = \sum_t \rho^{L-t} \left(\thestim_t\mytranspose  \thestim_p\right) \theresp_t
		\label{eq:Hebblistproberecency}
\end{equation}
The last term shows that the strength of the association stored in $\myassoc$
decays exponentially as a function of how far in the past the association was
learned.

One of the longest standing questions in memory research is whether we forget
over time due to the passage of time \emph{per se} or due to intervening
events.  To make an analogy, suppose one leaves an iron bar outside in the
northeastern United States and measures the amount of rust on the bar once per
year.  One will find that the amount of rust on the bar increases with each
passing year.  Knowing nothing of chemistry, one might be tempted to conclude
that rust is caused by the passage of time \emph{per se}.   In the case of the
iron bar, we know this account is incorrect; had the bar been kept in a
vacuum, it would not rust at all no matter how long one waits.   

In the case of memory, there is little question that many factors affect
forgetting above and beyond any effect due to time \emph{per se}.  One could
adapt Eq.~\ref{eq:Hebbforget} to accommodate these factors by allowing
$\forgetparam$
to change as a function of variables available at time $t$.   Considering
$\myassoc$ as a set of synapses, one might also construct alternative rules
for forgetting that allow effects specific to a particular cue and/or a
particular response.  However, as we will see, there are more fundamental
issues with this simple conception of memory as association, so we will not
dwell further on this point here.

\subsubsection{Forgetting \emph{via} stimulus sampling}
Weakening of associations, operationalized as a gradual decrease in the
strength of synapses, is not the only way to instantiate forgetting in a
simple neural network.  Consider Eq.~\ref{eq:Hebblistproberecency}.  The term
due to weakening of the synapses, $\forgetparam^{L-t}$ appears with a term
relating the similarity of the probe $\thestim_p$ to each of the cue stimuli
in the list $\thestim_t$.  If one provided a probe stimulus that was similar
but not identical to one of the cue stimuli in the list, one would expect this
to have a measurable effect on memory.  For instance, suppose the cue stimulus
in an animal conditioning experiment is a pure tone of 440~Hz.  One would
expect that the set of features caused by a similar tone (e.g., 441~Hz) to be
greater than the set of features caused by a less similar tone (e.g., 550~Hz).
Because this would manifest as changes in the $\thestim_t\mytranspose
\thestim_p$ terms in Eqs.~\ref{eq:Hebblistprobe}~and
~\ref{eq:Hebblistproberecency} we would expect this to result in more
conditioned responding to probes similar to the studied conditioned stimulus.
Indeed, it has long been known that one can observe this phenomenon, referred
to as stimulus generalization, in animal conditioning experiments
\cite{Hull47}.  

One can use stimulus generalization to construct associative models of
forgetting.  Stimulus sampling theory \cite{Este50,Este55b,Este55a} makes a distinction between the
``nominal stimulus'' that the experimenter provides and the ``functional
stimulus'' that the research participant experiences.  
To be more concrete,
consider a simple conditioning experiment in which the conditioned stimulus is
a 440~Hz tone.  The nominal stimulus is the tone itself.  A careful
experimenter can ensure that the nominal stimulus on each presentation is
physically identical.  However, no matter how careful the experimenter may be,
the functional stimulus experienced by the participant may be meaningfully
different from one presentation to the next.  For instance, an animal in a
Skinner box may have a slightly different posture from one presentation of the
nominal stimulus to the next.  Or, perhaps the animal is more or less
attentive to different properties of the nominal stimulus from one presentation to
the next.  In stimulus sampling theory the nominal stimulus presented by the
experimenter specifies a set of features that \emph{could} be experienced by
the participant.   On a particular trial, the participant samples from that
set of stimulus features to obtain the functional stimulus, which is used to
support learning.


It has been said (in a quotation that is often attributed to Heraclitus), that
``It is impossible to step into the same river twice.''  The identity and
position of the molecules of water changes continuously from moment to moment.
Suppose one steps into a river on two occasions,
$t_1$ and $t_2$.  Although the river at $t_1$ is not identical to the river at
$t_2$, it is reasonable to say that the similarity of the two
rivers, all else equal, is a decreasing function of $t_2 - t_1$.
\citeA{Este55a} proposed that, all else equal,  the functional stimulus
caused by presentations of the same nominal stimulus at $t_1$ and $t_2$ is
also  a monotonically decreasing function of $t_2 - t_1$.  
Let us write the functional stimulus caused by nominal stimulus $\alpha$ at
time $t$ as $\thestim_{\alpha,t}$.  One can incorporate this assumption into an
associative model to enable an account of forgetting without a decrease in the
strength of learned associations.  Suppose one learns an association $\theresp
\thestim_{\alpha,t_1}\mytranspose$.  Probing with $\thestim_{\alpha,t_2}$
thus gives $\theresp$ times a function that decreases with $t_2 - t_1$.

Remarkably, the assumption of gradually-changing stimulus features from
stimulus sampling theory from the 1950s has received support from recent
neurophysiological studies, at least for some kinds of stimuli.   
For instance, recent recordings from mouse piriform cortex studied the set of
neurons activated by odors during conditioning \cite{SchnEtal21}.
The piriform cortex is the first cortical region that receives input from the
olfactory bulb, making it roughly analogous to primary visual cortex for
visual images or primary auditory cortex for
auditory stimuli.\footnote{One may even argue that piriform cortex is more
peripheral than these regions.  Information from the retina projects to visual
cortex only after passing through a brain region called the thalamus which
receives information from many sensory modalities. For instance, information
from the ear passes through  thalamus on the way to auditory cortex.  In
contrast, the piriform cortex is directly connected to the olfactory bulb. }
Because it is so closely related to the sensory receptor itself, it makes
sense to think of the activation across piriform cortex as a direct
representation of the sensory stimulus.

The particular recording method that \citeA{SchnEtal21} used allows for stable
recordings of the same neurons over weeks and months.  At each stage of the
experiment, different odors evoked distinct neural populations.    However,
the populations that each odor evoked changed continuously over every time
period studied.  That is, at each time $t$, one could distinguish
$\thestim_{\alpha,t}$ from $\thestim_{\beta,t}$.  However, $\thestim_{\alpha,
t_1} \mytranspose \thestim_{\alpha, t_2}$ was a decreasing function of $t_2 -
t_1$ for all pairs of times considered.  
Recalling Heraclitus, one might say that the mouse could not smell the same
odor twice.  This neural phenomenon, referred to as representational drift, is a
topic of ongoing research \cite{MauEtal20,RuleEtal19}.
Representational drift  has been reported, at least under some circumstances,
in visual cortex \cite{DeitEtal20},  posterior parietal cortex \cite{RuleEtal20},
hippocampus \cite{MannEtal07,MankEtal12,CaiEtal16,RubiEtal15}, and prefrontal cortex \cite{HymaEtal12}, as well as piriform cortex.   

\section{\label{sec:ststcms} Short-term memory and temporal context models }
The Hebbian associative model from the previous section describes associations
between pairs of stimuli.  Given a probe stimulus, the model provides a
response as output.  Although simple and tractable, this model glosses over
some fundamental questions about human memory.   This section studies models
developed largely in response to the free recall task, which has been an
important driver of models of human memory since the 1960s.  

In free recall, the participant is presented with a list of
stimuli -- typically words -- one at a time.  The participant's task is to
recall as many stimuli as possible from the list.  In the free recall task,
the participant may recall the words in the order they come to mind (this is
in contrast with serial recall where the stimuli must be recalled in the order
in which they were presented).   
There are many variants of the free recall task.  In delayed free recall, a
distractor task of up to a minute intervenes between the last item in the list
and the beginning of the recall period.  In the list-before-last paradigm, the
participant does not recall the most recent list, but the previous list.   In
some experiments,
participants are given a final free recall task at the end of the experimental
session in which the participant is instructed to recall as many words as
possible from all of the preceding lists.

The first problem for the simple Hebbian model that free recall presents is
how the task is accomplished at all. The Hebbian model requires a probe to
generate a response.  What is the probe in free recall?  Because the
instructions are so general whatever prompts recall must be internal to the
participant.  The second challenge for the simple Hebbian model is
overwhelming evidence that functional associations are not limited to
adjacent items, but are instead distributed very broadly over many items.
These findings -- reviewed in the next subsection -- have led to a very
different conception of memory.  Rather than a collection of items and
associations among them, models originating from the free recall task have
postulated temporally-sensitive memory representations that carry information
about many items extended over macroscopic periods of time.  

\begin{figure}
	\centering
	\includegraphics[width=0.8\textwidth]{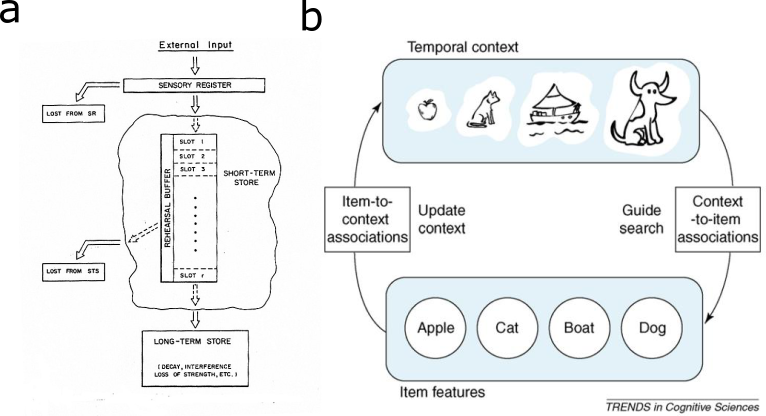}
	\caption{
		Schematic diagrams for short-term/long-term
		memory and temporal context models.
		\textbf{a.}  Models based on a distinction between short-term
		memory and long-term memory assign different properties to
		these different stores.  Short-term store  consists of a
		rehearsal buffer that contains a small integer number of
		items with high precision.  Long-term store holds a very large
		number of memory traces with less precision.
		After Atkinson \& Shiffrin (1968).
		\textbf{b.}   In
		temporal context models, the
		currently-experienced item activates a set of features on the
		item layer (bottom). 
		After an item is presented, it activates features that remain
		active in a gradually-changing state of temporal context  (top). 
		The context layer cues retrieval \emph{via} context-to-item
		associations.  The item layer can cause recovery of a previous
		state of temporal context associated with that item (not
		shown).
		After Polyn \& Kahana (2008).
		\label{fig:STSCMR}
	}
\end{figure}
\nocite{PolyKaha08}

\subsection{The recency effect and two-store models}

\begin{figure}
	\centering
	\includegraphics[width=0.8\textwidth]{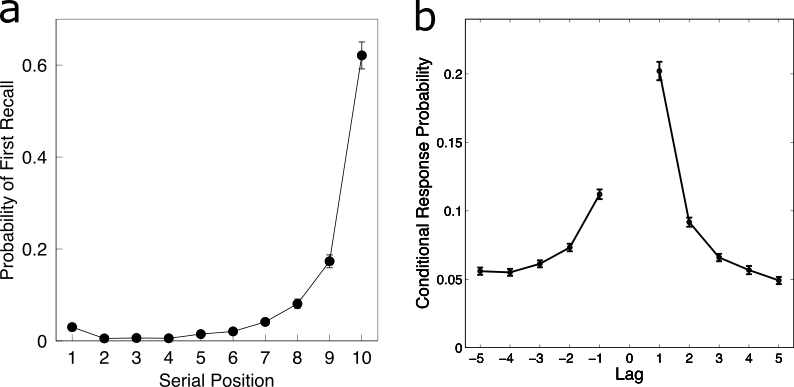}
	\caption{The recency and contiguity effects in free recall.  In the
	free recall task, participants are presented with a series of stimuli,
	usually words, and are then asked to recall as many words from the list as
	possible in the order they come to mind.
	\textbf{a.}  The recency effect measured by the probability of first
	recall.  The x-axis plots serial position within a list of ten words.
	The y-axis gives the probability that the first word the participants
	said came from each position within the list.  In this experiment
	there is a dramatic recency effect -- words from the end of the list
	are much more likely to be recalled first than words from the
	beginning or middle of the list.  After Howard, et al., (2008).
	\textbf{b.} The contiguity effect in free recall.  Given that a
	participant has just recalled word $i$ from the list, what is the
	probability that the next word recalled comes from position $i +
	\textnormal{lag}$?  All else equal, participants show a robust
	tendency to recall words from nearby positions within the list
	together in recall.  The data in this figure is averaged over many
	experiments.
	After Kahana (2012).
	\label{fig:RecencyContiguity}
	}
\end{figure}

\nocite{HowaEtal08,Kaha12}

The recency effect refers to the finding that, all else equal, memory is
better for information that was presented more recently.  In free recall, this
manifests as an increase in the tendency to initiate recall at the end of the
list (Fig.~\ref{fig:RecencyContiguity}) as well as higher probability of
recall overall.  The recency effect can be observed in all of the experimental
paradigms that people study with human participants. 

The recency effect is especially pronounced in immediate free recall, in which the recall test proceeds just after the last item in the
list \cite{Murd62}.  In delayed free recall, a delay interval is included during which participants typically perform a distractor task (to prevent them
from simply repeating the items in the list to themselves) prior to recalling
the words from the list.  In delayed free recall the recency effect is sharply
attenuated.  However, the probability of recall of early items from the list
is barely affected relative to immediate free recall
\cite{GlanCuni66,PostPhil65}.  In contrast, many other
variables (e.g., presenting the words faster or slower, choosing words that are semantically related, having medial temporal lobe amnesia) have a big effect on recall of items from the beginning and middle of the list, but barely any effect on the recency effect \cite{Glan72}.  These observations led researchers to propose that the recency effect draws on a specialized memory store, referred to as short-term store (STS) or short-term memory \cite{AtkiShif68,RaaiShif80}.

The view that memory was divided into distinct stores was hugely influential
in the 1970s and 1980s and remains so today.  The basic idea  (Figure~\ref{fig:STSCMR}a)
is that STS can store a small number of items with very high accuracy.
Items that are in STS at the time of test are recalled rapidly and with high
precision.  In addition, a subset of
items are passed from STS to a long-term store (LTS).  
LTS  does not have
capacity limitations and can store information  for a much longer duration.
The longer an item spends in STS during study, the greater the probability it
is transferred to LTS.  A key property of STS is that it is subject to
strategic control according to the goals of the participant.  For instance, if
participants are rewarded based on how many words starting with the letter
\textsc{q} they correctly recall, we might assume that words that start with a
different letter are less likely to enter STS and would be forgotten very
quickly.  

If one specifies a strategy for retaining information in STS it is
straightforward to work out (or simulate, if the strategy is very
complicated) the probability that an item is in STS at the time of test.
For instance, suppose that each item in a long list enters STS with certainty
displacing a random item in STS.  If the
short-term store can hold $N$ items, where $N$ is much smaller than the number
of items in the list then the probability that an item already in STS is
replaced by an incoming item $1/N$.  The probability that the item already in
STS persists in STS after a new item enters STS is thus $1-1/N$.  At the end of a list
of $L$ items, the probability that the $i$th item is still in STS at the time
of test is
$\left(1-1/N\right)^{L-i}$, leading to a recency effect.  
Note that although this function decays exponentially, recency due to STS has different
properties than recency due to  exponential weight decay
(Eq.~\ref{eq:Hebblistproberecency}).  First, the quantity that is decaying is
a probability rather than a strength \emph{per se}.  This probability gives
the proportion of trials where the item is available for recall from STS; on
trials where the item is not available, there is zero probability of retrieval
from STS.  This is distinct from a situation where the weights give a small
but reliable signal.  Second, although the probability of any one item
remaining in STS may be a decreasing function, it should be kept in mind that
the number of items in STS depends only on its capacity $N$ (assuming the list
has more than $N$ items).

One can similarly work out probabilities for the amount of time a word spends
in STS (recall that the probability of transfer to LTS goes up with time spent
in STS).  Coupled with a specfication of LTS one can make predictions for many
observable properties of memory retrieval resulting in a very detailed
description of immediate and delayed free recall, including but not limited to
the recency effect.  

A major challenge to the two-store account of recency came from a modification
to the free recall paradigm referred to as continual distractor free recall
(CDFR).
Recall that in immediate free recall the recall test follows shortly
after the last item in the list.   According to STS-based accounts the recency
effect in immediate free recall happens because the items from the end of the
list are still available in STS.  In delayed free recall, a distractor task
follows the last item on the list before the recall test.  The recency effect
is attenuated in delayed free recall.  According to STS-based accounts, this
is a consequence of the distractor task pushing list items out of STS.  In
CDFR, a distractor task follows each item in the list, not just the last item.  
Perhaps surprisingly, there is a pronounced recency effect in continual
distractor free recall relative to delayed free recall
\cite{BjorWhit74,GlenEtal80}.  
This finding was not
predicted by the STS-based account of recency and is difficult to reconcile
with an account of recency solely based on STS \cite{DaveEtal05,LehmMalm12}.

\subsection{The contiguity effect across delays}

As a thought experiment, try the following memory experiment on yourself.
Answer the following question: \textsc{What did you most recently have for
breakfast?}\footnote{If you  are eating breakfast while reading this you can
substitute the question \textsc{What did you most recently have for dinner?}}
Most people, when answering this question, do not merely generate a verbal
response (e.g., ``toast'') but experience a vivid recollection of the event in
the process of answering the question.  For instance, while writing this (in
the afternoon) in answering the question about breakfast I spontaneously
remembered where I sat down (kitchen table with the window to my right), the
hopeful look on my dog's face, and the news I read
on my phone.  I can take another moment and  search my
memory to vividly remember
events that happened shortly before eating breakfast (putting the coffee on
the stove, putting bread in the toaster) and shortly after (finishing my
coffee in the backyard with my dog). 

The ``kind of memory'' that supports vivid recollection of events from one's
life is referred to as episodic memory \cite{Tulv83}.
Episodic memory has
been extensively studied over the last several decades.  For the present
purposes we note that episodic memory is believed to be closely related to a
phenomenon referred to as the contiguity effect.  
In free recall, the
contiguity effect (Fig.~\ref{fig:RecencyContiguity}b) manifests as the finding
that (all else equal) if a participant has just recalled a word from the list,
the next word that participant recalls tends to come from a nearby position in
the list \cite{Kaha96}.   In memory experiments with a probe (e.g., cued recall), the
contiguity effect manifests as the finding that the probe tends to bring to
mind other items that were close together in time.  For instance in cued
recall, when a participant recalls a word that was not the correct response to
the probe, that erroneous word tends to come from a pair that was presented
nearby in the list.  The contiguity effect is not limited to experiments with
words as stimuli and is indeed quite general \cite{HealEtal18}.  

Note that the episodic memory for today's breakfast illustrates the contiguity
effect.  Sitting down at the table, giving my dog a piece of 
sausage and reading about terrible events unfolding overseas were not actually
simultaneous but were relatively close together in time (probably tens of
seconds).  The other events I retrieved -- putting the bread in the toaster and
finishing the coffee in the backyard -- were each separated by several minutes
from breakfast \emph{per se}.  Consistent with this intuition, the contiguity
effect is observed in the laboratory in CDFR experiments where the items are
separated by tens of seconds.   The contiguity effect can also be observed
over much longer time scales -- hundreds of seconds in final free recall
\cite{HowaEtal08}, hours in experiments using mobile phones to administer a
list as participants went through their daily lives \cite{CortEtal17} and even much
longer periods of time in retrieving news events \cite{UitvHeal19}.  

One may think of the contiguity effect as analogous to the recency effect, but
taken from a different temporal reference frame.  The recency effect describes
the availability of items in memory as a function of their temporal proximity
to the present.  In contrast, the contiguity effect describes the availability
of items in memory as a function of their temporal proximity to a remembered
moment from the past.    This analogy between recency and contiguity suggested
a different class of models for memory, which we turn to in the next
subsection. 

\subsection{Temporal context models}
\label{sec:tcms}
\newcommand{\context}{\ensuremath{\mathbf{c}}}
\newcommand{\theitem}{\ensuremath{\mathbf{f}}}
\newcommand{\MCF}{\ensuremath{\mathbf{M}^{CF}}}
\renewcommand{\cIN}{\ensuremath{\mathbf{c}^{\textnormal{IN}}}}
\renewcommand{\fIN}{\ensuremath{\mathbf{f}^{\textnormal{IN}}}}
\newcommand{\jbitparam}{\ensuremath{\gamma}}
\newcommand{\taumin}{\ensuremath{\taustar}_{\textnormal{min}}}
In this subsection we describe the memory representations of a class of models
referred to as temporal context models
\cite<TCMs,>{HowaKaha02a,SedeEtal08,PolyEtal09}.  These models were
originally developed to account for recency and contiguity effects in free
recall.  TCMs have since been applied to other episodic memory tasks, and even
memory tasks that are not considered to tap episodic memory \cite{Loga21}.
In this subsection we will describe the basic properties of these models and
how they result in properties of memory.  We will discuss neuroscientific work
inspired by TCMs before describing some fundamental limitations that follow
from the form of temporal context.

Temporal context models make three important conceptual changes relative to the models we
have considered thus far in this chapter.  First, these models hypothesize a
vector representation of temporal context that changes gradually from
moment-to-moment.  We will specify this in more detail
below.  For now, we note that the temporal context vector shares at least some
features with the content of short term store.  Second, temporal context models do not attribute behavioral associations between items -- such as the contiguity
effect -- to direct connections formed between item
representations (as in Eq.~\ref{eq:Hebb}).  Rather, functional associations in temporal
context models are mediated by items' effects on temporal context and a
temporal context's ability to cue retrieval of items.  Third, temporal context
models assume that it is possible to reinstate a previous state of temporal
context.  This ``jump back in time'' is hypothesized to be associated with the
experience of episodic memory.

\subsubsection{Two interacting vector spaces: items and contexts}

In TCMs, there are two  interconnected vector spaces (Fig.~\ref{fig:STSCMR}b).
One vector space, which we will sometimes refer to as the item space, is
activated by items that are currently available, either by virtue of having
been presented by the experimenter or by virtue of having been recalled by the
participant.  
We refer to the cognitive representation of specific items as vectors
$\thestim$ and the vector corresponding to the item presented at time step $t$
as $\thestim_t$.  The other vector space, which we
will sometimes refer to as the context space maintains a state of
temporal context.  We will refer to the state of temporal context at time $t$
as $\context_t$.  Temporal  context is affected by items; the input at time
$t$, $\cIN_t$  is caused by $\thestim_t$, the item available at time $t$.  

Temporal context evolves gradually, retaining information contributed by
recent items:
\begin{equation}
	\context_t = \forgetparam \context_{t-1} + \cIN_t
	\label{eq:EvEq}
\end{equation}
That is, at each time step $t$, the new state of temporal context is given by
$\rho$ times the previous state of temporal context, plus the input caused by
$\thestim_t$, $\cIN_t$.  
As before, $0 < \forgetparam < 1$ so that 
in some formulations, $\forgetparam$ is allowed to vary as a function of time
(for instance to normalize the context vector) and/or can vary for different
components of the context vector as attention to different features changes
(e.g., due to different encoding tasks).   We assume that on the initial
presentation of an item in a randomly-assembled list of words, the inputs
caused by each item $\cIN$ are uncorrelated with one another and treat them as
random vectors.  Equation~\ref{eq:EvEq} shows that information caused by a
particular item persists after it is presented.  Recursively unwinding
Eq.~\ref{eq:EvEq} we find
\begin{equation}
	\context_t = \sum_{\tau=0}^{\infty} \forgetparam^{t-\tau}
	\cIN_{t-\tau} .
	\label{eq:EvEqunfolded}
\end{equation}
That is, the input pattern $\cIN$ caused by an item decays exponentially as
additional items are presented.

At any particular moment, recall is cued by the current state of temporal
context \emph{via} an associative matrix $\MCF$ that connects the context
layer (containing context vectors $\context$)  to the item layer (containing
item vectors $\thestim$).  Analogous to our simple Hebbian model
(Eq.~\ref{eq:Hebb}), the basic formulation provides an outer product
association between the context available prior to presentation of the current item and
the item itself
\begin{equation}
	\Delta \MCF = \theitem_t \context_{t-1}\mytranspose 
	\label{eq:MCF}
\end{equation}
This shift in indices ensures that the temporal context that cues $\theitem_t$
does not include information $\cIN_t$ that item itself caused. 

Equation~\ref{eq:MCF} resembles Eq.~\ref{eq:Hebb} in that it associates
two patterns \emph{via} an outer product. However, rather than associating two
items $\thestim$ and $\theresp$, $\MCF$ associates a context vector to an
item vector.   The context-to-item association means that a probe context
activates each item in the list to the extent the probe context resembles that
items' encoding context.   By analogy to Eq.~\ref{eq:Hebblistprobe}, 
\begin{equation}
	\MCF \context_p = \sum_t \left(\context_{t-1}\mytranspose\context_p\right)  \theitem_t 
	\label{eq:MCFcontextp}
\end{equation}
Because context changes gradually, this typically results in a weighted sum of
many items.  Temporal context models use a retrieval rule to probabilistically
select an item for recall.   These mechanisms are sometimes quite elaborate;
the key feature they share is that the probability of recalling a particular
item at a particular retrieval attempt depends not only on the degree to which
it is activated, but also on the activation of the other items in the list.
That is to say, items compete to be retrieved. 

\subsubsection{Recency effect}

We are in a position at this stage to understand why TCMs predict recency
effects in immediate and delayed free recall.
Combining Eq.~\ref{eq:EvEqunfolded} and Eq.~\ref{eq:MCFcontextp} we find, under the
assumption that the $\cIN$ during initial study of a random list are
orthorgonal to one another, that probing with the context available at the end
of the list, $\context_L$, gives back the items from the list weighted
exponentially:
\begin{equation}
		\MCF \context_L \propto \sum_t \rho^{L-t+1} \theitem_t
		\label{eq:TCMIFR}
\end{equation}
The exponential decay clearly provides a large advantage to items from the
end of the list, leading  naturally to a robust recency effect.  Introducing a
delay $D$ takes $\context_L \rightarrow \rho^{D} \context_L +
\textnormal{distractors}$, where the distractors ought to be orthogonal to the
list items.  This reduces the difference in activation between the last items
in the list and earlier items, resulting in a decrease in the magnitude of the recency
effect.

\subsection{Contiguity effect}
Thus far we have considered only the case where the input patterns $\cIN$
caused by the items in the list  are orthogonal to one another.  In this
subsection we study the effects of relaxing this assumption.  To make the
ideas clear, let's repeat an item at the end of a very long list of unrepeated
items and see how the resulting context cues the neighbors of the repeated
item.  We consider two possibilities.  In the first case, the repeated item
simply causes the same input that it did during the initial presentation of
the list. In the second case we consider the case that the repeated item
recovers the temporal context available when it was intially presented; that
the repeated item causes a jump back in time.  We will find that these two
hypotheses result in very different qualitative properties.

Let us label the time index at which an item is repeated as $r$, the position
at which the repeated item was initially presented as $i$ and study the
ability of $\cIN_r$ to cue items near $i$, $\theitem_{i + \textnormal{lag}}$.
We assume that $r$ is far in the future so that we can neglect
$\context_{i+\textnormal{lag}}\mytranspose \context_{r-1}$  and restrict our
attention to $\theitem_{i+\textnormal{lag}}\mytranspose \MCF \cIN_r$.
Suppose that the repeated item simply causes the same input at time step $r$
that it did when it was initially presented at time step $i$.  Because
$\cIN{i}$ persisted after time step $i$ (see Eqs.~\ref{eq:EvEq},
\ref{eq:EvEqunfolded}), this results in similarity to the context states that
followed time step $i$.   This similarity decreases exponentially with
$\textnormal{lag} > 0$.  Put another way, because  temporal context contains
information from recently presented items, $\cIN_i$ is similar to the temporal
context of items for which $i$ was in the recent past.  However, the same is
not true for items that \emph{preceded} item $i$.  For $\textnormal{lag} \leq
0$, information retrieved by item $i$ is not in the recent past -- item $i$ has
not been presented yet and there is no way the participant should be able to
predict a word in a random list.  Putting these considerations together, we
find that if $\cIN_r = \cIN_i$:
\begin{equation}
	\theitem_{i + \textnormal{lag}}\mytranspose \MCF \cIN_i  =
	\left\{\begin{array}{lr}
		0  & \textnormal{lag} \leq 0 \\
		\forgetparam^{\textnormal{lag}}    & \textnormal{lag} > 0
        \end{array}\right.
	\label{eq:TCMasymm}
\end{equation}
That is, if at time step $r$, the item at time step $i$ simply recovers the same input it caused
during encoding, $\cIN_r = \cIN_i$, this results in an asymmetric functional association
to its the neighbors. 

Now let's consider the case in which the repeated item recovers the state of
context available when it was initially presented, $\cIN_r = \context_{i-1}$.
This context includes information caused by the items that preceded item $i$.
This information also persists in temporal context after item $i$ was
presented.  Noting that the inner product is symmetric,
$\mathbf{v}\mytranspose \mathbf{u} =  \mathbf{u}\mytranspose \mathbf{v}$, we
conclude that in this case
\begin{equation}
	\theitem_{i + \textnormal{lag}}\mytranspose \MCF \context_i  \propto
			\forgetparam^{|\textnormal{lag}|}.
	\label{eq:TCMsymm}
\end{equation}
That is to say, retrieving the previous state of temporal context results in a
symmetric association that falls off exponentially as a function of
$|\textnormal{lag}|$.

In most free recall experiments, the shape of the contiguity effect includes
a contiguity effect in both the backward and forward direction, with a
reliable advantage for forward transitions (Fig.~\ref{fig:RecencyContiguity}b
is representative).  In TCMs, the pattern retrieved by item $i$ when it is
re-experienced at time step $r$ is a mixture of these two patterns:
\begin{equation}
	\cIN_r = \left(1-\jbitparam\right)\cIN_i + \jbitparam \context_i .
	\label{eq:JBIT}
\end{equation}
The value of $\jbitparam$ can be estimated from the data and is believed to
vary not only from participant to participant but also from one retrieval to
the next.  This makes sense of the finding that episodic memory
retrieval -- presumably related to the recovery of a previous state of temporal
context -- does not always succeed.   This property of episodic memory is
familiar to anyone who has bumped into a familiar person in a public place
(e.g., a grocery store) \ldots but been unable to actually remember any
details of the person's actual identity.  

\subsection{Neural evidence for temporal context models}

Temporal context models have benefitted from a relatively close connection to
work in cognitive neuroscience.  After all, if the long-term goal of this kind of
modeling is to develop a more-or-less literal model of the computations that
take place in the brain during memory encoding and retrieval it is essential
to compare hypotheses to the activity of neurons in the brain.   We briefly
point to three pieces of evidence that speak to the utility of TCMs in making
sense of human and also animal neuroscience.

First, the division of $\cIN$ into two components with distinct properties
(Eq.~\ref{eq:JBIT}) has been very productive in explaining otherwise isolated
findings in neuropsychology and cognitive neuroimaging.  To take a simple
example, imagine if it were possible to alter $\jbitparam$ across experimental
groups.  A group with a lower value of $\jbitparam$ ought to have difficulties
with vivid episodic memory recall, but also show a more asymmetric
contiguity effect in free recall.    This finding has been observed with
patients with medial temporal lobe amnesia \cite{PaloEtal19}, electrical stimulation
to the entorhinal cortex \cite{GoyaEtal18}, and participants who are
experiencing cognitive declines with aging, perhaps leading to Alzheimer's
disease \cite{QuenEtal15,TalaEtal21}.  Moreover, according to the models,
retrieved temporal context ought to be preferentially involved in particular
sorts of memory.  Consider an experiment where participants learn pairs
separated by long periods of time, \textsc{absence hollow \ldots hollow
pupil}.  If the second presentation of \textsc{hollow} can cause recovery of
its previous context (i.e., the $\cIN$ caused by \textsc{absence}), then
\textsc{absence} in effect becomes part of the temporal context for
\textsc{pupil}.  If $\jbitparam=0$, the model can still learn the pairwise
associations using the forward part of the contiguity effect.  Indeed, normal
human participants generalize \textsc{absence pupil} associations even though
\textsc{absence} and \textsc{pupil} were never experienced nearby in time.  As
it turns out, lesions to a brain region called the hippocampus -- which is
believed to be important in episodic memory -- cause a deficit in these
bridging or ``transitive'' associations in rodents while leaving the pairwise
associations unaffected \cite{BunsEich96}, just as if the hippocampus is
responsible for causing a recovery of temporal context.  A number of
neuroimaging studies have studied similar experimental paradigms in humans,
showing that the hippocampus and hippocampal-prefrontal interactions are
important in these transitive associations 
\cite{ZeitEtal12}.

One can also measure direct neural predictions from TCMs.  The most
characteristic prediction is the existence of a temporal context vector
$\context$, which should show temporal autocorrelation extending over
macroscopic periods of time -- at least tens of seconds.   One can construct a
vector of brain activity using many different methods.  For instance, it is
practical to record simultaneously from many individual neurons at once.
Taking the number of spikes for each of $N$ neurons averaged over, say, a one
second interval gives an $N$-dimensional vector.  One can then compute a
temporal autocorrelation function by comparing response vectors from
neighboring time points.  This type of analysis has shown robust evidence for
signals that are autocorrelated over seconds, minutes, and even hours or days
in a number of brain regions, notably the hippocampus and prefrontal cortex
\cite{MankEtal12,HymaEtal12,CaiEtal16}.  These studies have focused on rodents
because of the array of systems neuroscience tools that can be brought to bear
in rodents, but analogous results have been found with human fMRI
\cite{HsieEtal14}.

The most characteristic prediction of TCMs is that the state of temporal
context should be recovered when an episodic memory is retrieved
(Eq.~\ref{eq:TCMsymm}).  When item $i$ is repeated at some later time step
$r$, and causes an episodic memory, the context at time step $r$ should
resemble the context \emph{prior} to the context at time step $i$.  This is
non-trivial; any neural information that was caused by item $i$ during study
can only be observed after its original presentation.  There is evidence from
invasive  human recordings of this phenomenon in several human memory
paradigms \cite{MannEtal11,YaffEtal14,FolkEtal18}, fMRI studies of free recall  \cite{ChanEtal17}, and real-world memory extended over hours and days and weeks \cite{NielEtal15}.

\subsection{\label{sec:scaleinv} Memory is scale-invariant; exponential functions are not}
In our discussion of models of short-term memory, we noted that the failure of
short-term memory models to account for the long-term recency effect  and
long-term contiguity effects was a serious problem for those models.  It is
true that  TCMs are better able to account for those phenomena.  In STS-based models, the probability that an item is perfectly represented in STS 
falls off exponentially.  As time passes, STS provides zero information about
the item on an increasingly high proportion of trials.  In contrast, in TCMs
the information about an item falls off exponentially with time, but is
reliable across trials.  With a bit of resourcefullness and a few free
parameters, one can exploit this property to provide a reasonable fit to
experimental data from continuous distractor free recall.  But this account is
still theoretically unsatisfactory, as we shall see shortly.

\begin{figure}
\includegraphics[width=0.9\textwidth]{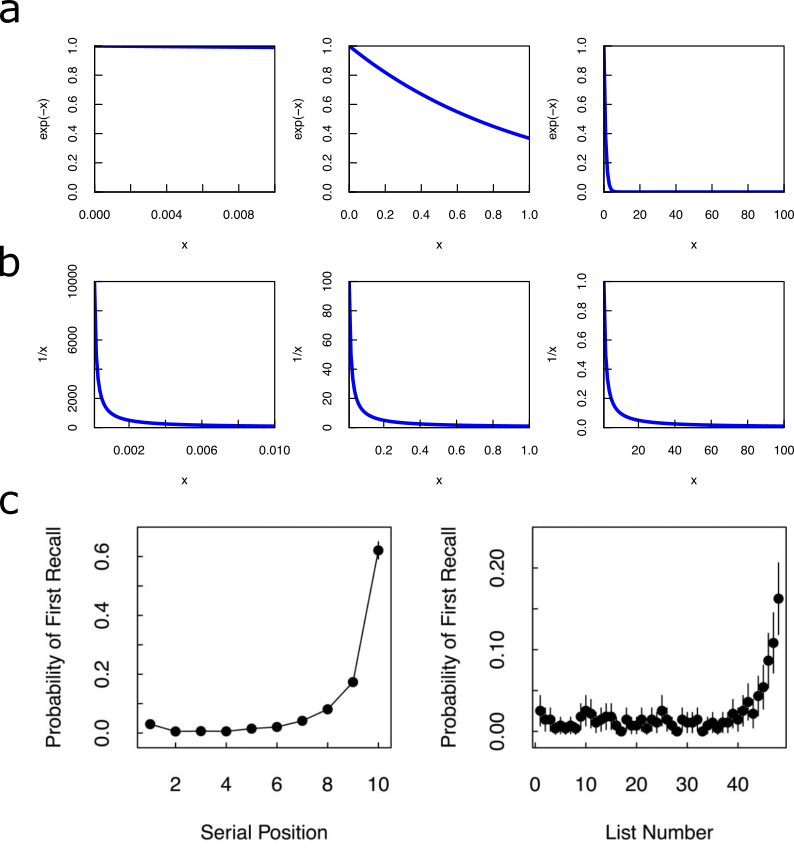}
	\caption{
		Scale-invariant memory. \textbf{a-b.}  Consider taking a variable $x$ and
		rescaling it $x \rightarrow ax$.  \textbf{a:} An 
		exponential function $e^{-x}$ zoomed in over different ranges
		of $x$.
		\textbf{b:} A power law function $x^{-1}$ zoomed in over
		different ranges of 
		$x$.   Starting from the middle panel, where $x$
		is shown over the range zero to 1, the left panels show the
		functions rescaled by zooming in on $x$ by a factor of 100;
		the right panels show the functions zoomed out by a factor of
		100.  Note that the exponential function has very
		different properties across scales.  In contrast the power law function has
		the same shape up to a scaling factor (note the change in the
		y axis) regardless of the scale over which it is examined.
		\textbf{c.} The recency effect in
		human memory persists across time scales.  Left: memory tested
		on the scale of seconds.  Right: memory tested on the scale of
		minutes.  Participants studied lists of words.   The left panel
		shows the probability that the first word that came to mind in
		a free recall task came from each position within the list.
		After learning 48~lists, participants were asked to recall all
		the words they could remember from all the lists in the
		experimental session.  The right panel plots the probability
		that the first word they recalled came from each \emph{list}
		in the session.
		 Note that the function has a similar shape
		across very different time scales.  After Howard, et al.,
		(2008).
		\label{fig:scale-invariance}
	}
\end{figure}

As discussed above, a great deal of evidence suggests that recency and
contiguity effects not only persist across a delay interval in CDFR, but are
observable at an extremely wide range of time scales
(Figure~\ref{fig:scale-invariance}c provides a particularly striking example).
This suggests that the memory representations governing recency and contiguity
effects are scale-invariant \cite{ChatBrow08}.  A function is said to be
scale-invariant if it is unaffected by rescaling the input up to a scaling
factor.  That is, a function $y(x)$ is said to be scale-invariant if
stretching or compressing its input by a constant, $x \rightarrow ax$, results
in the same function up to a constant term that depends only on $a$: $y(ax) =
f(a) y(x)$.  This property is true of power law functions that govern, say,
electrical potential as a
function of distance from a charged particle, or the gravitational field as a
function of distance from a massive object in Newtonian gravity.  We can
easily convince ourselves of this property by noting that if  $y(x) = x^{-1}$,
then $y(ax) = a^{-1} y(x)$, satisfying the constraint.
Figure~\ref{fig:scale-invariance}b illustrates this property for $y(x) =
x^{-1}$ by
rescaling the $x$ axis.

The exponential functions generated by TCMs are decidedly not scale-invariant.
Note that $\forgetparam^x = e^{-x}$ if we choose $\forgetparam = 1/e$.  More generally,
$\forgetparam^{x} = e^{-sx}$ if $\forgetparam= e^{-s}$ so that $s = - \log
\forgetparam$.  Thus, choosing a $\forgetparam$ is equivalent to specifying a
rate constant $s$ (or a time constant $1/s$) for an exponentially decaying
function.  Figure~\ref{fig:scale-invariance}a shows the function $y(x) =
e^{-x}$ rescaled over the same range of values as the power law function.
When $x$ is much less than one (left panel), the exponential function appears
linear.  This follows from the Taylor series expansion of the exponential
function:
\begin{equation}
	e^{- \Delta} = 1 - \Delta + \ldots
\end{equation}
where additional terms include higher powers of $\Delta$ multiplied by
$e^{-x}$.  As we zoom out (right panel), the exponential function comes to
approximate a delta function centered at zero.   Note that in both of these
two regimes $x \ll 1$ and $x \gg 1$, the exponential function is useless for
expressing a recency effect.   Mapping $x$ to recency, when $x$ is small,
there is no forgetting because all points are associated with a high nearly
constant value.  When $x$ is large, almost all points (excluding zero) are
mapped to a low nearly constant value.  

This rescaling is not an academic exercise.  CDFR approximates rescaling of
experience.  Insertion of a delay of duration $D$ between each item and at the
end of the list approximates taking $\forgetparam \rightarrow \forgetparam^D$,
so that the relative delay between serial positions relative to the time of
retrieval becomes effectively larger.   From this it is clear that, although
one may be able to approximate experimental data in restricted cases, the
machinery of the temporal context vector specified by Eq.~\ref{eq:EvEq} is not
scale-invariant and will eventually break down.  

\section{\label{sec:sith} Scale-invariant temporal history}

Thus far, we have considered models based on  more or less complicated
implementations of the idea of association.  In the case of the Hebbian
association model, the association is distributed across the entries in a
matrix corresponding roughly to the set of synapses between items.  In
temporal context models, associations between items are mediated by temporal
context, a representation of the recent past in which previous events decay
gradually.  These models share an implicit assumption that the goal of memory
is to express relationships as a scalar value.  That is, we can talk about the
relationship between, say \textsc{absence} and \textsc{hollow} only in terms
of the magnitude of the connection between them.  Given two pairs,
\textsc{absence---hollow} and \textsc{pupil---river}, the simple Hebbian model
does not have any mechanism to convey information about whether one pair was
learned before or after the second pair.  Yes, one might note that the
\textsc{absence---hollow} association is stronger than the
\textsc{pupil---river} association and use this to infer that
\textsc{absence---hollow} was more recent, but this inference would break
down if, for instance, the participant was paying less attention when
\textsc{pupil---river} was presented, or if \textsc{absence---hollow} was
presented multiple times.

Similar arguments apply to TCMs.  Although temporal relationships can be
inferred indirectly from the magnitude of the associations between multiple
words, there is no explicit information about the direction of time contained
in $\context_t$.  Consider two context vectors $\context_t$ and $\context_{t+
\textnormal{lag}}$.  The direction of the difference between these two vectors, 
$\context_{t+\textnormal{lag}} - \context_t$, depends on the particular choice
of items presented during the interval specified by $\textnormal{lag}$ rather
than the time \emph{per se}.  Moreover, as with simple Hebbian models,
repeated items can make even the magnitude of these vectors ambiguous.
The goal of the representation used in this section is to build a replacement
for the temporal context vector.  We desire that this representation carries
explicit information about temporal relationships.  We also desire that this
representation can be used to build scale-invariant models of memory.

\begin{figure}
	\centering
	\begin{tabular}{lclc}
		\textbf{\sf \Large a} && 
		\textbf{\sf \Large b} \\
		&	\includegraphics[width=0.3\textwidth]{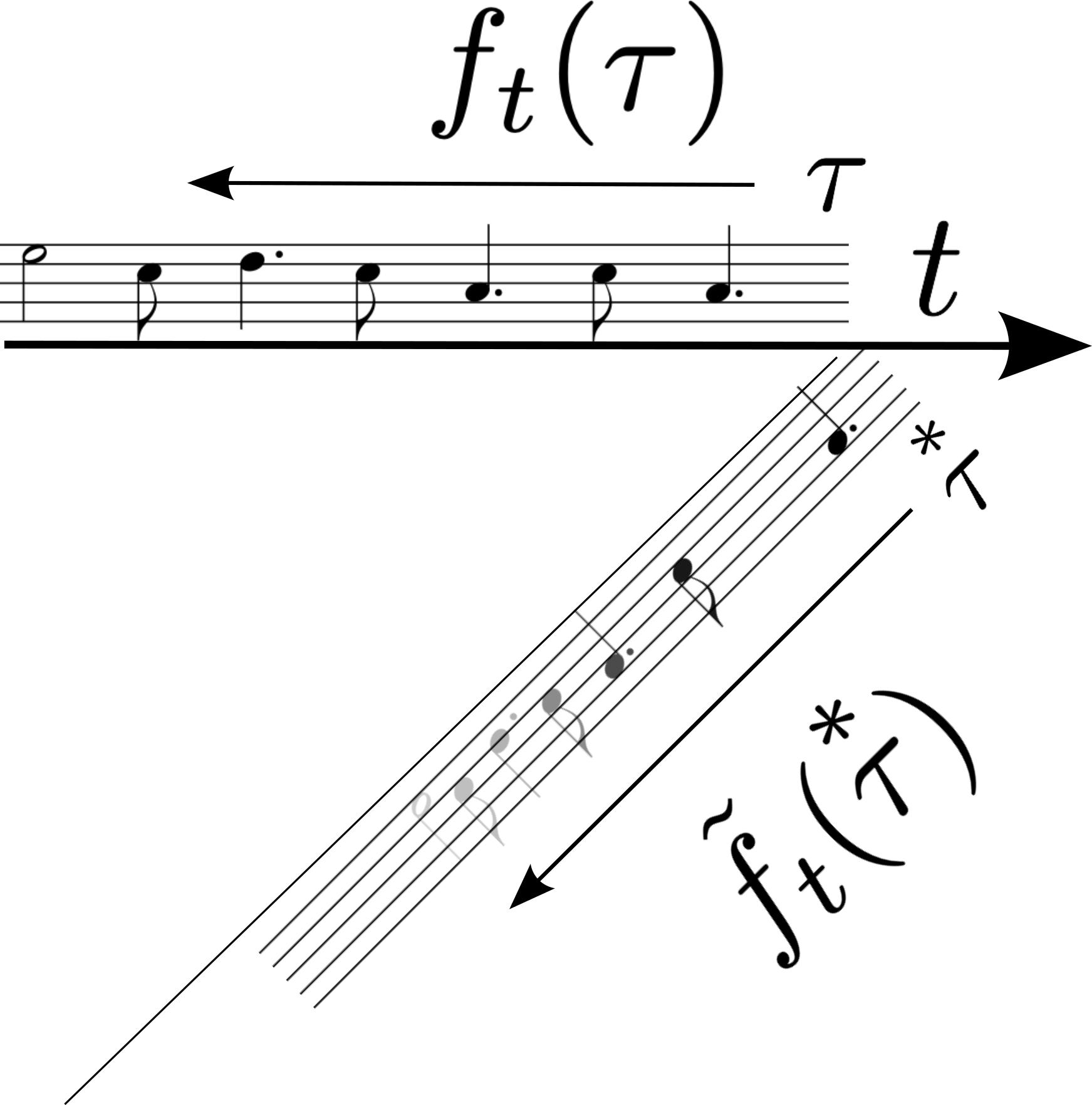}
		&& \includegraphics[width=0.3\textwidth]{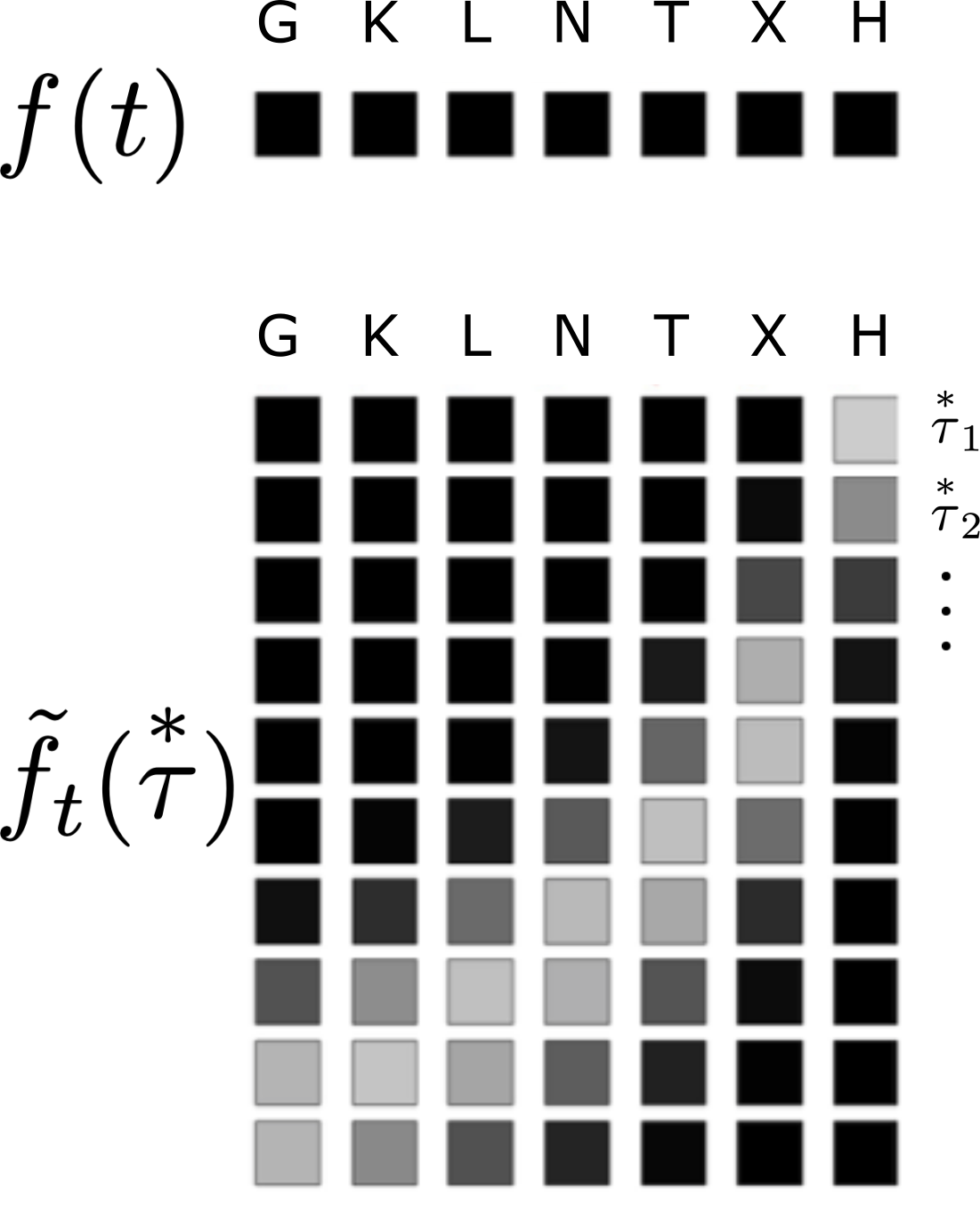}
	\end{tabular}
	\caption{
		Scale-invariant temporal history. \textbf{a.}  Cartoon
		illustrating the goal of the scale-invariant temporal history.
		At time $t$, the history leading up to the present is given by
		$f_t(\tau)$.  The argument $\tau$ runs from zero to $\infty$.
		The goal of the representation of temporal history is to
		construct at each moment a record of
		the recent past as a scale-invariant temporal history.  This history is
		compressed in that it has less temporal resolution for
		events further in the past.
		\textbf{b.} Schematic of the temporal history at a single
		moment shortly following
		presentation of a list \textsc{g k l n t x h}.  Each box gives
		the activation of a
		``unit'' at time $t$. Lighter boxes indicate higher
		activation.  Black boxes indicate zero activation.  Top:  As in
		TCMs, the input pattern $f(t)$ is a vector over items.  Here
		we assume that each item has an orthogonal representation; the
		features are sorted on their order of past presentation for
		ease of visualization.  Because we take $t$ to be shortly
		after presentation of the last item in the list, there is no
		activation in $f(t)$.  
		Bottom: The scale-invariant representation retains
		information about the past leading up to the present.  Here
		``columns'' are organized so that they correspond to the same
		features as in $f(t)$.  Columns correspond to what
		information.  Rows correspond to  when information.  
		For instance, at the top row, only
		the column corresponding to \textsc{h}, the last item in the
		list, is active.  For rows representing information further in
		the past, several items are active (note that the peaks for
		\textsc{k} and \textsc{l} overlap).  The curvature in the peak
		of activation across
		the list items is a consequence of the logarithmic compression of the
		internal time axis.   The greyscale changes
		across rows for ease of visualization.  In actuality, the
		peak of a stimulus a time $\tau$ in the past goes 
		down like $\tau^{-1}$.
		\label{fig:timekeepsonslippin}
	}
\end{figure}

Understanding vectors as activated populations of neurons, the simple Hebbian
model and temporal context vectors distribute ``what'' information about the
stimuli that are experienced across populations of neurons. 
Different basis vectors of the space correspond to different properties of
stimuli.   The temporal context vector provides decaying ``what'' information
``smeared'' over the recent past.  The strategy of this approach is to
construct a population of neurons that  not only represent information about
what has happened in the recent past, but to distribute information about when
it happened across different neurons.  That is, our computational goal is to
estimate the recent past as a function of time.
Figure~\ref{fig:timekeepsonslippin} provides an illustration and introduces
notation.   In this section we describe a specific solution to this problem
that has found considerable empirical support from data from both psychology
and neuroscience.

Let us suppose that the world provides a continuous stream of input
$f(t)$. Like the set of vectors corresponding to a list of words,  $f$ is in
general vector-valued but we will suppress vector notation for now.  Consider
the problem of an observer having examined $f$ up to a particular point $t$.
We will refer to the history leading up to this moment $t$  as $f_t(\tau)$,
where $\tau$ runs from zero to $\infty$ and $\tau=0$ corresponds to the
present.  Our goal is to construct an estimate of the history leading up to 
time $t$ as $\ftilde_t(\taustar)$.    We desire that this estimate approximates
reality -- with error that is comparable across time scales -- and is also a
computation that could be implemented by neural circuits. The next subsection
introduces a specific method that has these properties \cite<proposed by>{ShanHowa12}. Subsequent subsections demonstrate that it is straightforward to build not only temporal context models out of this form of representation but other more ``cognitive'' models as well. Finally, we touch on a wealth of neuroscience work that suggests that populations of neurons like those proposed for $\ftilde_t(\taustar)$.

\subsection{Estimating temporal relationships using the  Laplace
transform\label{sec:ftildetheory}}

This section describes a method for estimating $\ftilde_t(\taustar)$ based on Laplace transforms that was proposed by \citeA{ShanHowa12}. 
First let us write a continuous version of Eq.~\ref{eq:EvEq}.
For reasons that will become clear, we change notation such that the temporal
context vector $\context_t$ is written as $F(t)$ and the input
to the context vector $\cIN_t$ is written as $f(t)$.    We take both of these
to be vector-valued but will suppress the vector notation for present.
Defining $s = -\log \forgetparam$, this is just a continuous version of
Eq.~\ref{eq:EvEq}:
\begin{equation}
	\frac{dF}{dt} = -s F + f(t)
	\label{eq:LeakyIntegrator}
\end{equation}
Solving Eq.~\ref{eq:LeakyIntegrator} we find, in the general case:
\begin{equation}
	F_t(s) = \int_{0}^{\infty} e^{-s \tau} f_t(\tau) d\tau
	\label{eq:Fst}
\end{equation}
Comparing this to Eq.~\ref{eq:EvEqunfolded} we see a close correspondence
between $\context_t$ and $F_t$ if we make the identification $\forgetparam =
e^{-s}$.  In contrast to the TCMs we discussed in section~\ref{sec:tcms}, we
do not understand $s$ as a parameter to be estimated from the data of a
particular experiment, but as a continuous variable.  To be concrete, we can
imagine that we have an ensemble of units, each with a different value of $s$.

\subsubsection{Continuous $s$ enables information about continuous time}
Treating $s$ as a continuous variable allows us to reconstruct information
about the value of  $f_t(\tau)$  at different values of $\tau$.  With any
particular value $s_1$, $F_t(s_1)$ captures information about the past history
$f_t(\tau)$ up to a time scale on the order of $\tau_1 = 1/s_1$.  If we chose
a different value $s_2$, $F_t(s_2)$ would capture information up to $\tau_2 =
1/s_2$.  For simplicity, lets assume that $\tau_1 < \tau_2$.  Consider the
properties of the exponential function illustrated in
Figure~\ref{fig:scale-invariance}.  For values of $\tau$ much less than
$\tau_1$, both $F_t(s_1)$ and $F_t(s_2)$ weight $f_t(\tau)$ by similar
amounts.  Similarly, for values of $\tau$ much greater than $\tau_2$, both of
the exponential functions have decayed to zero and neither $F_t(s_1)$ nor
$F_t(s_2)$ carries information about $f(\tau)$ in that interval.  However,
consider how the two values of $F$ vary as $\tau$ increases from $\tau_1$  to
$\tau_2$ (recall that $\tau_1 < \tau_2$).   As $\tau$ passes through $\tau_1$,
the contribution of $f_t(\tau)$ to $F_t(s_1)$  rapidly decreases.  However,
the exponential for $F_t(s_2)$ decays less steeply in this region, so that the
contribution of these values to $F_t(s_2)$ is greater.  We conclude that one
can infer something about the values of $f_t(\tau)$ in a region specified by  
$\tau_1$ and $\tau_2$ by observing the difference between $F_t(s_1)$ and
$F_t(s_2)$.  Given many values of $s$ we can infer $f_t(\tau)$ at many values
of $\tau$.   

More formally, we can note that $F_t(s)$ from Eq.~\ref{eq:Fst} describes the
real Laplace transform of $f_t(\tau)$.  The Laplace transform is invertible;
if we know the value of $F_t(s)$ precisely with every real value of $s$ from
0 to $\infty$, then we can specify $f_t(\tau)$ precisely for every value of
$\tau$ from 0 to $\infty$.  We will restrict our attention to real positive
values of $s$.\footnote{Negative real values of $s$ would be neurally
unreasonable.  We ignore complex $s$ for simplicity.   }

\subsubsection{Approximately inverting the Laplace transform}
Now that we've established that $F_t(s)$ carries information about the time of
past events $f_t(\tau)$, we need to determine how to extract that information.
Knowing that $F_t(s)$ is the real Laplace transform of $f_t(\tau)$ suggests a
strategy -- simply invert the Laplace transform.  
That is,  $F_t(s)$ provides a memory for the past leading up to the present
$f_t(\tau)$.   After inverting the Laplace transform, we would obtain an  estimate
of the actual history, which we write as $\ftilde_t(\taustar)$.  Over the
years, many methods for the inverse  Laplace transform have been proposed.  We
focus on the Post approximation \cite{Post30}, which is relatively straightforward to
implement in neural circuits and has some computational properties that are
advantageous in describing psychological and neurophysiological results.

To approximately invert the transform, we define a mapping $\taustar \equiv
k/s$, where $k$ is an integer to be approximated from the data.  At each
moment, the value of $\ftilde$ at each value of $\taustar$ is computed as
\begin{equation}
	\ftilde_t(\taustar) \equiv \Lk F_t(s) = C_k s^{k+1} \frac{d^k\ }{ds^k}
	\ 
	F_t(s)
	\label{eq:ftildedef}
\end{equation}
The derviative on the right hand side is to be taken in the neighborhood of
the value of $s=k/\taustar$.  $C_k$ is a constant that ensures that the sign
and magnitude of $\ftilde_t(\taustar)$  corresponds to the sign and magnitude
of $f_t(\tau)$.  
The operator $\Lk$ includes a computation of the $k$th derivative with respect
to $s$.\footnote{Given a discrete set of $s$ values, $\Lk$ can be understood
as a matrix ${L_{ij}}$ that maps $F(s_j)$ onto $\ftilde(\taustar_i)$, with a
matrix implementation of the discrete derivative. } 
In the limit as $k\rightarrow \infty$, the Post approximation
becomes the inverse transform and $\ftilde_t(\taustar=\tau) = f_t(\tau)$.
However, for finite $k$, there is a temporal blur introduced.
$\ftilde_t(\taustar)$ is equal to an average of $f_t(\tau)$ in the
neighborhood around $\tau = \taustar$.
Suppose $f_t(\tau)$ is a delta function at a particular time $\tau_o$ in the
past.  Then 
\begin{eqnarray}
	\ftilde_t(\taustar) &=& C_k  \ s^{k+1}\  \frac{d^k\ }{ds^k} e^{-s\tau_o}\\
			& = &  C_k \ s^{k+1} \tau_o^k\  e^{-s \tau_o} \\
			&=  &  C_k \ 
			\frac{1}{\taustar}  \ 
				\left(\frac{\tau_o}{\taustar}\right)^k \ 
					e^{-k \left(\frac{\tau_o}{\taustar}\right)}
					\label{eq:impulseresp}
\end{eqnarray}
The constant $C_k$ includes a factor of $-1^k$ so that the right hand side of this
expression is positive for all $k$.   
The function on the right-hand side of Eq.~\ref{eq:impulseresp} is a product of
a growing power law and a decreasing exponential, resulting in a function that
has a single peak.  Freezing time at a particular $\tau_o$ and looking across
all $\taustar$, the peak comes at $\taustar = \tau_o \frac{k}{k+1}$.
Fixing a particular $\taustar$ and observing it through time as $\tau_o$
changes, the peak comes at $\tau_o = \taustar$.
The most important property of this expression is that the  right hand side
depends on the time $\tau_o$ only through ratio $\tau_o/\taustar$.  
Because of the linearity of Eq.~\ref{eq:Fst} and the linearity of $\Lk$, we
can write an expression for any history $f_t(\tau)$ as
\begin{eqnarray}
	\ftilde_t(\taustar)& =& \int_0^\infty 
			 C_k\ \frac{1}{\taustar}
			 \left(\frac{\tau}{\taustar}\right)^k 
				e^{-k \frac{\tau}{\taustar}}\ 
 					f_t(\tau)\ d\tau\\
		&=& \int_0^\infty
				\frac{1}{\taustar}\
				\Phi_k\left(\frac{\tau}{\taustar}\right)
				\
			f_t(\tau)\ d\tau  \label{eq:phik}\\
		& = &	\int_0^\infty \Phi_k(x)\  f_t\left(\taustar x\right) \ dx
		\label{eq:phikscale}
\end{eqnarray}
Where we have  defined $\Phi_k(x) \equiv x^{k} e^{-kx}$
and changed variables to $x \equiv \frac{\tau}{\taustar}$ in the last line.

\subsubsection{\label{sec:real}A note on biological realism}
As we will see later, these equations provide a reasonable description not
only of a memory representation that can be used to describe behavior in a
range of memory tasks, but also of neurophysiological data from a number of
brain regions.  The equations are in principle computable by
neurons -- Eq.~\ref{eq:LeakyIntegrator} simply requires slow time constants and
it has long been known that the brain can compute derivatives needed to
implement $\Lk$.   How literally should one take these equations?   
There is certainly a level of precision at which these equations are not a
correct description of the firing rate of neurons.  The author of this chapter
encourages the reader to take these equations seriously, but not literally.  

For instance, Eq.~\ref{eq:LeakyIntegrator} describes an instantaneous reaction
to an input in continuous time.  If one understands $f(t)$ as a stimulus under
external control this cannot be literally true.  Moreover, there are a number
of ways in which the brain could implement the slow rate constants in
Eq.~\ref{eq:LeakyIntegrator}, including recurrent connections, metabotropic
glutamate receptors \cite{GuoEtal21} and feedback loops between spiking and
intrinsic currents \cite{EgorEtal02,TigaEtal15}.  These mechanisms would all
have slightly different properties that would deviate from
Eq.~\ref{eq:LeakyIntegrator}.  However the larger point that firing for a
population of neurons decays roughly exponentially following a triggering
stimulus with a broad range of time constants may still be true.

Similarly the inverse operator $\Lk$ cannot be literally true.  One major
issue is that 
$\Lk$ is a linear operator.  Taken literally, linearity of the right hand side
of Eq.~\ref{eq:ftildedef} would require that every bit of information about
the change in $f(t)$ is reflected, at least a little bit, in
$\ftilde(\taustar)$, which seems unreasonable.  Another serious problem is
that empirical values of $k$ estimated from neural data can be quite high
\cite{CaoEtal21}.  This is a computational problem in that computing the
$k$th derivative becomes more and more sensitive to noise as $k$ increases
\cite{ShanHowa12}.  In real cortical circuits, recurrent feedback involving
networks of inhibitory interneurons works to dampen noise \cite{FersMill00}.  
Nonetheless, $\Lk$ captures some important phenomena of neural firing that
should be taken seriously.  First, the weights of $\Lk$ do not reflect
any type of learning or experience with the stimuli.  They only extract
information embedded in a population with different decay rates.  Second, the
shape of the receptive fields $\Lk$ predicts for  $\ftilde$  seem to agree
reasonably well with experiment \cite{HowaEtal14}, at least in cases with a
few discrete stimuli presented widely separated in time.   Third, the idea of using
derivatives with respect to $s$ as a signal to infer the time of a
stimulus presentation is a sound idea, even if the brain doesn't literally use
the Post approximation with $k=38$ (or some other very large value of $k$) to
extract this information.

\subsubsection{A logarithmic scale for past time}

Note that although Eq.~\ref{eq:phik} is written as an integral transform of
$f_t(\tau)$, it is not necessary to retain a detailed memory of $f_t(\tau)$.
Updating Equation~\ref{eq:LeakyIntegrator} requires only the preceding value
$F_{t-dt}(s)$ and the momentary value $f(t)$; there is no need to retain prior
values of $f$ above and beyond the information present in $F_t(s)$.
Moreover $\ftilde_t(\taustar)$ can be computed from $F_t(s)$.  We thus have a
choice to make about how much information to retain in $F_t(s)$.  That is, the
brain can't actually  have an infinite number of values of $s$. And there is
no reason \emph{a priori} to assume that the $s$ values that are sampled
should be evenly spaced.  Because $\taustar \equiv k/s$, choosing how to
distribute the $s$ also specifies how to distribute the $\taustar$.  
Equations~\ref{eq:phik}~and~\ref{eq:phikscale} suggest a specific choice for 
sampling $\taustar$.

Consider $\ftilde$ at  two nearby values of $\taustar$, which
we'll refer to as $\taustar_o$ and $\taustar_o + \epsilon$.  If we observe 
$\ftilde_t(\taustar_o)$ and find that it is at a high value, we know 
that $\ftilde_t(\taustar_o+\epsilon)$ is also likely to be at a high value.  
Conversely, if we observe that $\ftilde_t(\taustar_o)$ is close to zero, we
know that  $\ftilde_t(\taustar_o+\epsilon)$ is also likely to be close to
zero.  Because they are affected by nearby points in time, these two values of
$\ftilde$ are correlated with one another.  
Each value of $\taustar$ we sample costs us something (e.g., metabolic energy
for a brain, availability of RAM in a computer simulation, etc).  In the limit as $\epsilon
\rightarrow 0$, there is no benefit to measuring $\ftilde$ at a second value.
As $\epsilon$ increases from zero, the two values of $\ftilde$ provide different
information about the past and there is some benefit to counteract the cost of
sampling a second value of $\taustar$.  However, the benefit from a particular
number $\epsilon$ depends on the choice of the first $\taustar$.  To get an
intuition into why this is so, suppose that we start with a specific
$\taustar$ and specific $\epsilon$, then we vary $\taustar$ while keeping
$\epsilon$ fixed.  As we increase $\taustar$, 
the impact of a fixed value of $\epsilon$ becomes less and less.  This is true
because $\Phi$ in Eq.~\ref{eq:phik} depends only on the ratio
$\frac{\tau}{\taustar}$ and the difference between $\frac{\tau}{\taustar}$ and
$\frac{\tau}{\taustar + \epsilon}$ grows smaller as $\taustar$ increases for
all $\tau$.  If we adopt the strategy of choosing
$\epsilon$ so that each additional value of $\taustar$ provides the same
benefit, we arrive at a sampling strategy where the difference between
adjacent values of $\taustar$ goes up linearly with the value
$\taustar$. One can formalize this further.\footnote{For instance, it can be
shown that if $\ftilde$ is driven by white noise, the mutual information
between two values of $\ftilde$ sampled over time depends on the ratio of
their $\taustar$s   \cite<see Appendix A.1 of>{ShanHowa13}.  } 

Setting the spacing between adjacent samples of $\taustar$ to be proportional to the
starting value of $\taustar$  leads immediately to several properties.  First,
the ratio between adjacent values must be a constant,
\begin{eqnarray}
	\taustar_{n+1} - \taustar_n  =  c \taustar_n \implies
	\frac{\taustar_{n+1}}{\taustar_n}     =  1+ c
	\label{eq:1plusc}
\end{eqnarray}
Second, the number of units one observes with a particular value of $\taustar$
should go down with that value of $\taustar$:
\begin{equation}
	\frac{dn}{d\taustar} = \frac{1}{\taustar}
	\label{eq:numberdensity}
\end{equation}
This expression diverges at zero, which is obviously not physical.
One solution is to fix some minimum value of $\taustar$ that can be sampled
$\taumin$.\footnote{If it is important to sample zero, one could use some
other sampling scheme for values below some threshold in order to arrive at
zero \cite{HowaShan18}. } Third, the samples of $\taustar$ are evenly spaced
as a function of the
logarithm of $\taustar$:
\begin{eqnarray}
	\taustar_n &=& \left(1+c\right)^n\taumin\\
	n &=& \log_{1+c} \taustar_n  -  \log_{1+c} \taumin
	\label{eq:logscale}
\end{eqnarray}

This cluster of properties are quite theoretically satisfying.  Many  sensory
receptors in the mammalian brain sample continuous dimensions at
logarithmically spaced intervals.  For instance, the density of receptors on
the retina has long been known to decrease linearly with distance from the
center of the retina, as in Eq.~\ref{eq:numberdensity}, a property that
appears to be respected throughout early stages of the visual system in the
brain.  Psychologically, the logarithmic sampling of time
(Eq.~\ref{eq:logscale}) provides a close correspondence with the Weber-Fechner law from psychophysics, which states that the magnitude of a perceptual variable goes up linearly with the logarithm of the physical
stimulus that causes it.  
The Weber-Fechner law holds (at least approximately
over some range) for a number of simple stimulus dimensions (e.g., loudness of
a tone, pitch of a tone, length of lines, etc) and has been argued to hold for
perception of temporal intervals as well.  It would be quite elegant if the
brain distributes receptors along a time axis using the same mathematical
expression as receptors along the retina, resulting in a similar perceptual
invariance.  It is especially satisfying that the arguments leading to
logarithmic distribution of ``time receptors'' made no reference to these
data.  Rather, Eqs.~\ref{eq:1plusc}-\ref{eq:logscale} were derived from a
property of the Post approximation coupled with the argument that the brain
ought to equalize redundancy among the receptors.

\subsection{Behavioral models using scale-invariant temporal history
\label{sec:ftildebehav} }

The scale invariant temporal history described in
section~\ref{sec:ftildetheory} can be used to construct a wide variety of
behavioral models of memory.  It is straightforward to extend temporal context
models by using $\ftilde_t(\taustar)$  in place of $\context_t$.  The primary
result is that one obtains scale-invariant recency and contiguity effects
(Fig.~\ref{fig:scale-invariance}).   However, the temporal history
$\ftilde_t(\taustar)$ can also be used to construct computational models of
very different tasks that can not be readily modeled using temporal context
models.  Some of these tasks are believed to rely on different ``kinds of
memory'' than free recall.

\newcommand{\scale}{\ensuremath{\mathcal{S}}}
\newcommand{\trans}{\ensuremath{\mathcal{T}}}
\newcommand{\SITHM}{\ensuremath{\mathbf{M}}}
\subsubsection{Scale-invariant temporal context models}
TCMs rely on the temporal autocorrelation of the temporal context vector in order to generate recency and contiguity effects -- that is, even in a list of random words, the expectation of
$\context_t\mytranspose\context_{t+\textnormal{lag}}$ falls off gradually like
$\rho^{\textnormal{lag}}$.  However, exponential functions set a strong scale.
One can readily build a temporal context model using $\ftilde_t(\taustar)$ in
place of $\context_t$.  Rather than $\MCF$ associating context vectors to
items, one constructs an associative matrix for each $\taustar$: 
\begin{equation}
	\frac{d \SITHM(\taustar)}{dt} = f(t) \ftilde_t\ \mytranspose(\taustar)
	\label{eq:SITHM}
\end{equation}
Recall that $F(s)$ at a particular $s$ is essentially a temporal context
vector with $\rho=e^{-s}$.  If one imagines $\MCF(s)$ as the $\MCF$ matrix one
would get for each value of $s$ as a function of $s$, then $\SITHM(\taustar)$ is
just that matrix valued function of $s$, but with
the inverse transform applied.\footnote{The transform here would be applied from
the right: $ \SITHM(\taustar) = \MCF(s) \left[ \Lk\right]\mytranspose$.}
One may visualize $\SITHM(\taustar)$  for a particular $\taustar$ as a set of
connections between a particular row in
Figure~\ref{fig:timekeepsonslippin}b and the vector $f$.   
One obtains a probe as  $\fIN \equiv \sum_n \SITHM(\taustar_n)
\ftilde_p(\taustar_n)$.  
Each list item is activated to the extent that the
units in the temporal history when it was presented are also active in the
probe.  
One may visualize this operation with respect to
Figure~\ref{fig:timekeepsonslippin}b as follows.  When a particular item is
activated in $f(t)$, there is a particular pattern $\ftilde_t(\taustar)$.
That item is activated according to the match between $\ftilde_t(\taustar)$
and the probe $\ftilde_p(\taustar)$, summing over rows (corresponding to the
inner product) and columns (corresponding to the sum over $\taustar_n$).
In the case of a long list of non-repeating words, it can be shown
that this association falls off like a power law function \cite{HowaEtal15}.
This property makes TCMs built in this way scale-invariant.  It is thus
straightforward to build genuinely scale-invariant recency and contiguity
effects.


TCMs built from a scale-invariant temporal history also have qualitatively
different properties than TCMs that use only a single-scale temporal context
vector.    Consider a situation in which two items, \textsc{a} and
\textsc{b} are presented at a temporal separation of $\tau$ seconds.  The
temporal context for \textsc{b} has \textsc{a} presented $\tau$ seconds in the
past.  Let us repeat \textsc{a} and observe the prediction for \textsc{b} as
\textsc{a} recedes into the past.  First, in the case of a single temporal
context vector, the temporal context for \textsc{b} is just $\rho^{\tau}
\cIN_{\textnormal{A}}$.  When \textsc{a} is repeated (neglecting retrieval of
temporal context)  it again contributes a $\cIN_{\textnormal{A}}$ term to
the temporal context vector and \textsc{b} is cued by an amount proportional
to $\rho^{\tau}$.  But now consider what happens in the time after \textsc{a}
was repeated.  In the time following  repetition of \textsc{a}, the
magnitude of the $\cIN_{\textnormal{A}}$ component of the temporal context
vector decreases exponentially.  As a consequence \textsc{b} is cued less and
less as \textsc{a} recedes into the past after its repetition.  The behavior
is very different if temporal context is constructed from $\ftilde(\taustar)$.
As before, the temporal context that cues \textsc{b} is the representation of
\textsc{a} presented $\tau$ seconds in the past.   However, this corresponds
to an $\ftilde$ in which units triggered by \textsc{a} with $\taustar$ near
$\tau$ are active.  When \textsc{a} is repeated (again neglecting recovery of
temporal context), it again triggers a sequence of cells.   A time $t$ after
repeating \textsc{a}, the units with $\taustar$ near $t$ are active.  But if
$t \ll \tau$, these are different units than the ones that cue \textsc{b}.  As
the repetition of \textsc{a} recedes into the past, \textsc{b} is cued more as
$t$ approaches $\tau$ and then less as the sequence
passes through the units that form the temporal context for \textsc{b}.
Although the consequences of this property on models of free recall would be expected to be
relatively subtle (there are many items composing the temporal context and
retrieval of temporal context), this property could be extremely useful in
other behavioral applications (e.g., serial recall).


\renewcommand{\output}{\ensuremath{\mathbf{f}^{\textnormal{o}}}}
\newcommand{\attention}{\ensuremath{G}}
\newcommand{\timeoutput}{\ensuremath{{t}}}

\subsubsection{Probing a representation of what happened when}
The simple Hebbian model described in section~\ref{sec:SST} is a special case
of a class of distributed memory models called global match models.  The
name ``global match'' refers to the property that the probe is compared to one
composite memory $\myassoc$ that contains a mixture of information from all
of the items in memory.  Other distributed memory models made different
assumptions.  For instance, multitrace models \cite<e.g.,>{Hint84,ShifStey97}
assumed that memory is composed of a list of traces which can be selectively
accessed based on the probes one provides as part of a query of memory.  
Each
trace is a set of features stored at a particular time, closely analogous to
$\thestim_t$ in the simple Hebbian model and TCMs.

The temporal context model sketched above using $\ftilde_p(\taustar)$ as a
probe has the spirit of a global match model.  One builds an associative
$\SITHM(\taustar_n)$ and then takes a sum over both what and when information
in constructing the output of memory, $\fIN = \sum_n \SITHM(\taustar_n)
\ftilde_p(\taustar_n)$.  However, there are other ways one might query
$\ftilde(\taustar)$ to construct behavioral models of different memory tasks.
Multitrace models keep different elements of memory separate in a list.
Because it maintains separable information about what happened when, one can
understand  $\ftilde_t(\taustar)$  as a  multitrace model, albeit one where
the traces become more blurred together as time recedes into the past
(Figure~\ref{fig:timekeepsonslippin}b).
Behavioral modeling work has shown that by querying this representation in
different ways, it's possible to construct quantitative behavioral models of
different working memory tasks.

It is well established that people and animals can direct attention to a
restricted region of visual space.  Suppose that a participant maintains
fixation at a particular spot in a visual display for a few seconds (in
experiments a small spot is usually provided).  Now suppose that the
participant learns that something important will be presented in a particular
region above and to the left of the location that is being fixated.   It can
be shown that the ability to perceive visual information is greater if a
stimulus is presented in that region relative to a region where nothing in
particular is expected.  This increased perceptual and neural gain is referred
to as ``attention''.  

One can model attention, directed to particular regions of past time; this
capability is important in constructing behavioral  models of working memory
tasks.     Let us suppose that one can direct attention to particular regions
of the timeline  and then compute a vector-valued
output  like so:
\begin{equation}
	\output = \sum_n \ftilde(\taustar_n) \attention(\taustar_n)  
\end{equation}
Here $\attention(\taustar)$ is an attentional weight that can highlight the
contributions of items at different points in the past.   It is not reasonable
to suppose that attention can take the form of any arbitrary function over
$\taustar$.  Let us suppose three constraints on the form of attention.
First, attention can point at only one circumscribed region at a time.
The function for attention should have one peak at a particular index $n$.
Second, attention can be deployed over a wide region or a more narrow region
depending on the task demands.  To be concrete,  given that attention is
directed to a particular index $n$, one may imagine that the participant can
control whether attention extends to many nearby indices, falling off
gradually, or only extends to a few nearby indices, falling off more sharply.  Notice that
because of the spread in $\Phi$ over  $\taustar$ (e.g., see
Fig.~\ref{fig:timekeepsonslippin}b), even if attention was nonzero for exactly one
index $\taustar_n$, this would still allow information from nearby time points
to contribute to $\output$.  These simple assumptions  allow us to construct
very different behavioral models from the same memory representation.  

This flexibility is useful in modeling  working memory tasks.   Working memory
is a term used to describe a form of memory that stores information with high
precision for a short time.  Working memory is an intellectual descendent of
computational models based on STS and is believed to rely on brain regions
distinct from the regions responsible for episodic memory tasks like delayed
and continuous distractor free recall.  The first of these working memory
tasks is referred to as probe recognition; the second is judgment of recency
(JOR).  
In both tasks, the participant is presented with a short list of
highly-memorable stimuli -- to be concrete let's assume that the stimuli are
letters of the alphabet presented visually on a computer screen.  In both
tasks, the lists are relatively short (say 10~items) and the memory test is
given immediately.   In both tasks, the stimuli are repeated many times over an
experimental session lasting tens of minutes.  In both tasks, the participant
is given a probe consisting of letters for the memory test. The only
(important) way the tasks differ is in the judgment the participant must make
in response to the memory probe.  In probe recognition, the participants' job
is to press a button to indicate whether a probe stimulus was in the most
recent list or not.  Because the stimuli are repeated across many lists, the
task is really to judge whether the probe was presented in a relatively broad
region of time.  In the short-term JOR task, participants are given a pair of
probe stimuli and asked to select the probe stimulus that was presented more
recently.  Because both of the probe items came from the most recent list,
short-term JOR requires more fine-grained judgments of the temporal record of
the probe stimuli.

Although the details are beyond the scope of this chapter \cite{TigaEtal19},
a carefuly study of  accuracy and the amount of
time it takes participants to respond shows that although both tasks show a
robust recency effect, the manner in which memory is accessed is quite
different.  The findings from both experiments can be accomodated by models in
which one makes a decision based on how well a probe overlaps with $\output$,
$\mathbf{f}_p \mytranspose \output$.
The important difference between the model for probe recognition and JOR is
how attention is deployed.   In the model of probe recognition, attention is
deployed broadly such that it's constant over the list.  The overlap with the
probe is thus stronger for more recent items and this strength falls off like
a power law (Eq.~\ref{eq:phik}).  This provides a respectable model of probe
recognition \cite<see especially>{DonkNoso12}.  In short-term JOR the
pattern of results has long suggested that participants use what's called a
self-terminating serial scanning model. We can build a serial scanning model
over the scale-invariant temporal memory by supposing that the participant
first sets attention to the recent present, such that only
$\attention(\taustar_1)$ is one.  The participant then compares this output to
the memory probes.  After some very brief time, attention is shifted to a
slightly less recent time point, for instance only $\attention(\taustar_2)$ is
non-zero.  The decision terminates when a match is found.  One can visualize
this process with the help of Figure~\ref{fig:timekeepsonslippin}b.  
After studying the list \textsc{g k l n t x h} suppose the correct answer is
\textsc{x}.  The participant will not find a match to \textsc{x} looking at
the first several rows.   The amount of time it takes to find a match and
initiate a decision depends on how far in the past \textsc{x} was presented.
If instead the correct answer was \textsc{t} one would have to scan over a
longer distance  to find information about that probe, predicting a
correspondingly longer response time.  There are many more detailed
quantitative predictions that follow from these models that can be worked out.

The important point here is that it is only possible to construct such
distinct behavioral models because $\ftilde_t(\taustar)$ has separable
information about what happened when.  If the information about the time of
past events was stored as a single number, as in the temporal context vector,
it is much more difficult to imagine an attentional model, and certainly not
one that aligns as well to our current understanding of visual attention.

%
%

\subsection{Evidence for scale-invariant temporal history in the brain \label{sec:timecells}}

\nocite{BolkEtal17,MauEtal18,MacDEtal11}

Taken literally, $\ftilde_t(\taustar)$ specifies the properties of a
population  of neurons.    There is now extensive evidence for these
predictions; populations of neurons referred to as ``time cells''  behave much
as one might expect if they were implementing $\ftilde_t(\taustar)$.  
Let us
take $\ftilde_t(\taustar)$  literally -- as a description of the firing rate of
a population of neurons, each indexed by a particular value of $\taustar$.
Time cells have now been observed in rodents
\cite{PastEtal08,MacDEtal11,MellEtal15,TigaEtal17b} and  non-human
primates \cite{JinEtal09,TigaEtal18a,CruzEtal20} and have even received preliminary
support from studies in humans \cite{UmbaEtal20}.  Although the label ``time cells'' is most
frequently applied to neurons in the hippocampus, populations with similar
properties have been observed in a variety of prefrontal regions as well as
striatum.  These regions are believed to support different forms of memory.
For instance, hippocampus is believed to support episodic memory, prefrontal
regions are believed to support working memory, and striatum is believed to
support implicit memory.  If indeed different regions supporting different
kinds of memory show firing consistent with properties of
$\ftilde_t(\taustar)$, then this supports the hypothesis that behavioral
models for different kinds of memory all rely on the same form of
representation.

Consider how cells representing $\ftilde_t(\taustar)$ would change their
firing as a function of time following a delta function input at $t=0$.  Each
cell would start with a firing rate near zero. As $t$ approaches each cell's
value of $\taustar$, the firing rate of that cell would begin to increase, and
then decrease again as
$t$ becomes much larger than that cell's $\taustar$.  Different cells have
different values of $\taustar$, so cells in the population would fire in
sequence.  The duration each cell spends firing depends linearly on its value
of $\taustar$;  cells that fire later in the sequence should also fire for a
longer time.  Moreover, $\taustar$s are sampled evenly over $\log$ rather than
linear time, resulting in a decreasing number of cells that peak later in the
sequence.  Moreover, if the population carries information about what happened
when, different stimuli should trigger distinguishable sequences.  All of
these properties have been quantitatively demonstrated in multiple brain
regions, including hippocampus and prefrontal regions in monkey and rodent.
Moreover, time cells are observed in a wide variety of behavioral tasks
\cite{MacDEtal11,TigaEtal17b,TigaEtal18a,CruzEtal20,MellEtal15,JinEtal09},
including in cases where the animal is given no task at all, but simply passively observes stimuli \cite{Goh21}.

\begin{figure}
	\begin{tabular}{lll}
		\includegraphics[height=0.3\textheight]{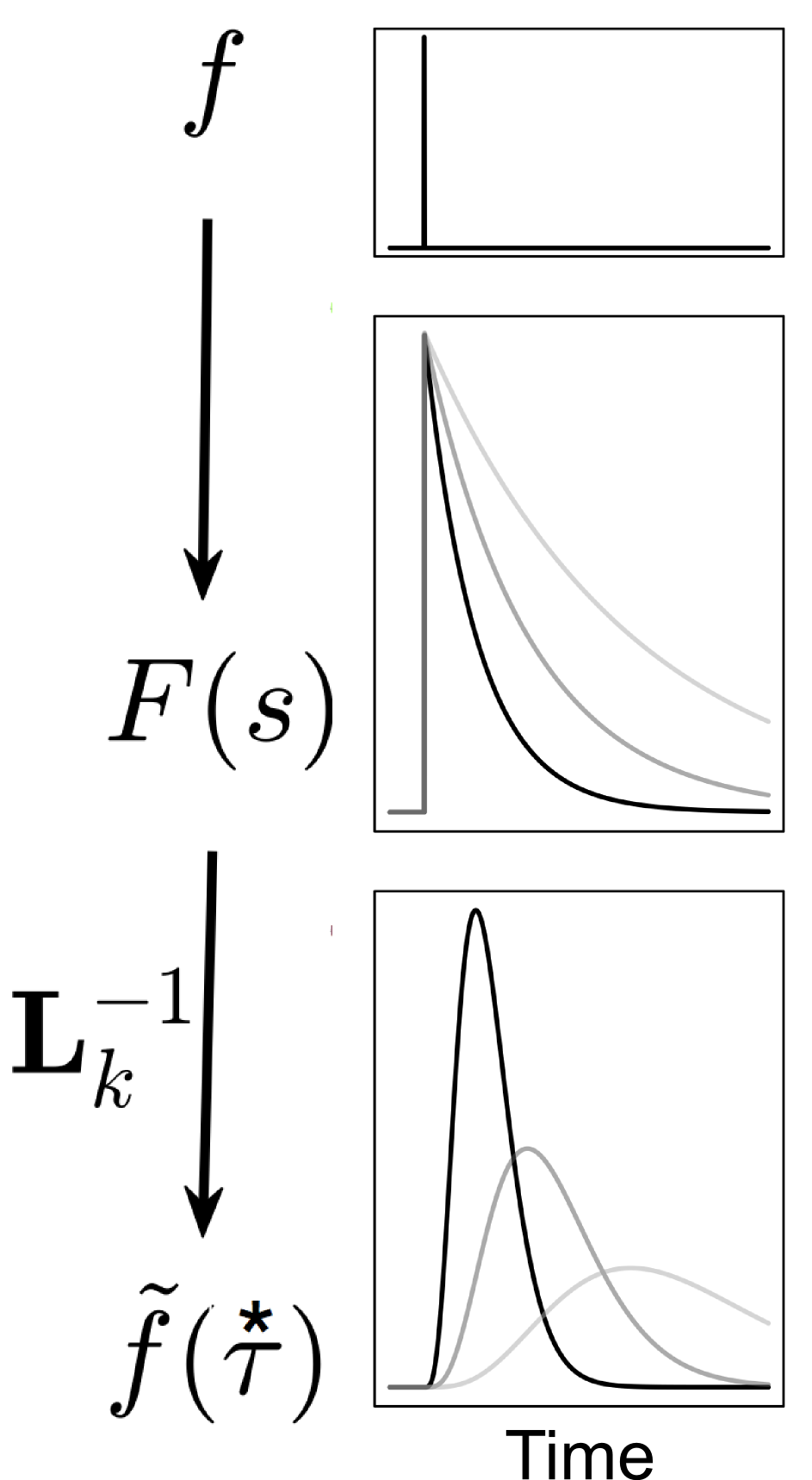}              
		&
	       	\includegraphics[height=0.3\textheight]{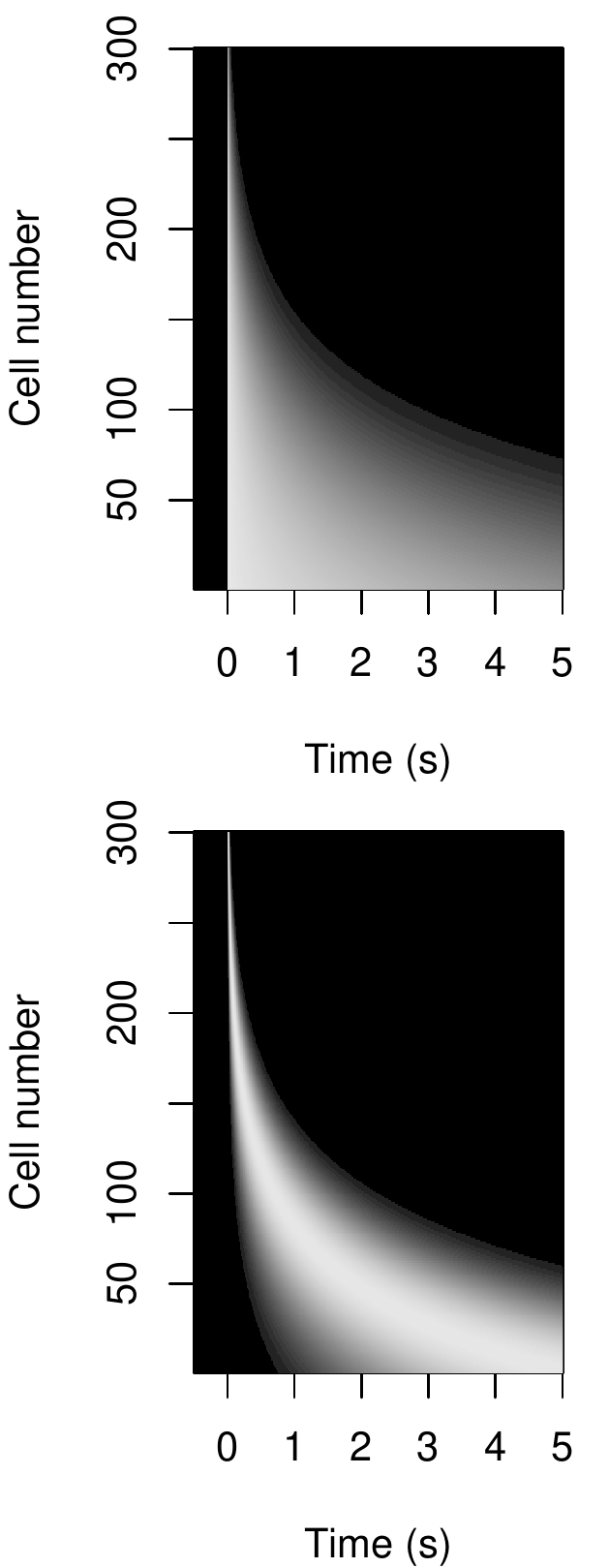}
	       	& 
	       \includegraphics[height=0.3\textheight]{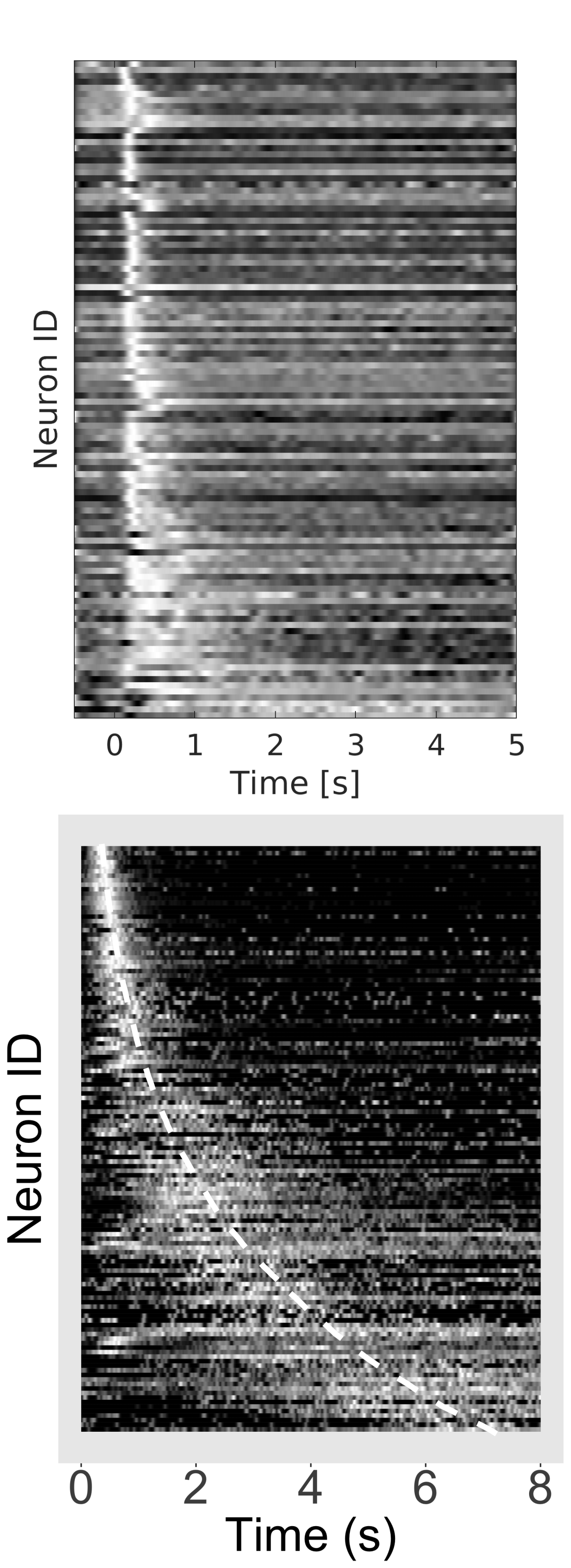}
	\end{tabular}
	\caption{ Laplace transform of the past captures properties of
	temporal context cells and time cells.  Left: Given a signal $f(t)$ as
	input, one can encode the real Laplace transform of the function
	leading up to the present using a bank of leaky integrators with rate
	constants $s$.  Given a delta function input at time zero, each
	integrator in $F(s)$ rises to one and then decays exponentially.  Each
	unit decays at a slightly different rate depending on that unit's
	value of $s$.  The leaky integrators provide input to another
	population $\ftilde$ constructed by approximating the
	inverse Laplace transform \emph{via} an operator $\Lk$.   Units in
	$\ftilde$  
	fire sequentially, with each cell peaking at a time controlled by the value of $s$ that provides input to it. 
	Middle: The two populations $F(s)$ (top) and
	$\ftilde$ (bottom) shown as heatmaps as a function of time
	to facilitate comparison with neurophysiological data.   Right: These
	representations resemble so-called ``temporal context cells'' in
	entorhinal cortex (top) and time cells in hippocampus (bottom).  Top
	after Bright, et al., (2020).  Bottom
	after Cao, et al.. (2021).  Ian Bright and Rui Cao helped with this
	figure. \label{fig:Laplace}
	}
\end{figure}
\nocite{BrigEtal20,CaoEtal21}

More recently, populations of neurons with properties like those predicted for
$F_t(s)$ have been observed in a brain region called the entorhinal cortex
\cite{TsaoEtal18,BrigEtal20}.  Because they so closely resemble components of
the temporal context vector (Eq.~\ref{eq:EvEq}), these kinds of cells have been
dubbed temporal context cells.    
The entorhinal cortex provides the major
projection to the hippocampus, where time cells were initially characterized.
Decades of neurophysiology, neuropsychology and cognitive
neuroscience have implicated the entorhinal cortex and hippocampus in human
episodic memory.  For instance, the famous amnesia patient Henry Molaison
(known prior to his death as H.M.) had bilateral damage to both the
hippocampus and entorhinal cortex.  Thus, a population of temporal context
cells, which resemble $F_t(s)$, project to a population of time cells, which
resemble $\ftilde_t(\taustar)$ in regions essential to human episodic memory.

\subsection{Going forward}

The convergence between theoretical considerations
(section~\ref{sec:ftildetheory}), behavioral models of memory
(section~\ref{sec:ftildebehav}) and neurophysiological findings
(section~\ref{sec:timecells}) seems very unlikely to happen by chance.  This
formalism could provide a foundation on which to build models of behavior and
cognition that are more or less literal descriptions of the computations
taking place in the brain.  Although a foundation may exist, the work of
constructing a complete theory of memory in the brain has barely begun.  Thus
far, the behavioral models that have been developed are sketches of important
effects.  A complete theory would require that these models be fleshed out to
provide a detailed description of behavior (like the models in
section~\ref{sec:ststcms}).  Development of such a theory would also require
careful neuroscientific studies across species and tasks informed by these
quantitative models of behavior.  Theoretically, the formalism for encoding
and inverting the Laplace transform of functions of time can be extended to
representing functions over other variables.  In this way it may prove
possible to connect computational models of memory to well-developed
computational models for spatial navigation, perception and simple
decision-making informed by neurobiological data.  

\section{\label{sec:related} Related literature}

This chapter necessarily touched on only a tiny fraction of the data and
computational models that have been used to understand human memory over the
years.  \citeA{Kaha12} provides a thorough introduction to behavioral models
of memory and important quantitative data from all the major human memory
paradigms.     

Stimulus sampling theory is much more rich than decribed in this chapter.   It
was rigorously developed by many researchers, with Stanford University
providing a focal point in the 1960s.  Students interested in stimulus
sampling theory should consider the following papers
\cite{AtkiEste62,Bowe67}.  

\citeA{AtkiShif68} is  a modeling \emph{tour de force} applying STS-based
behavioral models to many variants of cued and free recall.  It should be
considered required reading for mathematical psychologists interested in
modeling behavioral memory data.
\citeA{RaaiShif80} is a remarkably detailed desription of serial position
effects in free recall that relies heavily on ``fixed list context,'' an
important concept in models of this era that is not discussed here
\cite<see also>{CrisShif05}.

\citeA{Howa18} provides a high-level review of cognitive and neural data
related to the scale-invariant temporal history discussed in
section~\ref{sec:sith} \cite<see also>{HowaHass20}.
\citeA{HowaEtal15} built a number of simple cognitive models of behavioral
tasks corresponding to different ``kinds of memory''  and note how this
representation relates to distributed memory models.  \citeA{Lash51} provides
an eloquent critique of the limitations of  simple
associations  in describing memory that seems to anticipate many of the
properties of $\ftilde(\taustar)$ \cite<see also>{Jame90}.  There are also
interesting connections between the logarithmic temporal scale derived for
time here and measurement theory in mathematical psychology \cite<for an
overview see>{LuceSupp02} and exponential generalization \cite{Shep87}.



\bibliographystyle{apacite}
\bibliography{/Users/marc/doc/bibdesk}

\begin{thebibliography}{}

\bibitem [\protect \citeauthoryear {%
Anderson%
}{%
Anderson%
}{%
{\protect \APACyear {1972}}%
}]{%
Ande72}
\APACinsertmetastar {%
Ande72}%
\begin{APACrefauthors}%
Anderson, J\BPBI A.%
\end{APACrefauthors}%
\unskip\
\newblock
\APACrefYearMonthDay{1972}{}{}.
\newblock
{\BBOQ}\APACrefatitle {A simple neural network generating an interactive
  memory.} {A simple neural network generating an interactive memory.}{\BBCQ}
\newblock
\APACjournalVolNumPages{Mathematical Biosciences}{14}{}{197-220}.
\PrintBackRefs{\CurrentBib}

\bibitem [\protect \citeauthoryear {%
Anderson%
}{%
Anderson%
}{%
{\protect \APACyear {1973}}%
}]{%
Ande73}
\APACinsertmetastar {%
Ande73}%
\begin{APACrefauthors}%
Anderson, J\BPBI A.%
\end{APACrefauthors}%
\unskip\
\newblock
\APACrefYearMonthDay{1973}{}{}.
\newblock
{\BBOQ}\APACrefatitle {A theory for the recognition of items from short
  memorized lists.} {A theory for the recognition of items from short memorized
  lists.}{\BBCQ}
\newblock
\APACjournalVolNumPages{Psychological Review}{80}{}{417-438}.
\PrintBackRefs{\CurrentBib}

\bibitem [\protect \citeauthoryear {%
Atkinson%
\ \BBA {} Estes%
}{%
Atkinson%
\ \BBA {} Estes%
}{%
{\protect \APACyear {1962}}%
}]{%
AtkiEste62}
\APACinsertmetastar {%
AtkiEste62}%
\begin{APACrefauthors}%
Atkinson, R\BPBI C.%
\BCBT {}\ \BBA {} Estes, W\BPBI K.%
\end{APACrefauthors}%
\unskip\
\newblock
\APACrefYear{1962}.
\newblock
\APACrefbtitle {Stimulus sampling theory} {Stimulus sampling theory}\
  (\BNUM~48).
\newblock
\APACaddressPublisher{}{Citeseer}.
\PrintBackRefs{\CurrentBib}

\bibitem [\protect \citeauthoryear {%
Atkinson%
\ \BBA {} Shiffrin%
}{%
Atkinson%
\ \BBA {} Shiffrin%
}{%
{\protect \APACyear {1968}}%
}]{%
AtkiShif68}
\APACinsertmetastar {%
AtkiShif68}%
\begin{APACrefauthors}%
Atkinson, R\BPBI C.%
\BCBT {}\ \BBA {} Shiffrin, R\BPBI M.%
\end{APACrefauthors}%
\unskip\
\newblock
\APACrefYearMonthDay{1968}{}{}.
\newblock
{\BBOQ}\APACrefatitle {Human memory: A proposed system and its control
  processes} {Human memory: A proposed system and its control
  processes}.{\BBCQ}
\newblock
\BIn{} K\BPBI W.~Spence\ \BBA {} J\BPBI T.~Spence\ (\BEDS), \APACrefbtitle {The
  Psychology of Learning and Motivation} {The psychology of learning and
  motivation}\ (\BVOL~2, \BPG~89-105).
\newblock
\APACaddressPublisher{New York}{Academic Press}.
\PrintBackRefs{\CurrentBib}

\bibitem [\protect \citeauthoryear {%
Baddeley%
\ \BBA {} Hitch%
}{%
Baddeley%
\ \BBA {} Hitch%
}{%
{\protect \APACyear {1977}}%
}]{%
BaddHitc77}
\APACinsertmetastar {%
BaddHitc77}%
\begin{APACrefauthors}%
Baddeley, A\BPBI D.%
\BCBT {}\ \BBA {} Hitch, G\BPBI J.%
\end{APACrefauthors}%
\unskip\
\newblock
\APACrefYearMonthDay{1977}{}{}.
\newblock
{\BBOQ}\APACrefatitle {Recency reexamined} {Recency reexamined}.{\BBCQ}
\newblock
\BIn{} S.~Dornic\ (\BED), \APACrefbtitle {Attention and Performance {VI}}
  {Attention and performance {VI}}\ (\BPG~647-667).
\newblock
\APACaddressPublisher{Hillsdale, NJ}{Erlbaum}.
\PrintBackRefs{\CurrentBib}

\bibitem [\protect \citeauthoryear {%
Balsam%
\ \BBA {} Gallistel%
}{%
Balsam%
\ \BBA {} Gallistel%
}{%
{\protect \APACyear {2009}}%
}]{%
BalsGall09}
\APACinsertmetastar {%
BalsGall09}%
\begin{APACrefauthors}%
Balsam, P\BPBI D.%
\BCBT {}\ \BBA {} Gallistel, C\BPBI R.%
\end{APACrefauthors}%
\unskip\
\newblock
\APACrefYearMonthDay{2009}{}{}.
\newblock
{\BBOQ}\APACrefatitle {Temporal maps and informativeness in associative
  learning.} {Temporal maps and informativeness in associative
  learning.}{\BBCQ}
\newblock
\APACjournalVolNumPages{Trends in Neuroscience}{32}{2}{73--78}.
\PrintBackRefs{\CurrentBib}

\bibitem [\protect \citeauthoryear {%
Bjork%
\ \BBA {} Whitten%
}{%
Bjork%
\ \BBA {} Whitten%
}{%
{\protect \APACyear {1974}}%
}]{%
BjorWhit74}
\APACinsertmetastar {%
BjorWhit74}%
\begin{APACrefauthors}%
Bjork, R\BPBI A.%
\BCBT {}\ \BBA {} Whitten, W\BPBI B.%
\end{APACrefauthors}%
\unskip\
\newblock
\APACrefYearMonthDay{1974}{}{}.
\newblock
{\BBOQ}\APACrefatitle {Recency-sensitive retrieval processes in long-term free
  recall} {Recency-sensitive retrieval processes in long-term free
  recall}.{\BBCQ}
\newblock
\APACjournalVolNumPages{Cognitive Psychology}{6}{}{173-189}.
\PrintBackRefs{\CurrentBib}

\bibitem [\protect \citeauthoryear {%
Bolkan%
\ \protect \BOthers {.}}{%
Bolkan%
\ \protect \BOthers {.}}{%
{\protect \APACyear {2017}}%
}]{%
BolkEtal17}
\APACinsertmetastar {%
BolkEtal17}%
\begin{APACrefauthors}%
Bolkan, S\BPBI S.%
, Stujenske, J\BPBI M.%
, Parnaudeau, S.%
, Spellman, T\BPBI J.%
, Rauffenbart, C.%
, Abbas, A\BPBI I.%
\BDBL {}Kellendonk, C.%
\end{APACrefauthors}%
\unskip\
\newblock
\APACrefYearMonthDay{2017}{}{}.
\newblock
{\BBOQ}\APACrefatitle {Thalamic projections sustain prefrontal activity during
  working memory maintenance} {Thalamic projections sustain prefrontal activity
  during working memory maintenance}.{\BBCQ}
\newblock
\APACjournalVolNumPages{Nature Neuroscience}{20}{7}{987--996}.
\PrintBackRefs{\CurrentBib}

\bibitem [\protect \citeauthoryear {%
Bower%
}{%
Bower%
}{%
{\protect \APACyear {1967}}%
}]{%
Bowe67}
\APACinsertmetastar {%
Bowe67}%
\begin{APACrefauthors}%
Bower, G\BPBI H.%
\end{APACrefauthors}%
\unskip\
\newblock
\APACrefYearMonthDay{1967}{}{}.
\newblock
{\BBOQ}\APACrefatitle {A multicomponent theory of the memory trace} {A
  multicomponent theory of the memory trace}.{\BBCQ}
\newblock
\BIn{} K\BPBI W.~Spence\ \BBA {} J\BPBI T.~Spence\ (\BEDS), \APACrefbtitle {The
  Psychology of Learning and Motivation : Advances in Research and Theory} {The
  psychology of learning and motivation : Advances in research and theory}\
  (\BVOL~1, \BPG~229-325).
\newblock
\APACaddressPublisher{New York}{Academic Press}.
\PrintBackRefs{\CurrentBib}

\bibitem [\protect \citeauthoryear {%
Bright%
\ \protect \BOthers {.}}{%
Bright%
\ \protect \BOthers {.}}{%
{\protect \APACyear {2020}}%
}]{%
BrigEtal20}
\APACinsertmetastar {%
BrigEtal20}%
\begin{APACrefauthors}%
Bright, I\BPBI M.%
, Meister, M\BPBI L\BPBI R.%
, Cruzado, N\BPBI A.%
, Tiganj, Z.%
, Buffalo, E\BPBI A.%
\BCBL {}\ \BBA {} Howard, M\BPBI W.%
\end{APACrefauthors}%
\unskip\
\newblock
\APACrefYearMonthDay{2020}{}{}.
\newblock
{\BBOQ}\APACrefatitle {A temporal record of the past with a spectrum of time
  constants in the monkey entorhinal cortex} {A temporal record of the past
  with a spectrum of time constants in the monkey entorhinal cortex}.{\BBCQ}
\newblock
\APACjournalVolNumPages{Proceedings of the National Academy of
  Sciences}{117}{}{20274-20283}.
\PrintBackRefs{\CurrentBib}

\bibitem [\protect \citeauthoryear {%
Bunsey%
\ \BBA {} Eichenbaum%
}{%
Bunsey%
\ \BBA {} Eichenbaum%
}{%
{\protect \APACyear {1996}}%
}]{%
BunsEich96}
\APACinsertmetastar {%
BunsEich96}%
\begin{APACrefauthors}%
Bunsey, M.%
\BCBT {}\ \BBA {} Eichenbaum, H\BPBI B.%
\end{APACrefauthors}%
\unskip\
\newblock
\APACrefYearMonthDay{1996}{}{}.
\newblock
{\BBOQ}\APACrefatitle {Conservation of hippocampal memory function in rats and
  humans} {Conservation of hippocampal memory function in rats and
  humans}.{\BBCQ}
\newblock
\APACjournalVolNumPages{Nature}{379}{6562}{255-257}.
\PrintBackRefs{\CurrentBib}

\bibitem [\protect \citeauthoryear {%
Bush%
\ \BBA {} Mosteller%
}{%
Bush%
\ \BBA {} Mosteller%
}{%
{\protect \APACyear {1951}}%
}]{%
BushMost51}
\APACinsertmetastar {%
BushMost51}%
\begin{APACrefauthors}%
Bush, R\BPBI R.%
\BCBT {}\ \BBA {} Mosteller, F.%
\end{APACrefauthors}%
\unskip\
\newblock
\APACrefYearMonthDay{1951}{}{}.
\newblock
{\BBOQ}\APACrefatitle {A mathematical model for simple learning} {A
  mathematical model for simple learning}.{\BBCQ}
\newblock
\APACjournalVolNumPages{Psychological Review}{58}{}{313-323}.
\PrintBackRefs{\CurrentBib}

\bibitem [\protect \citeauthoryear {%
Cai%
\ \protect \BOthers {.}}{%
Cai%
\ \protect \BOthers {.}}{%
{\protect \APACyear {2016}}%
}]{%
CaiEtal16}
\APACinsertmetastar {%
CaiEtal16}%
\begin{APACrefauthors}%
Cai, D\BPBI J.%
, Aharoni, D.%
, Shuman, T.%
, Shobe, J.%
, Biane, J.%
, Song, W.%
\BDBL {}Silva, A.%
\end{APACrefauthors}%
\unskip\
\newblock
\APACrefYearMonthDay{2016}{}{}.
\newblock
{\BBOQ}\APACrefatitle {A shared neural ensemble links distinct contextual
  memories encoded close in time} {A shared neural ensemble links distinct
  contextual memories encoded close in time}.{\BBCQ}
\newblock
\APACjournalVolNumPages{Nature}{534}{7605}{115--118}.
\PrintBackRefs{\CurrentBib}

\bibitem [\protect \citeauthoryear {%
Cao%
, Bladon%
, Charczynski%
, Hasselmo%
\BCBL {}\ \BBA {} Howard%
}{%
Cao%
\ \protect \BOthers {.}}{%
{\protect \APACyear {2021}}%
}]{%
CaoEtal21}
\APACinsertmetastar {%
CaoEtal21}%
\begin{APACrefauthors}%
Cao, R.%
, Bladon, J\BPBI H.%
, Charczynski, S\BPBI J.%
, Hasselmo, M.%
\BCBL {}\ \BBA {} Howard, M.%
\end{APACrefauthors}%
\unskip\
\newblock
\APACrefYearMonthDay{2021}{}{}.
\newblock
{\BBOQ}\APACrefatitle {Internally generated time in the rodent hippocampus is
  logarithmically compressed} {Internally generated time in the rodent
  hippocampus is logarithmically compressed}.{\BBCQ}
\newblock
\APACjournalVolNumPages{bioRxiv}{2021.10.25.465750}{}{}.
\PrintBackRefs{\CurrentBib}

\bibitem [\protect \citeauthoryear {%
Chan%
, Applegate%
, Morton%
, Polyn%
\BCBL {}\ \BBA {} Norman%
}{%
Chan%
\ \protect \BOthers {.}}{%
{\protect \APACyear {2017}}%
}]{%
ChanEtal17}
\APACinsertmetastar {%
ChanEtal17}%
\begin{APACrefauthors}%
Chan, S\BPBI C.%
, Applegate, M\BPBI C.%
, Morton, N\BPBI W.%
, Polyn, S\BPBI M.%
\BCBL {}\ \BBA {} Norman, K\BPBI A.%
\end{APACrefauthors}%
\unskip\
\newblock
\APACrefYearMonthDay{2017}{}{}.
\newblock
{\BBOQ}\APACrefatitle {Lingering representations of stimuli influence recall
  organization} {Lingering representations of stimuli influence recall
  organization}.{\BBCQ}
\newblock
\APACjournalVolNumPages{Neuropsychologia}{97}{}{72--82}.
\PrintBackRefs{\CurrentBib}

\bibitem [\protect \citeauthoryear {%
Chater%
\ \BBA {} Brown%
}{%
Chater%
\ \BBA {} Brown%
}{%
{\protect \APACyear {2008}}%
}]{%
ChatBrow08}
\APACinsertmetastar {%
ChatBrow08}%
\begin{APACrefauthors}%
Chater, N.%
\BCBT {}\ \BBA {} Brown, G\BPBI D\BPBI A.%
\end{APACrefauthors}%
\unskip\
\newblock
\APACrefYearMonthDay{2008}{}{}.
\newblock
{\BBOQ}\APACrefatitle {From universal laws of cognition to specific cognitive
  models} {From universal laws of cognition to specific cognitive
  models}.{\BBCQ}
\newblock
\APACjournalVolNumPages{Cognitive Science}{32}{1}{36-67}.
\newblock
\begin{APACrefDOI} \doi{10.1080/03640210701801941} \end{APACrefDOI}
\PrintBackRefs{\CurrentBib}

\bibitem [\protect \citeauthoryear {%
Criss%
\ \BBA {} Shiffrin%
}{%
Criss%
\ \BBA {} Shiffrin%
}{%
{\protect \APACyear {2005}}%
}]{%
CrisShif05}
\APACinsertmetastar {%
CrisShif05}%
\begin{APACrefauthors}%
Criss, A\BPBI H.%
\BCBT {}\ \BBA {} Shiffrin, R\BPBI M.%
\end{APACrefauthors}%
\unskip\
\newblock
\APACrefYearMonthDay{2005}{}{}.
\newblock
{\BBOQ}\APACrefatitle {List discrimination in associative recognition and
  implications for representation} {List discrimination in associative
  recognition and implications for representation}.{\BBCQ}
\newblock
\APACjournalVolNumPages{Journal Experimental Psychology: Learning, Memory and
  Cogntion}{31}{6}{1199-212}.
\newblock
\begin{APACrefDOI} \doi{10.1037/0278-7393.31.6.1199} \end{APACrefDOI}
\PrintBackRefs{\CurrentBib}

\bibitem [\protect \citeauthoryear {%
Cruzado%
, Tiganj%
, Brincat%
, Miller%
\BCBL {}\ \BBA {} Howard%
}{%
Cruzado%
\ \protect \BOthers {.}}{%
{\protect \APACyear {2020}}%
}]{%
CruzEtal20}
\APACinsertmetastar {%
CruzEtal20}%
\begin{APACrefauthors}%
Cruzado, N\BPBI A.%
, Tiganj, Z.%
, Brincat, S\BPBI L.%
, Miller, E\BPBI K.%
\BCBL {}\ \BBA {} Howard, M\BPBI W.%
\end{APACrefauthors}%
\unskip\
\newblock
\APACrefYearMonthDay{2020}{}{}.
\newblock
{\BBOQ}\APACrefatitle {Conjunctive representation of what and when in monkey
  hippocampus and lateral prefrontal cortex during an associative memory task}
  {Conjunctive representation of what and when in monkey hippocampus and
  lateral prefrontal cortex during an associative memory task}.{\BBCQ}
\newblock
\APACjournalVolNumPages{Hippocampus}{30}{}{1332-1346}.
\PrintBackRefs{\CurrentBib}

\bibitem [\protect \citeauthoryear {%
Davelaar%
, Goshen-{G}ottstein%
, Ashkenazi%
, Haarmann%
\BCBL {}\ \BBA {} Usher%
}{%
Davelaar%
\ \protect \BOthers {.}}{%
{\protect \APACyear {2005}}%
}]{%
DaveEtal05}
\APACinsertmetastar {%
DaveEtal05}%
\begin{APACrefauthors}%
Davelaar, E\BPBI J.%
, Goshen-{G}ottstein, Y.%
, Ashkenazi, A.%
, Haarmann, H\BPBI J.%
\BCBL {}\ \BBA {} Usher, M.%
\end{APACrefauthors}%
\unskip\
\newblock
\APACrefYearMonthDay{2005}{}{}.
\newblock
{\BBOQ}\APACrefatitle {The demise of short-term memory revisited: empirical and
  computational investigations of recency effects} {The demise of short-term
  memory revisited: empirical and computational investigations of recency
  effects}.{\BBCQ}
\newblock
\APACjournalVolNumPages{Psychological Review}{112}{1}{3-42}.
\PrintBackRefs{\CurrentBib}

\bibitem [\protect \citeauthoryear {%
Deitch%
, Rubin%
\BCBL {}\ \BBA {} Ziv%
}{%
Deitch%
\ \protect \BOthers {.}}{%
{\protect \APACyear {2020}}%
}]{%
DeitEtal20}
\APACinsertmetastar {%
DeitEtal20}%
\begin{APACrefauthors}%
Deitch, D.%
, Rubin, A.%
\BCBL {}\ \BBA {} Ziv, Y.%
\end{APACrefauthors}%
\unskip\
\newblock
\APACrefYearMonthDay{2020}{}{}.
\newblock
{\BBOQ}\APACrefatitle {Representational drift in the mouse visual cortex}
  {Representational drift in the mouse visual cortex}.{\BBCQ}
\newblock
\APACjournalVolNumPages{bioRxiv}{}{}{}.
\PrintBackRefs{\CurrentBib}

\bibitem [\protect \citeauthoryear {%
Donkin%
\ \BBA {} Nosofsky%
}{%
Donkin%
\ \BBA {} Nosofsky%
}{%
{\protect \APACyear {2012}}%
}]{%
DonkNoso12}
\APACinsertmetastar {%
DonkNoso12}%
\begin{APACrefauthors}%
Donkin, C.%
\BCBT {}\ \BBA {} Nosofsky, R\BPBI M.%
\end{APACrefauthors}%
\unskip\
\newblock
\APACrefYearMonthDay{2012}{}{}.
\newblock
{\BBOQ}\APACrefatitle {A Power-Law Model of Psychological Memory Strength in
  Short- and Long-Term Recognition} {A power-law model of psychological memory
  strength in short- and long-term recognition}.{\BBCQ}
\newblock
\APACjournalVolNumPages{Psychological Science}{}{}{}.
\newblock
\begin{APACrefDOI} \doi{10.1177/0956797611430961} \end{APACrefDOI}
\PrintBackRefs{\CurrentBib}

\bibitem [\protect \citeauthoryear {%
Ebbinghaus%
}{%
Ebbinghaus%
}{%
{\protect \APACyear {{1885/1913}}}%
}]{%
Ebbinghaus}
\APACinsertmetastar {%
Ebbinghaus}%
\begin{APACrefauthors}%
Ebbinghaus, H.%
\end{APACrefauthors}%
\unskip\
\newblock
\APACrefYear{{1885/1913}}.
\newblock
\APACrefbtitle {{Memory: A contribution to experimental psychology}} {{Memory:
  A contribution to experimental psychology}}.
\newblock
\APACaddressPublisher{New York}{Teachers College, Columbia University}.
\PrintBackRefs{\CurrentBib}

\bibitem [\protect \citeauthoryear {%
Egorov%
, Hamam%
, Frans\'{e}n%
, Hasselmo%
\BCBL {}\ \BBA {} Alonso%
}{%
Egorov%
\ \protect \BOthers {.}}{%
{\protect \APACyear {2002}}%
}]{%
EgorEtal02}
\APACinsertmetastar {%
EgorEtal02}%
\begin{APACrefauthors}%
Egorov, A\BPBI V.%
, Hamam, B\BPBI N.%
, Frans\'{e}n, E.%
, Hasselmo, M\BPBI E.%
\BCBL {}\ \BBA {} Alonso, A\BPBI A.%
\end{APACrefauthors}%
\unskip\
\newblock
\APACrefYearMonthDay{2002}{}{}.
\newblock
{\BBOQ}\APACrefatitle {Graded persistent activity in entorhinal cortex
  neurons.} {Graded persistent activity in entorhinal cortex neurons.}{\BBCQ}
\newblock
\APACjournalVolNumPages{Nature}{420}{6912}{173-8}.
\PrintBackRefs{\CurrentBib}

\bibitem [\protect \citeauthoryear {%
Eichenbaum%
}{%
Eichenbaum%
}{%
{\protect \APACyear {2017}}%
}]{%
Eich17a}
\APACinsertmetastar {%
Eich17a}%
\begin{APACrefauthors}%
Eichenbaum, H.%
\end{APACrefauthors}%
\unskip\
\newblock
\APACrefYearMonthDay{2017}{}{}.
\newblock
{\BBOQ}\APACrefatitle {On the Integration of Space, Time, and Memory} {On the
  integration of space, time, and memory}.{\BBCQ}
\newblock
\APACjournalVolNumPages{Neuron}{95}{5}{1007-1018}.
\newblock
\begin{APACrefDOI} \doi{10.1016/j.neuron.2017.06.036} \end{APACrefDOI}
\PrintBackRefs{\CurrentBib}

\bibitem [\protect \citeauthoryear {%
Estes%
}{%
Estes%
}{%
{\protect \APACyear {1950}}%
}]{%
Este50}
\APACinsertmetastar {%
Este50}%
\begin{APACrefauthors}%
Estes, W\BPBI K.%
\end{APACrefauthors}%
\unskip\
\newblock
\APACrefYearMonthDay{1950}{}{}.
\newblock
{\BBOQ}\APACrefatitle {Toward a statistical theory of learning} {Toward a
  statistical theory of learning}.{\BBCQ}
\newblock
\APACjournalVolNumPages{Psychological Review}{57}{}{94-107}.
\PrintBackRefs{\CurrentBib}

\bibitem [\protect \citeauthoryear {%
Estes%
}{%
Estes%
}{%
{\protect \APACyear {1955}}%
{\protect \APACexlab {{\protect \BCnt {1}}}}}]{%
Este55b}
\APACinsertmetastar {%
Este55b}%
\begin{APACrefauthors}%
Estes, W\BPBI K.%
\end{APACrefauthors}%
\unskip\
\newblock
\APACrefYearMonthDay{1955{\protect \BCnt {1}}}{}{}.
\newblock
{\BBOQ}\APACrefatitle {Statistical theory of distributional phenomena in
  learning} {Statistical theory of distributional phenomena in
  learning}.{\BBCQ}
\newblock
\APACjournalVolNumPages{Psychological Review}{62}{}{369-377}.
\PrintBackRefs{\CurrentBib}

\bibitem [\protect \citeauthoryear {%
Estes%
}{%
Estes%
}{%
{\protect \APACyear {1955}}%
{\protect \APACexlab {{\protect \BCnt {2}}}}}]{%
Este55a}
\APACinsertmetastar {%
Este55a}%
\begin{APACrefauthors}%
Estes, W\BPBI K.%
\end{APACrefauthors}%
\unskip\
\newblock
\APACrefYearMonthDay{1955{\protect \BCnt {2}}}{}{}.
\newblock
{\BBOQ}\APACrefatitle {Statistical theory of spontaneous recovery and
  regression} {Statistical theory of spontaneous recovery and
  regression}.{\BBCQ}
\newblock
\APACjournalVolNumPages{Psychological Review}{62}{}{145-154}.
\PrintBackRefs{\CurrentBib}

\bibitem [\protect \citeauthoryear {%
Ferster%
\ \BBA {} Miller%
}{%
Ferster%
\ \BBA {} Miller%
}{%
{\protect \APACyear {2000}}%
}]{%
FersMill00}
\APACinsertmetastar {%
FersMill00}%
\begin{APACrefauthors}%
Ferster, D.%
\BCBT {}\ \BBA {} Miller, K\BPBI D.%
\end{APACrefauthors}%
\unskip\
\newblock
\APACrefYearMonthDay{2000}{}{}.
\newblock
{\BBOQ}\APACrefatitle {Neural mechanisms of orientation selectivity in the
  visual cortex} {Neural mechanisms of orientation selectivity in the visual
  cortex}.{\BBCQ}
\newblock
\APACjournalVolNumPages{Annual Review of Neuroscience}{23}{1}{441--471}.
\PrintBackRefs{\CurrentBib}

\bibitem [\protect \citeauthoryear {%
Folkerts%
, Rutishauser%
\BCBL {}\ \BBA {} Howard%
}{%
Folkerts%
\ \protect \BOthers {.}}{%
{\protect \APACyear {2018}}%
}]{%
FolkEtal18}
\APACinsertmetastar {%
FolkEtal18}%
\begin{APACrefauthors}%
Folkerts, S.%
, Rutishauser, U.%
\BCBL {}\ \BBA {} Howard, M.%
\end{APACrefauthors}%
\unskip\
\newblock
\APACrefYearMonthDay{2018}{}{}.
\newblock
{\BBOQ}\APACrefatitle {Human episodic memory retrieval is accompanied by a
  neural contiguity effect} {Human episodic memory retrieval is accompanied by
  a neural contiguity effect}.{\BBCQ}
\newblock
\APACjournalVolNumPages{Journal of Neuroscience}{38}{}{4200-4211}.
\PrintBackRefs{\CurrentBib}

\bibitem [\protect \citeauthoryear {%
Gallistel%
\ \BBA {} Gibbon%
}{%
Gallistel%
\ \BBA {} Gibbon%
}{%
{\protect \APACyear {2000}}%
}]{%
GallGibb00}
\APACinsertmetastar {%
GallGibb00}%
\begin{APACrefauthors}%
Gallistel, C\BPBI R.%
\BCBT {}\ \BBA {} Gibbon, J.%
\end{APACrefauthors}%
\unskip\
\newblock
\APACrefYearMonthDay{2000}{}{}.
\newblock
{\BBOQ}\APACrefatitle {Time, rate, and conditioning} {Time, rate, and
  conditioning}.{\BBCQ}
\newblock
\APACjournalVolNumPages{Psychological Review}{107}{2}{289-344}.
\PrintBackRefs{\CurrentBib}

\bibitem [\protect \citeauthoryear {%
Gibbon%
}{%
Gibbon%
}{%
{\protect \APACyear {1977}}%
}]{%
Gibb77}
\APACinsertmetastar {%
Gibb77}%
\begin{APACrefauthors}%
Gibbon, J.%
\end{APACrefauthors}%
\unskip\
\newblock
\APACrefYearMonthDay{1977}{}{}.
\newblock
{\BBOQ}\APACrefatitle {Scalar expectancy theory and {Weber's} law in animal
  timing} {Scalar expectancy theory and {Weber's} law in animal timing}.{\BBCQ}
\newblock
\APACjournalVolNumPages{Psychological Review}{84}{3}{279-325}.
\PrintBackRefs{\CurrentBib}

\bibitem [\protect \citeauthoryear {%
Gillund%
\ \BBA {} Shiffrin%
}{%
Gillund%
\ \BBA {} Shiffrin%
}{%
{\protect \APACyear {1984}}%
}]{%
GillShif84}
\APACinsertmetastar {%
GillShif84}%
\begin{APACrefauthors}%
Gillund, G.%
\BCBT {}\ \BBA {} Shiffrin, R\BPBI M.%
\end{APACrefauthors}%
\unskip\
\newblock
\APACrefYearMonthDay{1984}{}{}.
\newblock
{\BBOQ}\APACrefatitle {A retrieval model for both recognition and recall} {A
  retrieval model for both recognition and recall}.{\BBCQ}
\newblock
\APACjournalVolNumPages{Psychological Review}{91}{}{1-67}.
\PrintBackRefs{\CurrentBib}

\bibitem [\protect \citeauthoryear {%
Glanzer%
}{%
Glanzer%
}{%
{\protect \APACyear {1972}}%
}]{%
Glan72}
\APACinsertmetastar {%
Glan72}%
\begin{APACrefauthors}%
Glanzer, M.%
\end{APACrefauthors}%
\unskip\
\newblock
\APACrefYearMonthDay{1972}{}{}.
\newblock
{\BBOQ}\APACrefatitle {Storage mechanisms in recall} {Storage mechanisms in
  recall}.{\BBCQ}
\newblock
\BIn{} K\BPBI W.~Spence\ \BBA {} J\BPBI T.~Spence\ (\BEDS), \APACrefbtitle {The
  Psychology of Learning and Motivation} {The psychology of learning and
  motivation}\ (\BPG~129-193).
\newblock
\APACaddressPublisher{New York}{Academic Press}.
\PrintBackRefs{\CurrentBib}

\bibitem [\protect \citeauthoryear {%
Glanzer%
\ \BBA {} Cunitz%
}{%
Glanzer%
\ \BBA {} Cunitz%
}{%
{\protect \APACyear {1966}}%
}]{%
GlanCuni66}
\APACinsertmetastar {%
GlanCuni66}%
\begin{APACrefauthors}%
Glanzer, M.%
\BCBT {}\ \BBA {} Cunitz, A\BPBI R.%
\end{APACrefauthors}%
\unskip\
\newblock
\APACrefYearMonthDay{1966}{}{}.
\newblock
{\BBOQ}\APACrefatitle {Two storage mechanisms in free recall} {Two storage
  mechanisms in free recall}.{\BBCQ}
\newblock
\APACjournalVolNumPages{Journal of Verbal Learning and Verbal
  Behavior}{5}{}{351-360}.
\PrintBackRefs{\CurrentBib}

\bibitem [\protect \citeauthoryear {%
Glenberg%
\ \protect \BOthers {.}}{%
Glenberg%
\ \protect \BOthers {.}}{%
{\protect \APACyear {1980}}%
}]{%
GlenEtal80}
\APACinsertmetastar {%
GlenEtal80}%
\begin{APACrefauthors}%
Glenberg, A\BPBI M.%
, Bradley, M\BPBI M.%
, Stevenson, J\BPBI A.%
, Kraus, T\BPBI A.%
, Tkachuk, M\BPBI J.%
\BCBL {}\ \BBA {} Gretz, A\BPBI L.%
\end{APACrefauthors}%
\unskip\
\newblock
\APACrefYearMonthDay{1980}{}{}.
\newblock
{\BBOQ}\APACrefatitle {A two-process account of long-term serial position
  effects} {A two-process account of long-term serial position effects}.{\BBCQ}
\newblock
\APACjournalVolNumPages{Journal of Experimental Psychology: Human Learning and
  Memory}{6}{}{355-369}.
\PrintBackRefs{\CurrentBib}

\bibitem [\protect \citeauthoryear {%
Goh%
}{%
Goh%
}{%
{\protect \APACyear {2021}}%
}]{%
Goh21}
\APACinsertmetastar {%
Goh21}%
\begin{APACrefauthors}%
Goh, W\BPBI Z.%
\end{APACrefauthors}%
\unskip\
\newblock
\APACrefYear{2021}.
\unskip\
\newblock
\APACrefbtitle {Remembering the past to predict the future: A scale-invariant
  timeline for memory and anticipation} {Remembering the past to predict the
  future: A scale-invariant timeline for memory and anticipation}\
  \APACtypeAddressSchool {\BUPhD}{}{}.
\unskip\
\newblock
\APACaddressSchool {}{Boston University}.
\PrintBackRefs{\CurrentBib}

\bibitem [\protect \citeauthoryear {%
Goyal%
\ \protect \BOthers {.}}{%
Goyal%
\ \protect \BOthers {.}}{%
{\protect \APACyear {2018}}%
}]{%
GoyaEtal18}
\APACinsertmetastar {%
GoyaEtal18}%
\begin{APACrefauthors}%
Goyal, A.%
, Miller, J.%
, Watrous, A\BPBI J.%
, Lee, S\BPBI A.%
, Coffey, T.%
, Sperling, M\BPBI R.%
\BDBL {}others%
\end{APACrefauthors}%
\unskip\
\newblock
\APACrefYearMonthDay{2018}{}{}.
\newblock
{\BBOQ}\APACrefatitle {Electrical stimulation in hippocampus and entorhinal
  cortex impairs spatial and temporal memory} {Electrical stimulation in
  hippocampus and entorhinal cortex impairs spatial and temporal
  memory}.{\BBCQ}
\newblock
\APACjournalVolNumPages{Journal of Neuroscience}{}{}{3049--17}.
\PrintBackRefs{\CurrentBib}

\bibitem [\protect \citeauthoryear {%
Guo%
, Huson%
, Macosko%
\BCBL {}\ \BBA {} Regehr%
}{%
Guo%
\ \protect \BOthers {.}}{%
{\protect \APACyear {2021}}%
}]{%
GuoEtal21}
\APACinsertmetastar {%
GuoEtal21}%
\begin{APACrefauthors}%
Guo, C.%
, Huson, V.%
, Macosko, E\BPBI Z.%
\BCBL {}\ \BBA {} Regehr, W\BPBI G.%
\end{APACrefauthors}%
\unskip\
\newblock
\APACrefYearMonthDay{2021}{}{}.
\newblock
{\BBOQ}\APACrefatitle {Graded heterogeneity of metabotropic signaling underlies
  a continuum of cell-intrinsic temporal responses in unipolar brush cells}
  {Graded heterogeneity of metabotropic signaling underlies a continuum of
  cell-intrinsic temporal responses in unipolar brush cells}.{\BBCQ}
\newblock
\APACjournalVolNumPages{Nature Communications}{12}{1}{1--12}.
\PrintBackRefs{\CurrentBib}

\bibitem [\protect \citeauthoryear {%
Hasselmo%
\ \BBA {} Mc{C}lelland%
}{%
Hasselmo%
\ \BBA {} Mc{C}lelland%
}{%
{\protect \APACyear {1999}}%
}]{%
HassMcCl99}
\APACinsertmetastar {%
HassMcCl99}%
\begin{APACrefauthors}%
Hasselmo, M\BPBI E.%
\BCBT {}\ \BBA {} Mc{C}lelland, J\BPBI L.%
\end{APACrefauthors}%
\unskip\
\newblock
\APACrefYearMonthDay{1999}{}{}.
\newblock
{\BBOQ}\APACrefatitle {Neural Models of Memory} {Neural models of
  memory}.{\BBCQ}
\newblock
\APACjournalVolNumPages{Current Opinion in Neurobiology}{9}{}{184-188}.
\PrintBackRefs{\CurrentBib}

\bibitem [\protect \citeauthoryear {%
Hasselmo%
\ \BBA {} Wyble%
}{%
Hasselmo%
\ \BBA {} Wyble%
}{%
{\protect \APACyear {1997}}%
}]{%
HassWybl97}
\APACinsertmetastar {%
HassWybl97}%
\begin{APACrefauthors}%
Hasselmo, M\BPBI E.%
\BCBT {}\ \BBA {} Wyble, B\BPBI P.%
\end{APACrefauthors}%
\unskip\
\newblock
\APACrefYearMonthDay{1997}{}{}.
\newblock
{\BBOQ}\APACrefatitle {Free recall and recognition in a network model of the
  hippocampus: simulating effects of scopolamine on human memory function}
  {Free recall and recognition in a network model of the hippocampus:
  simulating effects of scopolamine on human memory function}.{\BBCQ}
\newblock
\APACjournalVolNumPages{Behavioural Brain Research}{89}{1-2}{1-34}.
\PrintBackRefs{\CurrentBib}

\bibitem [\protect \citeauthoryear {%
Healey%
, Long%
\BCBL {}\ \BBA {} Kahana%
}{%
Healey%
\ \protect \BOthers {.}}{%
{\protect \APACyear {2018}}%
}]{%
HealEtal18}
\APACinsertmetastar {%
HealEtal18}%
\begin{APACrefauthors}%
Healey, M\BPBI K.%
, Long, N\BPBI M.%
\BCBL {}\ \BBA {} Kahana, M\BPBI J.%
\end{APACrefauthors}%
\unskip\
\newblock
\APACrefYearMonthDay{2018}{}{}.
\newblock
{\BBOQ}\APACrefatitle {Contiguity in episodic memory} {Contiguity in episodic
  memory}.{\BBCQ}
\newblock
\APACjournalVolNumPages{Psychonomic bulletin \& review}{}{}{1--22}.
\PrintBackRefs{\CurrentBib}

\bibitem [\protect \citeauthoryear {%
D.~Hintzman%
}{%
D.~Hintzman%
}{%
{\protect \APACyear {1987}}%
}]{%
Hint87}
\APACinsertmetastar {%
Hint87}%
\begin{APACrefauthors}%
Hintzman, D.%
\end{APACrefauthors}%
\unskip\
\newblock
\APACrefYearMonthDay{1987}{}{}.
\newblock
{\BBOQ}\APACrefatitle {Recognition and recall in MINERVA 2: Analysis of the
  `recognition-failure' paradigm} {Recognition and recall in minerva 2:
  Analysis of the `recognition-failure' paradigm}.{\BBCQ}
\newblock
\BIn{} P.~Morris\ (\BED), \APACrefbtitle {Modelling cognition} {Modelling
  cognition}\ (\BPG~215-229).
\newblock
\APACaddressPublisher{New York}{Wiley}.
\PrintBackRefs{\CurrentBib}

\bibitem [\protect \citeauthoryear {%
D\BPBI L.~Hintzman%
}{%
D\BPBI L.~Hintzman%
}{%
{\protect \APACyear {1984}}%
}]{%
Hint84}
\APACinsertmetastar {%
Hint84}%
\begin{APACrefauthors}%
Hintzman, D\BPBI L.%
\end{APACrefauthors}%
\unskip\
\newblock
\APACrefYearMonthDay{1984}{}{}.
\newblock
{\BBOQ}\APACrefatitle {{MINERVA 2}: A simulation model of human memory}
  {{MINERVA 2}: A simulation model of human memory}.{\BBCQ}
\newblock
\APACjournalVolNumPages{Behavior Research Methods, Instruments \&
  Computers}{16}{2}{96-101}.
\PrintBackRefs{\CurrentBib}

\bibitem [\protect \citeauthoryear {%
Howard%
}{%
Howard%
}{%
{\protect \APACyear {2018}}%
}]{%
Howa18}
\APACinsertmetastar {%
Howa18}%
\begin{APACrefauthors}%
Howard, M\BPBI W.%
\end{APACrefauthors}%
\unskip\
\newblock
\APACrefYearMonthDay{2018}{}{}.
\newblock
{\BBOQ}\APACrefatitle {Memory as perception of the past: Compressed time in
  mind and brain.} {Memory as perception of the past: Compressed time in mind
  and brain.}{\BBCQ}
\newblock
\APACjournalVolNumPages{Trends in Cognitive Sciences}{22}{}{124-136}.
\PrintBackRefs{\CurrentBib}

\bibitem [\protect \citeauthoryear {%
Howard%
\ \BBA {} Hasselmo%
}{%
Howard%
\ \BBA {} Hasselmo%
}{%
{\protect \APACyear {2020}}%
}]{%
HowaHass20}
\APACinsertmetastar {%
HowaHass20}%
\begin{APACrefauthors}%
Howard, M\BPBI W.%
\BCBT {}\ \BBA {} Hasselmo, M\BPBI E.%
\end{APACrefauthors}%
\unskip\
\newblock
\APACrefYearMonthDay{2020}{}{}.
\newblock
{\BBOQ}\APACrefatitle {Cognitive computation using neural representations of
  time and space in the Laplace domain} {Cognitive computation using neural
  representations of time and space in the laplace domain}.{\BBCQ}
\newblock
\APACjournalVolNumPages{arXiv preprint arXiv:2003.11668}{}{}{}.
\PrintBackRefs{\CurrentBib}

\bibitem [\protect \citeauthoryear {%
Howard%
\ \BBA {} Kahana%
}{%
Howard%
\ \BBA {} Kahana%
}{%
{\protect \APACyear {2002}}%
}]{%
HowaKaha02a}
\APACinsertmetastar {%
HowaKaha02a}%
\begin{APACrefauthors}%
Howard, M\BPBI W.%
\BCBT {}\ \BBA {} Kahana, M\BPBI J.%
\end{APACrefauthors}%
\unskip\
\newblock
\APACrefYearMonthDay{2002}{}{}.
\newblock
{\BBOQ}\APACrefatitle {A Distributed Representation of Temporal Context} {A
  distributed representation of temporal context}.{\BBCQ}
\newblock
\APACjournalVolNumPages{Journal of Mathematical Psychology}{46}{3}{269-299}.
\PrintBackRefs{\CurrentBib}

\bibitem [\protect \citeauthoryear {%
Howard%
\ \protect \BOthers {.}}{%
Howard%
\ \protect \BOthers {.}}{%
{\protect \APACyear {2014}}%
}]{%
HowaEtal14}
\APACinsertmetastar {%
HowaEtal14}%
\begin{APACrefauthors}%
Howard, M\BPBI W.%
, Mac{D}onald, C\BPBI J.%
, Tiganj, Z.%
, Shankar, K\BPBI H.%
, Du, Q.%
, Hasselmo, M\BPBI E.%
\BCBL {}\ \BBA {} Eichenbaum, H.%
\end{APACrefauthors}%
\unskip\
\newblock
\APACrefYearMonthDay{2014}{}{}.
\newblock
{\BBOQ}\APACrefatitle {A unified mathematical framework for coding time, space,
  and sequences in the hippocampal region} {A unified mathematical framework
  for coding time, space, and sequences in the hippocampal region}.{\BBCQ}
\newblock
\APACjournalVolNumPages{Journal of Neuroscience}{34}{13}{4692-707}.
\newblock
\begin{APACrefDOI} \doi{10.1523/JNEUROSCI.5808-12.2014} \end{APACrefDOI}
\PrintBackRefs{\CurrentBib}

\bibitem [\protect \citeauthoryear {%
Howard%
\ \BBA {} Shankar%
}{%
Howard%
\ \BBA {} Shankar%
}{%
{\protect \APACyear {2018}}%
}]{%
HowaShan18}
\APACinsertmetastar {%
HowaShan18}%
\begin{APACrefauthors}%
Howard, M\BPBI W.%
\BCBT {}\ \BBA {} Shankar, K\BPBI H.%
\end{APACrefauthors}%
\unskip\
\newblock
\APACrefYearMonthDay{2018}{}{}.
\newblock
{\BBOQ}\APACrefatitle {Neural Scaling Laws for an Uncertain World} {Neural
  scaling laws for an uncertain world}.{\BBCQ}
\newblock
\APACjournalVolNumPages{Psychologial Review}{125}{}{47-58}.
\newblock
\begin{APACrefDOI} \doi{10.1037/rev0000081} \end{APACrefDOI}
\PrintBackRefs{\CurrentBib}

\bibitem [\protect \citeauthoryear {%
Howard%
, Shankar%
, Aue%
\BCBL {}\ \BBA {} Criss%
}{%
Howard%
\ \protect \BOthers {.}}{%
{\protect \APACyear {2015}}%
}]{%
HowaEtal15}
\APACinsertmetastar {%
HowaEtal15}%
\begin{APACrefauthors}%
Howard, M\BPBI W.%
, Shankar, K\BPBI H.%
, Aue, W.%
\BCBL {}\ \BBA {} Criss, A\BPBI H.%
\end{APACrefauthors}%
\unskip\
\newblock
\APACrefYearMonthDay{2015}{}{}.
\newblock
{\BBOQ}\APACrefatitle {A distributed representation of internal time} {A
  distributed representation of internal time}.{\BBCQ}
\newblock
\APACjournalVolNumPages{Psychological Review}{122}{1}{24-53}.
\PrintBackRefs{\CurrentBib}

\bibitem [\protect \citeauthoryear {%
Howard%
, Youker%
\BCBL {}\ \BBA {} Venkatadass%
}{%
Howard%
\ \protect \BOthers {.}}{%
{\protect \APACyear {2008}}%
}]{%
HowaEtal08}
\APACinsertmetastar {%
HowaEtal08}%
\begin{APACrefauthors}%
Howard, M\BPBI W.%
, Youker, T\BPBI E.%
\BCBL {}\ \BBA {} Venkatadass, V.%
\end{APACrefauthors}%
\unskip\
\newblock
\APACrefYearMonthDay{2008}{}{}.
\newblock
{\BBOQ}\APACrefatitle {The persistence of memory: Contiguity effects across
  several minutes} {The persistence of memory: Contiguity effects across
  several minutes}.{\BBCQ}
\newblock
\APACjournalVolNumPages{Psychonomic Bulletin \& Review}{15}{PMC2493616}{58-63}.
\PrintBackRefs{\CurrentBib}

\bibitem [\protect \citeauthoryear {%
Hsieh%
, Gruber%
, Jenkins%
\BCBL {}\ \BBA {} Ranganath%
}{%
Hsieh%
\ \protect \BOthers {.}}{%
{\protect \APACyear {2014}}%
}]{%
HsieEtal14}
\APACinsertmetastar {%
HsieEtal14}%
\begin{APACrefauthors}%
Hsieh, L\BHBI T.%
, Gruber, M\BPBI J.%
, Jenkins, L\BPBI J.%
\BCBL {}\ \BBA {} Ranganath, C.%
\end{APACrefauthors}%
\unskip\
\newblock
\APACrefYearMonthDay{2014}{}{}.
\newblock
{\BBOQ}\APACrefatitle {Hippocampal Activity Patterns Carry Information about
  Objects in Temporal Context} {Hippocampal activity patterns carry information
  about objects in temporal context}.{\BBCQ}
\newblock
\APACjournalVolNumPages{Neuron}{81}{5}{1165--1178}.
\PrintBackRefs{\CurrentBib}

\bibitem [\protect \citeauthoryear {%
Hull%
}{%
Hull%
}{%
{\protect \APACyear {1939}}%
}]{%
Hull39}
\APACinsertmetastar {%
Hull39}%
\begin{APACrefauthors}%
Hull, C\BPBI L.%
\end{APACrefauthors}%
\unskip\
\newblock
\APACrefYearMonthDay{1939}{}{}.
\newblock
{\BBOQ}\APACrefatitle {The problem of stimulus equivalence in behavior theory.}
  {The problem of stimulus equivalence in behavior theory.}{\BBCQ}
\newblock
\APACjournalVolNumPages{Psychological Review}{46}{1}{9}.
\PrintBackRefs{\CurrentBib}

\bibitem [\protect \citeauthoryear {%
Hull%
}{%
Hull%
}{%
{\protect \APACyear {1947}}%
}]{%
Hull47}
\APACinsertmetastar {%
Hull47}%
\begin{APACrefauthors}%
Hull, C\BPBI L.%
\end{APACrefauthors}%
\unskip\
\newblock
\APACrefYearMonthDay{1947}{}{}.
\newblock
{\BBOQ}\APACrefatitle {The problem of primary stimulus generalization} {The
  problem of primary stimulus generalization}.{\BBCQ}
\newblock
\APACjournalVolNumPages{Psychological Review}{54}{}{120-134}.
\PrintBackRefs{\CurrentBib}

\bibitem [\protect \citeauthoryear {%
Humphreys%
, Bain%
\BCBL {}\ \BBA {} Pike%
}{%
Humphreys%
\ \protect \BOthers {.}}{%
{\protect \APACyear {1989}}%
}]{%
HumpEtal89b}
\APACinsertmetastar {%
HumpEtal89b}%
\begin{APACrefauthors}%
Humphreys, M\BPBI S.%
, Bain, J\BPBI D.%
\BCBL {}\ \BBA {} Pike, R.%
\end{APACrefauthors}%
\unskip\
\newblock
\APACrefYearMonthDay{1989}{}{}.
\newblock
{\BBOQ}\APACrefatitle {Different ways to cue a coherent memory system: A theory
  for episodic, semantic, and procedural tasks} {Different ways to cue a
  coherent memory system: A theory for episodic, semantic, and procedural
  tasks}.{\BBCQ}
\newblock
\APACjournalVolNumPages{Psychological Review}{96}{}{208-233}.
\PrintBackRefs{\CurrentBib}

\bibitem [\protect \citeauthoryear {%
Hyman%
, Ma%
, Balaguer-Ballester%
, Durstewitz%
\BCBL {}\ \BBA {} Seamans%
}{%
Hyman%
\ \protect \BOthers {.}}{%
{\protect \APACyear {2012}}%
}]{%
HymaEtal12}
\APACinsertmetastar {%
HymaEtal12}%
\begin{APACrefauthors}%
Hyman, J\BPBI M.%
, Ma, L.%
, Balaguer-Ballester, E.%
, Durstewitz, D.%
\BCBL {}\ \BBA {} Seamans, J\BPBI K.%
\end{APACrefauthors}%
\unskip\
\newblock
\APACrefYearMonthDay{2012}{}{}.
\newblock
{\BBOQ}\APACrefatitle {Contextual encoding by ensembles of medial prefrontal
  cortex neurons} {Contextual encoding by ensembles of medial prefrontal cortex
  neurons}.{\BBCQ}
\newblock
\APACjournalVolNumPages{Proceedings of the National Academy of Sciences
  USA}{109}{}{5086-91}.
\newblock
\begin{APACrefDOI} \doi{10.1073/pnas.1114415109} \end{APACrefDOI}
\PrintBackRefs{\CurrentBib}

\bibitem [\protect \citeauthoryear {%
James%
}{%
James%
}{%
{\protect \APACyear {1890}}%
}]{%
Jame90}
\APACinsertmetastar {%
Jame90}%
\begin{APACrefauthors}%
James, W.%
\end{APACrefauthors}%
\unskip\
\newblock
\APACrefYear{1890}.
\newblock
\APACrefbtitle {The principles of psychology} {The principles of psychology}.
\newblock
\APACaddressPublisher{New York}{Holt}.
\PrintBackRefs{\CurrentBib}

\bibitem [\protect \citeauthoryear {%
Jin%
, Fujii%
\BCBL {}\ \BBA {} Graybiel%
}{%
Jin%
\ \protect \BOthers {.}}{%
{\protect \APACyear {2009}}%
}]{%
JinEtal09}
\APACinsertmetastar {%
JinEtal09}%
\begin{APACrefauthors}%
Jin, D\BPBI Z.%
, Fujii, N.%
\BCBL {}\ \BBA {} Graybiel, A\BPBI M.%
\end{APACrefauthors}%
\unskip\
\newblock
\APACrefYearMonthDay{2009}{}{}.
\newblock
{\BBOQ}\APACrefatitle {Neural representation of time in cortico-basal ganglia
  circuits} {Neural representation of time in cortico-basal ganglia
  circuits}.{\BBCQ}
\newblock
\APACjournalVolNumPages{Proceedings of the National Academy of
  Sciences}{106}{45}{19156--19161}.
\PrintBackRefs{\CurrentBib}

\bibitem [\protect \citeauthoryear {%
Kahana%
}{%
Kahana%
}{%
{\protect \APACyear {1996}}%
}]{%
Kaha96}
\APACinsertmetastar {%
Kaha96}%
\begin{APACrefauthors}%
Kahana, M\BPBI J.%
\end{APACrefauthors}%
\unskip\
\newblock
\APACrefYearMonthDay{1996}{}{}.
\newblock
{\BBOQ}\APACrefatitle {Associative retrieval processes in free recall}
  {Associative retrieval processes in free recall}.{\BBCQ}
\newblock
\APACjournalVolNumPages{Memory \& Cognition}{24}{}{103-109}.
\PrintBackRefs{\CurrentBib}

\bibitem [\protect \citeauthoryear {%
Kahana%
}{%
Kahana%
}{%
{\protect \APACyear {2012}}%
}]{%
Kaha12}
\APACinsertmetastar {%
Kaha12}%
\begin{APACrefauthors}%
Kahana, M\BPBI J.%
\end{APACrefauthors}%
\unskip\
\newblock
\APACrefYear{2012}.
\newblock
\APACrefbtitle {Foundations of human memory} {Foundations of human memory}.
\newblock
\APACaddressPublisher{}{OUP USA}.
\PrintBackRefs{\CurrentBib}

\bibitem [\protect \citeauthoryear {%
Killeen%
\ \BBA {} Fetterman%
}{%
Killeen%
\ \BBA {} Fetterman%
}{%
{\protect \APACyear {1988}}%
}]{%
KillFett88}
\APACinsertmetastar {%
KillFett88}%
\begin{APACrefauthors}%
Killeen, P\BPBI R.%
\BCBT {}\ \BBA {} Fetterman, J\BPBI G.%
\end{APACrefauthors}%
\unskip\
\newblock
\APACrefYearMonthDay{1988}{}{}.
\newblock
{\BBOQ}\APACrefatitle {{A behavioral theory of timing.}} {{A behavioral theory
  of timing.}}{\BBCQ}
\newblock
\APACjournalVolNumPages{Psychological Review}{95}{2}{274--295}.
\PrintBackRefs{\CurrentBib}

\bibitem [\protect \citeauthoryear {%
Lashley%
}{%
Lashley%
}{%
{\protect \APACyear {1951}}%
}]{%
Lash51}
\APACinsertmetastar {%
Lash51}%
\begin{APACrefauthors}%
Lashley, K\BPBI S.%
\end{APACrefauthors}%
\unskip\
\newblock
\APACrefYearMonthDay{1951}{}{}.
\newblock
{\BBOQ}\APACrefatitle {The problem of serial order in behavior} {The problem of
  serial order in behavior}.{\BBCQ}
\newblock
\BIn{} L\BPBI A.~Jeffress\ (\BED), \APACrefbtitle {Cerebral Mechanisms in
  Behavior; the Hixon symposium} {Cerebral mechanisms in behavior; the hixon
  symposium}\ (\BPG~112-146).
\newblock
\APACaddressPublisher{Oxford}{Wiley}.
\PrintBackRefs{\CurrentBib}

\bibitem [\protect \citeauthoryear {%
Lehman%
\ \BBA {} Malmberg%
}{%
Lehman%
\ \BBA {} Malmberg%
}{%
{\protect \APACyear {2012}}%
}]{%
LehmMalm12}
\APACinsertmetastar {%
LehmMalm12}%
\begin{APACrefauthors}%
Lehman, M.%
\BCBT {}\ \BBA {} Malmberg, K\BPBI J.%
\end{APACrefauthors}%
\unskip\
\newblock
\APACrefYearMonthDay{2012}{}{}.
\newblock
{\BBOQ}\APACrefatitle {A Buffer Model of Memory Encoding and Temporal
  Correlations in Retrieval} {A buffer model of memory encoding and temporal
  correlations in retrieval}.{\BBCQ}
\newblock
\APACjournalVolNumPages{Psychological Review}{}{}{}.
\newblock
\begin{APACrefDOI} \doi{10.1037/a0030851} \end{APACrefDOI}
\PrintBackRefs{\CurrentBib}

\bibitem [\protect \citeauthoryear {%
Logan%
}{%
Logan%
}{%
{\protect \APACyear {2021}}%
}]{%
Loga21}
\APACinsertmetastar {%
Loga21}%
\begin{APACrefauthors}%
Logan, G\BPBI D.%
\end{APACrefauthors}%
\unskip\
\newblock
\APACrefYearMonthDay{2021}{}{}.
\newblock
{\BBOQ}\APACrefatitle {Serial order in perception, memory, and action.} {Serial
  order in perception, memory, and action.}{\BBCQ}
\newblock
\APACjournalVolNumPages{Psychological Review}{128}{1}{1}.
\PrintBackRefs{\CurrentBib}

\bibitem [\protect \citeauthoryear {%
Luce%
\ \BBA {} Suppes%
}{%
Luce%
\ \BBA {} Suppes%
}{%
{\protect \APACyear {2002}}%
}]{%
LuceSupp02}
\APACinsertmetastar {%
LuceSupp02}%
\begin{APACrefauthors}%
Luce, R\BPBI D.%
\BCBT {}\ \BBA {} Suppes, P.%
\end{APACrefauthors}%
\unskip\
\newblock
\APACrefYearMonthDay{2002}{}{}.
\newblock
{\BBOQ}\APACrefatitle {Representational measurement theory} {Representational
  measurement theory}.{\BBCQ}
\newblock
\BIn{} J.~Wixted\ \BBA {} H.~Pashler\ (\BEDS), \APACrefbtitle {Stevens Handbook
  of Experimental Psychology, 3rd Edition} {Stevens handbook of experimental
  psychology, 3rd edition}\ (\BVOL\ 4: Methodology in Experimental Psychology,
  \BPG~1-41).
\newblock
\APACaddressPublisher{}{Wiley Online Library}.
\PrintBackRefs{\CurrentBib}

\bibitem [\protect \citeauthoryear {%
Mac{D}onald%
, Lepage%
, Eden%
\BCBL {}\ \BBA {} Eichenbaum%
}{%
Mac{D}onald%
\ \protect \BOthers {.}}{%
{\protect \APACyear {2011}}%
}]{%
MacDEtal11}
\APACinsertmetastar {%
MacDEtal11}%
\begin{APACrefauthors}%
Mac{D}onald, C\BPBI J.%
, Lepage, K\BPBI Q.%
, Eden, U\BPBI T.%
\BCBL {}\ \BBA {} Eichenbaum, H.%
\end{APACrefauthors}%
\unskip\
\newblock
\APACrefYearMonthDay{2011}{}{}.
\newblock
{\BBOQ}\APACrefatitle {Hippocampal ``Time Cells'' Bridge the Gap in Memory for
  Discontiguous Events} {Hippocampal ``time cells'' bridge the gap in memory
  for discontiguous events}.{\BBCQ}
\newblock
\APACjournalVolNumPages{Neuron}{71}{4}{737-749}.
\PrintBackRefs{\CurrentBib}

\bibitem [\protect \citeauthoryear {%
Mack%
, Cinel%
, Davies%
, Harding%
\BCBL {}\ \BBA {} Ward%
}{%
Mack%
\ \protect \BOthers {.}}{%
{\protect \APACyear {2017}}%
}]{%
CortEtal17}
\APACinsertmetastar {%
CortEtal17}%
\begin{APACrefauthors}%
Mack, C\BPBI C.%
, Cinel, C.%
, Davies, N.%
, Harding, M.%
\BCBL {}\ \BBA {} Ward, G.%
\end{APACrefauthors}%
\unskip\
\newblock
\APACrefYearMonthDay{2017}{}{}.
\newblock
{\BBOQ}\APACrefatitle {Serial position, output order, and list length effects
  for words presented on smartphones over very long intervals} {Serial
  position, output order, and list length effects for words presented on
  smartphones over very long intervals}.{\BBCQ}
\newblock
\APACjournalVolNumPages{Journal of Memory and Language}{97}{}{61--80}.
\PrintBackRefs{\CurrentBib}

\bibitem [\protect \citeauthoryear {%
Mankin%
\ \protect \BOthers {.}}{%
Mankin%
\ \protect \BOthers {.}}{%
{\protect \APACyear {2012}}%
}]{%
MankEtal12}
\APACinsertmetastar {%
MankEtal12}%
\begin{APACrefauthors}%
Mankin, E\BPBI A.%
, Sparks, F\BPBI T.%
, Slayyeh, B.%
, Sutherland, R\BPBI J.%
, Leutgeb, S.%
\BCBL {}\ \BBA {} Leutgeb, J\BPBI K.%
\end{APACrefauthors}%
\unskip\
\newblock
\APACrefYearMonthDay{2012}{}{}.
\newblock
{\BBOQ}\APACrefatitle {Neuronal code for extended time in the hippocampus}
  {Neuronal code for extended time in the hippocampus}.{\BBCQ}
\newblock
\APACjournalVolNumPages{Proceedings of the National Academy of
  Sciences}{109}{}{19462-7}.
\newblock
\begin{APACrefDOI} \doi{10.1073/pnas.1214107109} \end{APACrefDOI}
\PrintBackRefs{\CurrentBib}

\bibitem [\protect \citeauthoryear {%
Manning%
, Polyn%
, Litt%
, Baltuch%
\BCBL {}\ \BBA {} Kahana%
}{%
Manning%
\ \protect \BOthers {.}}{%
{\protect \APACyear {2011}}%
}]{%
MannEtal11}
\APACinsertmetastar {%
MannEtal11}%
\begin{APACrefauthors}%
Manning, J\BPBI R.%
, Polyn, S\BPBI M.%
, Litt, B.%
, Baltuch, G.%
\BCBL {}\ \BBA {} Kahana, M\BPBI J.%
\end{APACrefauthors}%
\unskip\
\newblock
\APACrefYearMonthDay{2011}{}{}.
\newblock
{\BBOQ}\APACrefatitle {Oscillatory patterns in temporal lobe reveal context
  reinstatement during memory search} {Oscillatory patterns in temporal lobe
  reveal context reinstatement during memory search}.{\BBCQ}
\newblock
\APACjournalVolNumPages{Proceedings of the National Academy of Science,
  {USA}}{108}{31}{12893-7}.
\PrintBackRefs{\CurrentBib}

\bibitem [\protect \citeauthoryear {%
Manns%
, Howard%
\BCBL {}\ \BBA {} Eichenbaum%
}{%
Manns%
\ \protect \BOthers {.}}{%
{\protect \APACyear {2007}}%
}]{%
MannEtal07}
\APACinsertmetastar {%
MannEtal07}%
\begin{APACrefauthors}%
Manns, J\BPBI R.%
, Howard, M\BPBI W.%
\BCBL {}\ \BBA {} Eichenbaum, H\BPBI B.%
\end{APACrefauthors}%
\unskip\
\newblock
\APACrefYearMonthDay{2007}{}{}.
\newblock
{\BBOQ}\APACrefatitle {Gradual changes in hippocampal activity support
  remembering the order of events} {Gradual changes in hippocampal activity
  support remembering the order of events}.{\BBCQ}
\newblock
\APACjournalVolNumPages{Neuron}{56}{3}{530-540}.
\PrintBackRefs{\CurrentBib}

\bibitem [\protect \citeauthoryear {%
Mau%
, Hasselmo%
\BCBL {}\ \BBA {} Cai%
}{%
Mau%
\ \protect \BOthers {.}}{%
{\protect \APACyear {2020}}%
}]{%
MauEtal20}
\APACinsertmetastar {%
MauEtal20}%
\begin{APACrefauthors}%
Mau, W.%
, Hasselmo, M\BPBI E.%
\BCBL {}\ \BBA {} Cai, D\BPBI J.%
\end{APACrefauthors}%
\unskip\
\newblock
\APACrefYearMonthDay{2020}{}{}.
\newblock
{\BBOQ}\APACrefatitle {The brain in motion: How ensemble fluidity drives
  memory-updating and flexibility} {The brain in motion: How ensemble fluidity
  drives memory-updating and flexibility}.{\BBCQ}
\newblock
\APACjournalVolNumPages{Elife}{9}{}{e63550}.
\PrintBackRefs{\CurrentBib}

\bibitem [\protect \citeauthoryear {%
Mau%
\ \protect \BOthers {.}}{%
Mau%
\ \protect \BOthers {.}}{%
{\protect \APACyear {2018}}%
}]{%
MauEtal18}
\APACinsertmetastar {%
MauEtal18}%
\begin{APACrefauthors}%
Mau, W.%
, Sullivan, D\BPBI W.%
, Kinsky, N\BPBI R.%
, Hasselmo, M\BPBI E.%
, Howard, M\BPBI W.%
\BCBL {}\ \BBA {} Eichenbaum, H.%
\end{APACrefauthors}%
\unskip\
\newblock
\APACrefYearMonthDay{2018}{}{}.
\newblock
{\BBOQ}\APACrefatitle {The same hippocampal {CA1} population simultaneously
  codes temporal information over multiple timescales} {The same hippocampal
  {CA1} population simultaneously codes temporal information over multiple
  timescales}.{\BBCQ}
\newblock
\APACjournalVolNumPages{Current Biology}{28}{}{1499-1508}.
\PrintBackRefs{\CurrentBib}

\bibitem [\protect \citeauthoryear {%
McClelland%
, McNaughton%
\BCBL {}\ \BBA {} O'Reilly%
}{%
McClelland%
\ \protect \BOthers {.}}{%
{\protect \APACyear {1995}}%
}]{%
McClEtal95}
\APACinsertmetastar {%
McClEtal95}%
\begin{APACrefauthors}%
McClelland, J\BPBI L.%
, McNaughton, B\BPBI L.%
\BCBL {}\ \BBA {} O'Reilly, R\BPBI C.%
\end{APACrefauthors}%
\unskip\
\newblock
\APACrefYearMonthDay{1995}{}{}.
\newblock
{\BBOQ}\APACrefatitle {Why there are complementary learning systems in the
  hippocampus and neocortex: insights from the successes and failures of
  connectionist models of learning and memory} {Why there are complementary
  learning systems in the hippocampus and neocortex: insights from the
  successes and failures of connectionist models of learning and
  memory}.{\BBCQ}
\newblock
\APACjournalVolNumPages{Psychological Review}{102}{3}{419-57}.
\PrintBackRefs{\CurrentBib}

\bibitem [\protect \citeauthoryear {%
Mello%
, Soares%
\BCBL {}\ \BBA {} Paton%
}{%
Mello%
\ \protect \BOthers {.}}{%
{\protect \APACyear {2015}}%
}]{%
MellEtal15}
\APACinsertmetastar {%
MellEtal15}%
\begin{APACrefauthors}%
Mello, G\BPBI B.%
, Soares, S.%
\BCBL {}\ \BBA {} Paton, J\BPBI J.%
\end{APACrefauthors}%
\unskip\
\newblock
\APACrefYearMonthDay{2015}{}{}.
\newblock
{\BBOQ}\APACrefatitle {A scalable population code for time in the striatum} {A
  scalable population code for time in the striatum}.{\BBCQ}
\newblock
\APACjournalVolNumPages{Current Biology}{25}{9}{1113--1122}.
\PrintBackRefs{\CurrentBib}

\bibitem [\protect \citeauthoryear {%
Metcalfe%
}{%
Metcalfe%
}{%
{\protect \APACyear {1985}}%
}]{%
Metc85}
\APACinsertmetastar {%
Metc85}%
\begin{APACrefauthors}%
Metcalfe, J.%
\end{APACrefauthors}%
\unskip\
\newblock
\APACrefYearMonthDay{1985}{}{}.
\newblock
{\BBOQ}\APACrefatitle {Levels of processing, encoding specificity, elaboration,
  and {CHARM}} {Levels of processing, encoding specificity, elaboration, and
  {CHARM}}.{\BBCQ}
\newblock
\APACjournalVolNumPages{Psychological Review}{92}{}{1-38}.
\PrintBackRefs{\CurrentBib}

\bibitem [\protect \citeauthoryear {%
Miller%
}{%
Miller%
}{%
{\protect \APACyear {1956}}%
}]{%
Mill56}
\APACinsertmetastar {%
Mill56}%
\begin{APACrefauthors}%
Miller, G\BPBI A.%
\end{APACrefauthors}%
\unskip\
\newblock
\APACrefYearMonthDay{1956}{}{}.
\newblock
{\BBOQ}\APACrefatitle {The magical number seven, plus or minus two: some limits
  on our capacity for processing information} {The magical number seven, plus
  or minus two: some limits on our capacity for processing information}.{\BBCQ}
\newblock
\APACjournalVolNumPages{Psychological Review}{63}{}{81-97}.
\PrintBackRefs{\CurrentBib}

\bibitem [\protect \citeauthoryear {%
Mnih%
\ \protect \BOthers {.}}{%
Mnih%
\ \protect \BOthers {.}}{%
{\protect \APACyear {2015}}%
}]{%
MnihEtal15}
\APACinsertmetastar {%
MnihEtal15}%
\begin{APACrefauthors}%
Mnih, V.%
, Kavukcuoglu, K.%
, Silver, D.%
, Rusu, A\BPBI A.%
, Veness, J.%
, Bellemare, M\BPBI G.%
\BDBL {}others%
\end{APACrefauthors}%
\unskip\
\newblock
\APACrefYearMonthDay{2015}{}{}.
\newblock
{\BBOQ}\APACrefatitle {Human-level control through deep reinforcement learning}
  {Human-level control through deep reinforcement learning}.{\BBCQ}
\newblock
\APACjournalVolNumPages{Nature}{518}{7540}{529--533}.
\PrintBackRefs{\CurrentBib}

\bibitem [\protect \citeauthoryear {%
Murdock%
}{%
Murdock%
}{%
{\protect \APACyear {1962}}%
}]{%
Murd62}
\APACinsertmetastar {%
Murd62}%
\begin{APACrefauthors}%
Murdock, B\BPBI B.%
\end{APACrefauthors}%
\unskip\
\newblock
\APACrefYearMonthDay{1962}{}{}.
\newblock
{\BBOQ}\APACrefatitle {The serial position effect of free recall} {The serial
  position effect of free recall}.{\BBCQ}
\newblock
\APACjournalVolNumPages{Journal of Experimental Psychology}{64}{}{482-488}.
\PrintBackRefs{\CurrentBib}

\bibitem [\protect \citeauthoryear {%
Murdock%
}{%
Murdock%
}{%
{\protect \APACyear {1982}}%
}]{%
Murd82}
\APACinsertmetastar {%
Murd82}%
\begin{APACrefauthors}%
Murdock, B\BPBI B.%
\end{APACrefauthors}%
\unskip\
\newblock
\APACrefYearMonthDay{1982}{}{}.
\newblock
{\BBOQ}\APACrefatitle {A theory for the storage and retrieval of item and
  associative information} {A theory for the storage and retrieval of item and
  associative information}.{\BBCQ}
\newblock
\APACjournalVolNumPages{Psychological Review}{89}{}{609-626}.
\PrintBackRefs{\CurrentBib}

\bibitem [\protect \citeauthoryear {%
Murdock%
}{%
Murdock%
}{%
{\protect \APACyear {1997}}%
}]{%
Murd97}
\APACinsertmetastar {%
Murd97}%
\begin{APACrefauthors}%
Murdock, B\BPBI B.%
\end{APACrefauthors}%
\unskip\
\newblock
\APACrefYearMonthDay{1997}{}{}.
\newblock
{\BBOQ}\APACrefatitle {Context and mediators in a theory of distributed
  associative memory {(TODAM2)}} {Context and mediators in a theory of
  distributed associative memory {(TODAM2)}}.{\BBCQ}
\newblock
\APACjournalVolNumPages{Psychological Review}{104}{2}{839-862}.
\PrintBackRefs{\CurrentBib}

\bibitem [\protect \citeauthoryear {%
Nielson%
, Smith%
, Sreekumar%
, Dennis%
\BCBL {}\ \BBA {} Sederberg%
}{%
Nielson%
\ \protect \BOthers {.}}{%
{\protect \APACyear {2015}}%
}]{%
NielEtal15}
\APACinsertmetastar {%
NielEtal15}%
\begin{APACrefauthors}%
Nielson, D\BPBI M.%
, Smith, T\BPBI A.%
, Sreekumar, V.%
, Dennis, S.%
\BCBL {}\ \BBA {} Sederberg, P\BPBI B.%
\end{APACrefauthors}%
\unskip\
\newblock
\APACrefYearMonthDay{2015}{}{}.
\newblock
{\BBOQ}\APACrefatitle {Human hippocampus represents space and time during
  retrieval of real-world memories} {Human hippocampus represents space and
  time during retrieval of real-world memories}.{\BBCQ}
\newblock
\APACjournalVolNumPages{Proceedings of the National Academy of
  Sciences}{112}{35}{11078--11083}.
\PrintBackRefs{\CurrentBib}

\bibitem [\protect \citeauthoryear {%
Norman%
\ \BBA {} O'Reilly%
}{%
Norman%
\ \BBA {} O'Reilly%
}{%
{\protect \APACyear {2003}}%
}]{%
NormORei03}
\APACinsertmetastar {%
NormORei03}%
\begin{APACrefauthors}%
Norman, K\BPBI A.%
\BCBT {}\ \BBA {} O'Reilly, R\BPBI C.%
\end{APACrefauthors}%
\unskip\
\newblock
\APACrefYearMonthDay{2003}{}{}.
\newblock
{\BBOQ}\APACrefatitle {Modeling hippocampal and neocortical contributions to
  recognition memory: a complementary-learning-systems approach.} {Modeling
  hippocampal and neocortical contributions to recognition memory: a
  complementary-learning-systems approach.}{\BBCQ}
\newblock
\APACjournalVolNumPages{Psychological Review}{110}{4}{611-46}.
\PrintBackRefs{\CurrentBib}

\bibitem [\protect \citeauthoryear {%
Palombo%
, Di~Lascio%
, Howard%
\BCBL {}\ \BBA {} Verfaellie%
}{%
Palombo%
\ \protect \BOthers {.}}{%
{\protect \APACyear {2019}}%
}]{%
PaloEtal19}
\APACinsertmetastar {%
PaloEtal19}%
\begin{APACrefauthors}%
Palombo, D\BPBI J.%
, Di~Lascio, J\BPBI M.%
, Howard, M\BPBI W.%
\BCBL {}\ \BBA {} Verfaellie, M.%
\end{APACrefauthors}%
\unskip\
\newblock
\APACrefYearMonthDay{2019}{}{}.
\newblock
{\BBOQ}\APACrefatitle {Medial temporal lobe amnesia is associated with a
  deficit in recovering temporal context} {Medial temporal lobe amnesia is
  associated with a deficit in recovering temporal context}.{\BBCQ}
\newblock
\APACjournalVolNumPages{Journal of cognitive neuroscience}{31}{2}{236--248}.
\PrintBackRefs{\CurrentBib}

\bibitem [\protect \citeauthoryear {%
Pastalkova%
, Itskov%
, Amarasingham%
\BCBL {}\ \BBA {} Buzsaki%
}{%
Pastalkova%
\ \protect \BOthers {.}}{%
{\protect \APACyear {2008}}%
}]{%
PastEtal08}
\APACinsertmetastar {%
PastEtal08}%
\begin{APACrefauthors}%
Pastalkova, E.%
, Itskov, V.%
, Amarasingham, A.%
\BCBL {}\ \BBA {} Buzsaki, G.%
\end{APACrefauthors}%
\unskip\
\newblock
\APACrefYearMonthDay{2008}{}{}.
\newblock
{\BBOQ}\APACrefatitle {Internally generated cell assembly sequences in the rat
  hippocampus.} {Internally generated cell assembly sequences in the rat
  hippocampus.}{\BBCQ}
\newblock
\APACjournalVolNumPages{Science}{321}{5894}{1322-7}.
\PrintBackRefs{\CurrentBib}

\bibitem [\protect \citeauthoryear {%
Polyn%
\ \BBA {} Kahana%
}{%
Polyn%
\ \BBA {} Kahana%
}{%
{\protect \APACyear {2008}}%
}]{%
PolyKaha08}
\APACinsertmetastar {%
PolyKaha08}%
\begin{APACrefauthors}%
Polyn, S\BPBI M.%
\BCBT {}\ \BBA {} Kahana, M\BPBI J.%
\end{APACrefauthors}%
\unskip\
\newblock
\APACrefYearMonthDay{2008}{}{}.
\newblock
{\BBOQ}\APACrefatitle {Memory search and the neural representation of context.}
  {Memory search and the neural representation of context.}{\BBCQ}
\newblock
\APACjournalVolNumPages{Trends in Cognitive Science}{12}{1}{24-30}.
\PrintBackRefs{\CurrentBib}

\bibitem [\protect \citeauthoryear {%
Polyn%
, Norman%
\BCBL {}\ \BBA {} Kahana%
}{%
Polyn%
\ \protect \BOthers {.}}{%
{\protect \APACyear {2009}}%
}]{%
PolyEtal09}
\APACinsertmetastar {%
PolyEtal09}%
\begin{APACrefauthors}%
Polyn, S\BPBI M.%
, Norman, K\BPBI A.%
\BCBL {}\ \BBA {} Kahana, M\BPBI J.%
\end{APACrefauthors}%
\unskip\
\newblock
\APACrefYearMonthDay{2009}{}{}.
\newblock
{\BBOQ}\APACrefatitle {A context maintenance and retrieval model of
  organizational processes in free recall} {A context maintenance and retrieval
  model of organizational processes in free recall}.{\BBCQ}
\newblock
\APACjournalVolNumPages{Psychological Review}{116}{}{129-156}.
\PrintBackRefs{\CurrentBib}

\bibitem [\protect \citeauthoryear {%
Post%
}{%
Post%
}{%
{\protect \APACyear {1930}}%
}]{%
Post30}
\APACinsertmetastar {%
Post30}%
\begin{APACrefauthors}%
Post, E.%
\end{APACrefauthors}%
\unskip\
\newblock
\APACrefYearMonthDay{1930}{}{}.
\newblock
{\BBOQ}\APACrefatitle {Generalized Differentiation} {Generalized
  differentiation}.{\BBCQ}
\newblock
\APACjournalVolNumPages{Transactions of the American Mathematical
  Society}{32}{}{723-781}.
\PrintBackRefs{\CurrentBib}

\bibitem [\protect \citeauthoryear {%
Postman%
\ \BBA {} Phillips%
}{%
Postman%
\ \BBA {} Phillips%
}{%
{\protect \APACyear {1965}}%
}]{%
PostPhil65}
\APACinsertmetastar {%
PostPhil65}%
\begin{APACrefauthors}%
Postman, L.%
\BCBT {}\ \BBA {} Phillips, L\BPBI W.%
\end{APACrefauthors}%
\unskip\
\newblock
\APACrefYearMonthDay{1965}{}{}.
\newblock
{\BBOQ}\APACrefatitle {Short-term temporal changes in free recall} {Short-term
  temporal changes in free recall}.{\BBCQ}
\newblock
\APACjournalVolNumPages{Quarterly Journal of Experimental
  Psychology}{17}{}{132-138}.
\PrintBackRefs{\CurrentBib}

\bibitem [\protect \citeauthoryear {%
Quenon%
, de Xivry%
, Hanseeuw%
\BCBL {}\ \BBA {} Ivanoiu%
}{%
Quenon%
\ \protect \BOthers {.}}{%
{\protect \APACyear {2015}}%
}]{%
QuenEtal15}
\APACinsertmetastar {%
QuenEtal15}%
\begin{APACrefauthors}%
Quenon, L.%
, de Xivry, J\BHBI J\BPBI O.%
, Hanseeuw, B.%
\BCBL {}\ \BBA {} Ivanoiu, A.%
\end{APACrefauthors}%
\unskip\
\newblock
\APACrefYearMonthDay{2015}{}{}.
\newblock
{\BBOQ}\APACrefatitle {Investigating Associative Learning Effects in Patients
  with Prodromal Alzheimer's Disease Using the Temporal Context Model}
  {Investigating associative learning effects in patients with prodromal
  alzheimer's disease using the temporal context model}.{\BBCQ}
\newblock
\APACjournalVolNumPages{Journal of the International Neuropsychological
  Society}{21}{09}{699--708}.
\PrintBackRefs{\CurrentBib}

\bibitem [\protect \citeauthoryear {%
Raaijmakers%
\ \BBA {} Shiffrin%
}{%
Raaijmakers%
\ \BBA {} Shiffrin%
}{%
{\protect \APACyear {1980}}%
}]{%
RaaiShif80}
\APACinsertmetastar {%
RaaiShif80}%
\begin{APACrefauthors}%
Raaijmakers, J\BPBI G\BPBI W.%
\BCBT {}\ \BBA {} Shiffrin, R\BPBI M.%
\end{APACrefauthors}%
\unskip\
\newblock
\APACrefYearMonthDay{1980}{}{}.
\newblock
{\BBOQ}\APACrefatitle {{SAM: A} theory of probabilistic search of associative
  memory} {{SAM: A} theory of probabilistic search of associative
  memory}.{\BBCQ}
\newblock
\BIn{} G\BPBI H.~Bower\ (\BED), \APACrefbtitle {The psychology of learning and
  motivation: Advances in research and theory} {The psychology of learning and
  motivation: Advances in research and theory}\ (\BVOL~14, \BPG~207-262).
\newblock
\APACaddressPublisher{New York}{Academic Press}.
\PrintBackRefs{\CurrentBib}

\bibitem [\protect \citeauthoryear {%
Rescorla%
\ \BBA {} Wagner%
}{%
Rescorla%
\ \BBA {} Wagner%
}{%
{\protect \APACyear {1972}}%
}]{%
RescWagn72}
\APACinsertmetastar {%
RescWagn72}%
\begin{APACrefauthors}%
Rescorla, R\BPBI A.%
\BCBT {}\ \BBA {} Wagner, A\BPBI R.%
\end{APACrefauthors}%
\unskip\
\newblock
\APACrefYearMonthDay{1972}{}{}.
\newblock
{\BBOQ}\APACrefatitle {A theory of {P}avlovian conditioning: Variations in the
  efectivenesss of reinforcement and nonreinforcement} {A theory of {P}avlovian
  conditioning: Variations in the efectivenesss of reinforcement and
  nonreinforcement}.{\BBCQ}
\newblock
\BIn{} A\BPBI H.~Black\ \BBA {} W\BPBI F.~Prokasy\ (\BEDS), \APACrefbtitle
  {Classical conditioning {II}: Current research and theory.} {Classical
  conditioning {II}: Current research and theory.}
\newblock
\APACaddressPublisher{New York}{Appleton-Century-Crofts}.
\PrintBackRefs{\CurrentBib}

\bibitem [\protect \citeauthoryear {%
Rubin%
, Geva%
, Sheintuch%
\BCBL {}\ \BBA {} Ziv%
}{%
Rubin%
\ \protect \BOthers {.}}{%
{\protect \APACyear {2015}}%
}]{%
RubiEtal15}
\APACinsertmetastar {%
RubiEtal15}%
\begin{APACrefauthors}%
Rubin, A.%
, Geva, N.%
, Sheintuch, L.%
\BCBL {}\ \BBA {} Ziv, Y.%
\end{APACrefauthors}%
\unskip\
\newblock
\APACrefYearMonthDay{2015}{}{}.
\newblock
{\BBOQ}\APACrefatitle {Hippocampal ensemble dynamics timestamp events in
  long-term memory} {Hippocampal ensemble dynamics timestamp events in
  long-term memory}.{\BBCQ}
\newblock
\APACjournalVolNumPages{eLife}{4}{}{e12247}.
\PrintBackRefs{\CurrentBib}

\bibitem [\protect \citeauthoryear {%
Rule%
\ \protect \BOthers {.}}{%
Rule%
\ \protect \BOthers {.}}{%
{\protect \APACyear {2020}}%
}]{%
RuleEtal20}
\APACinsertmetastar {%
RuleEtal20}%
\begin{APACrefauthors}%
Rule, M\BPBI E.%
, Loback, A\BPBI R.%
, Raman, D\BPBI V.%
, Driscoll, L\BPBI N.%
, Harvey, C\BPBI D.%
\BCBL {}\ \BBA {} O'Leary, T.%
\end{APACrefauthors}%
\unskip\
\newblock
\APACrefYearMonthDay{2020}{}{}.
\newblock
{\BBOQ}\APACrefatitle {Stable task information from an unstable neural
  population} {Stable task information from an unstable neural
  population}.{\BBCQ}
\newblock
\APACjournalVolNumPages{Elife}{9}{}{e51121}.
\PrintBackRefs{\CurrentBib}

\bibitem [\protect \citeauthoryear {%
Rule%
, O'Leary%
\BCBL {}\ \BBA {} Harvey%
}{%
Rule%
\ \protect \BOthers {.}}{%
{\protect \APACyear {2019}}%
}]{%
RuleEtal19}
\APACinsertmetastar {%
RuleEtal19}%
\begin{APACrefauthors}%
Rule, M\BPBI E.%
, O'Leary, T.%
\BCBL {}\ \BBA {} Harvey, C\BPBI D.%
\end{APACrefauthors}%
\unskip\
\newblock
\APACrefYearMonthDay{2019}{}{}.
\newblock
{\BBOQ}\APACrefatitle {Causes and consequences of representational drift}
  {Causes and consequences of representational drift}.{\BBCQ}
\newblock
\APACjournalVolNumPages{Current opinion in neurobiology}{58}{}{141--147}.
\PrintBackRefs{\CurrentBib}

\bibitem [\protect \citeauthoryear {%
Schoonover%
, Ohashi%
, Axel%
\BCBL {}\ \BBA {} Fink%
}{%
Schoonover%
\ \protect \BOthers {.}}{%
{\protect \APACyear {2021}}%
}]{%
SchnEtal21}
\APACinsertmetastar {%
SchnEtal21}%
\begin{APACrefauthors}%
Schoonover, C\BPBI E.%
, Ohashi, S\BPBI N.%
, Axel, R.%
\BCBL {}\ \BBA {} Fink, A\BPBI J.%
\end{APACrefauthors}%
\unskip\
\newblock
\APACrefYearMonthDay{2021}{}{}.
\newblock
{\BBOQ}\APACrefatitle {Representational drift in primary olfactory cortex}
  {Representational drift in primary olfactory cortex}.{\BBCQ}
\newblock
\APACjournalVolNumPages{Nature}{}{}{1--6}.
\PrintBackRefs{\CurrentBib}

\bibitem [\protect \citeauthoryear {%
Schultz%
, Dayan%
\BCBL {}\ \BBA {} Montague%
}{%
Schultz%
\ \protect \BOthers {.}}{%
{\protect \APACyear {1997}}%
}]{%
SchuEtal97}
\APACinsertmetastar {%
SchuEtal97}%
\begin{APACrefauthors}%
Schultz, W.%
, Dayan, P.%
\BCBL {}\ \BBA {} Montague, P\BPBI R.%
\end{APACrefauthors}%
\unskip\
\newblock
\APACrefYearMonthDay{1997}{}{}.
\newblock
{\BBOQ}\APACrefatitle {A neural substrate of prediction and reward} {A neural
  substrate of prediction and reward}.{\BBCQ}
\newblock
\APACjournalVolNumPages{Science}{275}{}{1593-1599}.
\PrintBackRefs{\CurrentBib}

\bibitem [\protect \citeauthoryear {%
Sederberg%
, Howard%
\BCBL {}\ \BBA {} Kahana%
}{%
Sederberg%
\ \protect \BOthers {.}}{%
{\protect \APACyear {2008}}%
}]{%
SedeEtal08}
\APACinsertmetastar {%
SedeEtal08}%
\begin{APACrefauthors}%
Sederberg, P\BPBI B.%
, Howard, M\BPBI W.%
\BCBL {}\ \BBA {} Kahana, M\BPBI J.%
\end{APACrefauthors}%
\unskip\
\newblock
\APACrefYearMonthDay{2008}{}{}.
\newblock
{\BBOQ}\APACrefatitle {A context-based theory of recency and contiguity in free
  recall} {A context-based theory of recency and contiguity in free
  recall}.{\BBCQ}
\newblock
\APACjournalVolNumPages{Psychological Review}{115}{}{893-912}.
\PrintBackRefs{\CurrentBib}

\bibitem [\protect \citeauthoryear {%
Shankar%
\ \BBA {} Howard%
}{%
Shankar%
\ \BBA {} Howard%
}{%
{\protect \APACyear {2010}}%
}]{%
ShanHowa10}
\APACinsertmetastar {%
ShanHowa10}%
\begin{APACrefauthors}%
Shankar, K\BPBI H.%
\BCBT {}\ \BBA {} Howard, M\BPBI W.%
\end{APACrefauthors}%
\unskip\
\newblock
\APACrefYearMonthDay{2010}{}{}.
\newblock
{\BBOQ}\APACrefatitle {Timing using temporal context} {Timing using temporal
  context}.{\BBCQ}
\newblock
\APACjournalVolNumPages{Brain Research}{1365}{}{3-17}.
\PrintBackRefs{\CurrentBib}

\bibitem [\protect \citeauthoryear {%
Shankar%
\ \BBA {} Howard%
}{%
Shankar%
\ \BBA {} Howard%
}{%
{\protect \APACyear {2012}}%
}]{%
ShanHowa12}
\APACinsertmetastar {%
ShanHowa12}%
\begin{APACrefauthors}%
Shankar, K\BPBI H.%
\BCBT {}\ \BBA {} Howard, M\BPBI W.%
\end{APACrefauthors}%
\unskip\
\newblock
\APACrefYearMonthDay{2012}{}{}.
\newblock
{\BBOQ}\APACrefatitle {A scale-invariant internal representation of time} {A
  scale-invariant internal representation of time}.{\BBCQ}
\newblock
\APACjournalVolNumPages{Neural Computation}{24}{1}{134-193}.
\PrintBackRefs{\CurrentBib}

\bibitem [\protect \citeauthoryear {%
Shankar%
\ \BBA {} Howard%
}{%
Shankar%
\ \BBA {} Howard%
}{%
{\protect \APACyear {2013}}%
}]{%
ShanHowa13}
\APACinsertmetastar {%
ShanHowa13}%
\begin{APACrefauthors}%
Shankar, K\BPBI H.%
\BCBT {}\ \BBA {} Howard, M\BPBI W.%
\end{APACrefauthors}%
\unskip\
\newblock
\APACrefYearMonthDay{2013}{}{}.
\newblock
{\BBOQ}\APACrefatitle {Optimally fuzzy temporal memory} {Optimally fuzzy
  temporal memory}.{\BBCQ}
\newblock
\APACjournalVolNumPages{Journal of Machine Learning Research}{14}{}{3753-3780}.
\PrintBackRefs{\CurrentBib}

\bibitem [\protect \citeauthoryear {%
Shepard%
}{%
Shepard%
}{%
{\protect \APACyear {1987}}%
}]{%
Shep87}
\APACinsertmetastar {%
Shep87}%
\begin{APACrefauthors}%
Shepard, R\BPBI N.%
\end{APACrefauthors}%
\unskip\
\newblock
\APACrefYearMonthDay{1987}{}{}.
\newblock
{\BBOQ}\APACrefatitle {Toward a universal law of generalization for
  psychological science} {Toward a universal law of generalization for
  psychological science}.{\BBCQ}
\newblock
\APACjournalVolNumPages{Science}{237}{4820}{1317--1323}.
\PrintBackRefs{\CurrentBib}

\bibitem [\protect \citeauthoryear {%
Shiffrin%
\ \BBA {} Steyvers%
}{%
Shiffrin%
\ \BBA {} Steyvers%
}{%
{\protect \APACyear {1997}}%
}]{%
ShifStey97}
\APACinsertmetastar {%
ShifStey97}%
\begin{APACrefauthors}%
Shiffrin, R\BPBI M.%
\BCBT {}\ \BBA {} Steyvers, M.%
\end{APACrefauthors}%
\unskip\
\newblock
\APACrefYearMonthDay{1997}{}{}.
\newblock
{\BBOQ}\APACrefatitle {A Model for Recognition Memory: {REM} --- Retrieving
  Effectively From Memory} {A model for recognition memory: {REM} ---
  retrieving effectively from memory}.{\BBCQ}
\newblock
\APACjournalVolNumPages{Psychonomic Bulletin and Review}{4}{}{145-166}.
\PrintBackRefs{\CurrentBib}

\bibitem [\protect \citeauthoryear {%
Silver%
\ \protect \BOthers {.}}{%
Silver%
\ \protect \BOthers {.}}{%
{\protect \APACyear {2016}}%
}]{%
SilvEtal16}
\APACinsertmetastar {%
SilvEtal16}%
\begin{APACrefauthors}%
Silver, D.%
, Huang, A.%
, Maddison, C\BPBI J.%
, Guez, A.%
, Sifre, L.%
, Van Den~Driessche, G.%
\BDBL {}others%
\end{APACrefauthors}%
\unskip\
\newblock
\APACrefYearMonthDay{2016}{}{}.
\newblock
{\BBOQ}\APACrefatitle {Mastering the game of Go with deep neural networks and
  tree search} {Mastering the game of go with deep neural networks and tree
  search}.{\BBCQ}
\newblock
\APACjournalVolNumPages{Nature}{529}{7587}{484--489}.
\PrintBackRefs{\CurrentBib}

\bibitem [\protect \citeauthoryear {%
Sutton%
\ \BBA {} Barto%
}{%
Sutton%
\ \BBA {} Barto%
}{%
{\protect \APACyear {1981}}%
}]{%
SuttBart81}
\APACinsertmetastar {%
SuttBart81}%
\begin{APACrefauthors}%
Sutton, R\BPBI S.%
\BCBT {}\ \BBA {} Barto, A\BPBI G.%
\end{APACrefauthors}%
\unskip\
\newblock
\APACrefYearMonthDay{1981}{}{}.
\newblock
{\BBOQ}\APACrefatitle {Toward a modern theory of adaptive networks: Expectation
  and Prediction} {Toward a modern theory of adaptive networks: Expectation and
  prediction}.{\BBCQ}
\newblock
\APACjournalVolNumPages{Psychological Review}{88}{}{135-171}.
\PrintBackRefs{\CurrentBib}

\bibitem [\protect \citeauthoryear {%
Talamonti%
, Koscik%
, Johnson%
\BCBL {}\ \BBA {} Bruno%
}{%
Talamonti%
\ \protect \BOthers {.}}{%
{\protect \APACyear {2021}}%
}]{%
TalaEtal21}
\APACinsertmetastar {%
TalaEtal21}%
\begin{APACrefauthors}%
Talamonti, D.%
, Koscik, R.%
, Johnson, S.%
\BCBL {}\ \BBA {} Bruno, D.%
\end{APACrefauthors}%
\unskip\
\newblock
\APACrefYearMonthDay{2021}{}{}.
\newblock
{\BBOQ}\APACrefatitle {Temporal contiguity and ageing: The role of memory
  organization in cognitive decline} {Temporal contiguity and ageing: The role
  of memory organization in cognitive decline}.{\BBCQ}
\newblock
\APACjournalVolNumPages{Journal of Neuropsychology}{15}{}{53--65}.
\PrintBackRefs{\CurrentBib}

\bibitem [\protect \citeauthoryear {%
Tiganj%
, Cromer%
, Roy%
, Miller%
\BCBL {}\ \BBA {} Howard%
}{%
Tiganj%
\ \protect \BOthers {.}}{%
{\protect \APACyear {2018}}%
}]{%
TigaEtal18a}
\APACinsertmetastar {%
TigaEtal18a}%
\begin{APACrefauthors}%
Tiganj, Z.%
, Cromer, J\BPBI A.%
, Roy, J\BPBI E.%
, Miller, E\BPBI K.%
\BCBL {}\ \BBA {} Howard, M\BPBI W.%
\end{APACrefauthors}%
\unskip\
\newblock
\APACrefYearMonthDay{2018}{}{}.
\newblock
{\BBOQ}\APACrefatitle {Compressed timeline of recent experience in monkey
  {lPFC}} {Compressed timeline of recent experience in monkey {lPFC}}.{\BBCQ}
\newblock
\APACjournalVolNumPages{Journal of Cognitive Neuroscience}{30}{}{935-950}.
\PrintBackRefs{\CurrentBib}

\bibitem [\protect \citeauthoryear {%
Tiganj%
, Cruzado%
\BCBL {}\ \BBA {} Howard%
}{%
Tiganj%
\ \protect \BOthers {.}}{%
{\protect \APACyear {2019}}%
}]{%
TigaEtal19}
\APACinsertmetastar {%
TigaEtal19}%
\begin{APACrefauthors}%
Tiganj, Z.%
, Cruzado, N\BPBI A.%
\BCBL {}\ \BBA {} Howard, M\BPBI W.%
\end{APACrefauthors}%
\unskip\
\newblock
\APACrefYearMonthDay{2019}{}{}.
\newblock
{\BBOQ}\APACrefatitle {Towards a neural-level cognitive architecture: modeling
  behavior in working memory tasks with neurons} {Towards a neural-level
  cognitive architecture: modeling behavior in working memory tasks with
  neurons}.{\BBCQ}
\newblock
\BIn{} A.~Goel, C.~Seifert\BCBL {}\ \BBA {} C.~Freksa\ (\BEDS), \APACrefbtitle
  {Proceedings of the 41st Annual Conference of the Cognitive Science Society}
  {Proceedings of the 41st annual conference of the cognitive science society}\
  (\BPG~1118-1123).
\newblock
\APACaddressPublisher{Montreal}{Cognitive Science Society}.
\PrintBackRefs{\CurrentBib}

\bibitem [\protect \citeauthoryear {%
Tiganj%
, Hasselmo%
\BCBL {}\ \BBA {} Howard%
}{%
Tiganj%
\ \protect \BOthers {.}}{%
{\protect \APACyear {2015}}%
}]{%
TigaEtal15}
\APACinsertmetastar {%
TigaEtal15}%
\begin{APACrefauthors}%
Tiganj, Z.%
, Hasselmo, M\BPBI E.%
\BCBL {}\ \BBA {} Howard, M\BPBI W.%
\end{APACrefauthors}%
\unskip\
\newblock
\APACrefYearMonthDay{2015}{}{}.
\newblock
{\BBOQ}\APACrefatitle {A Simple biophysically plausible model for long time
  constants in single neurons} {A simple biophysically plausible model for long
  time constants in single neurons}.{\BBCQ}
\newblock
\APACjournalVolNumPages{Hippocampus}{25}{1}{27-37}.
\PrintBackRefs{\CurrentBib}

\bibitem [\protect \citeauthoryear {%
Tiganj%
, Kim%
, Jung%
\BCBL {}\ \BBA {} Howard%
}{%
Tiganj%
\ \protect \BOthers {.}}{%
{\protect \APACyear {2017}}%
}]{%
TigaEtal17b}
\APACinsertmetastar {%
TigaEtal17b}%
\begin{APACrefauthors}%
Tiganj, Z.%
, Kim, J.%
, Jung, M\BPBI W.%
\BCBL {}\ \BBA {} Howard, M\BPBI W.%
\end{APACrefauthors}%
\unskip\
\newblock
\APACrefYearMonthDay{2017}{}{}.
\newblock
{\BBOQ}\APACrefatitle {Sequential firing codes for time in rodent {mPFC}}
  {Sequential firing codes for time in rodent {mPFC}}.{\BBCQ}
\newblock
\APACjournalVolNumPages{Cerebral Cortex}{27}{}{5663-5671}.
\PrintBackRefs{\CurrentBib}

\bibitem [\protect \citeauthoryear {%
Trutti%
, Verschooren%
, Forstmann%
\BCBL {}\ \BBA {} Boag%
}{%
Trutti%
\ \protect \BOthers {.}}{%
{\protect \APACyear {2021}}%
}]{%
TrutEtal21}
\APACinsertmetastar {%
TrutEtal21}%
\begin{APACrefauthors}%
Trutti, A\BPBI C.%
, Verschooren, S.%
, Forstmann, B\BPBI U.%
\BCBL {}\ \BBA {} Boag, R\BPBI J.%
\end{APACrefauthors}%
\unskip\
\newblock
\APACrefYearMonthDay{2021}{}{}.
\newblock
{\BBOQ}\APACrefatitle {Understanding subprocesses of working memory through the
  lens of model-based cognitive neuroscience} {Understanding subprocesses of
  working memory through the lens of model-based cognitive
  neuroscience}.{\BBCQ}
\newblock
\APACjournalVolNumPages{Current Opinion in Behavioral Sciences}{38}{}{57--65}.
\PrintBackRefs{\CurrentBib}

\bibitem [\protect \citeauthoryear {%
Tsao%
\ \protect \BOthers {.}}{%
Tsao%
\ \protect \BOthers {.}}{%
{\protect \APACyear {2018}}%
}]{%
TsaoEtal18}
\APACinsertmetastar {%
TsaoEtal18}%
\begin{APACrefauthors}%
Tsao, A.%
, Sugar, J.%
, Lu, L.%
, Wang, C.%
, Knierim, J\BPBI J.%
, Moser, M\BHBI B.%
\BCBL {}\ \BBA {} Moser, E\BPBI I.%
\end{APACrefauthors}%
\unskip\
\newblock
\APACrefYearMonthDay{2018}{}{}.
\newblock
{\BBOQ}\APACrefatitle {Integrating time from experience in the lateral
  entorhinal cortex} {Integrating time from experience in the lateral
  entorhinal cortex}.{\BBCQ}
\newblock
\APACjournalVolNumPages{Nature}{561}{}{57-62}.
\PrintBackRefs{\CurrentBib}

\bibitem [\protect \citeauthoryear {%
Tulving%
}{%
Tulving%
}{%
{\protect \APACyear {1983}}%
}]{%
Tulv83}
\APACinsertmetastar {%
Tulv83}%
\begin{APACrefauthors}%
Tulving, E.%
\end{APACrefauthors}%
\unskip\
\newblock
\APACrefYear{1983}.
\newblock
\APACrefbtitle {Elements of Episodic Memory} {Elements of episodic memory}.
\newblock
\APACaddressPublisher{New York}{Oxford}.
\PrintBackRefs{\CurrentBib}

\bibitem [\protect \citeauthoryear {%
Uitvlugt%
\ \BBA {} Healey%
}{%
Uitvlugt%
\ \BBA {} Healey%
}{%
{\protect \APACyear {2019}}%
}]{%
UitvHeal19}
\APACinsertmetastar {%
UitvHeal19}%
\begin{APACrefauthors}%
Uitvlugt, M\BPBI G.%
\BCBT {}\ \BBA {} Healey, M\BPBI K.%
\end{APACrefauthors}%
\unskip\
\newblock
\APACrefYearMonthDay{2019}{}{}.
\newblock
{\BBOQ}\APACrefatitle {Temporal proximity links unrelated news events in
  memory} {Temporal proximity links unrelated news events in memory}.{\BBCQ}
\newblock
\APACjournalVolNumPages{Psychological science}{30}{1}{92--104}.
\PrintBackRefs{\CurrentBib}

\bibitem [\protect \citeauthoryear {%
Umbach%
\ \protect \BOthers {.}}{%
Umbach%
\ \protect \BOthers {.}}{%
{\protect \APACyear {2020}}%
}]{%
UmbaEtal20}
\APACinsertmetastar {%
UmbaEtal20}%
\begin{APACrefauthors}%
Umbach, G.%
, Kantak, P.%
, Jacobs, J.%
, Kahana, M\BPBI J.%
, Pfeiffer, B\BPBI E.%
, Sperling, M.%
\BCBL {}\ \BBA {} Lega, B.%
\end{APACrefauthors}%
\unskip\
\newblock
\APACrefYearMonthDay{2020}{}{}.
\newblock
{\BBOQ}\APACrefatitle {Time cells in the human hippocampus and entorhinal
  cortex support episodic memory} {Time cells in the human hippocampus and
  entorhinal cortex support episodic memory}.{\BBCQ}
\newblock
\APACjournalVolNumPages{Proceedings of the National Academy of
  Sciences}{117}{}{28463--28474}.
\PrintBackRefs{\CurrentBib}

\bibitem [\protect \citeauthoryear {%
Yaffe%
\ \protect \BOthers {.}}{%
Yaffe%
\ \protect \BOthers {.}}{%
{\protect \APACyear {2014}}%
}]{%
YaffEtal14}
\APACinsertmetastar {%
YaffEtal14}%
\begin{APACrefauthors}%
Yaffe, R\BPBI B.%
, Kerr, M\BPBI S\BPBI D.%
, Damera, S.%
, Sarma, S\BPBI V.%
, Inati, S\BPBI K.%
\BCBL {}\ \BBA {} Zaghloul, K\BPBI A.%
\end{APACrefauthors}%
\unskip\
\newblock
\APACrefYearMonthDay{2014}{}{}.
\newblock
{\BBOQ}\APACrefatitle {Reinstatement of distributed cortical oscillations
  occurs with precise spatiotemporal dynamics during successful memory
  retrieval} {Reinstatement of distributed cortical oscillations occurs with
  precise spatiotemporal dynamics during successful memory retrieval}.{\BBCQ}
\newblock
\APACjournalVolNumPages{Proceedings of the National Academy of
  Sciences}{111}{52}{18727-32}.
\newblock
\begin{APACrefDOI} \doi{10.1073/pnas.1417017112} \end{APACrefDOI}
\PrintBackRefs{\CurrentBib}

\bibitem [\protect \citeauthoryear {%
Zeithamova%
, Dominick%
\BCBL {}\ \BBA {} Preston%
}{%
Zeithamova%
\ \protect \BOthers {.}}{%
{\protect \APACyear {2012}}%
}]{%
ZeitEtal12}
\APACinsertmetastar {%
ZeitEtal12}%
\begin{APACrefauthors}%
Zeithamova, D.%
, Dominick, A\BPBI L.%
\BCBL {}\ \BBA {} Preston, A\BPBI R.%
\end{APACrefauthors}%
\unskip\
\newblock
\APACrefYearMonthDay{2012}{}{}.
\newblock
{\BBOQ}\APACrefatitle {Hippocampal and ventral medial prefrontal activation
  during retrieval-mediated learning supports novel inference} {Hippocampal and
  ventral medial prefrontal activation during retrieval-mediated learning
  supports novel inference}.{\BBCQ}
\newblock
\APACjournalVolNumPages{Neuron}{75}{1}{168-179}.
\PrintBackRefs{\CurrentBib}

\end{thebibliography}

\end{document}